\documentclass{article}

\usepackage{amsthm}
\RequirePackage[authoryear]{natbib}
\usepackage[latin1]{inputenc} 
\usepackage[fleqn]{amsmath}
\usepackage{amsfonts}
\usepackage{amssymb}
\usepackage{graphicx}
\usepackage{listings}

\usepackage{relsize}
\usepackage{cite}
\usepackage{multirow}
\usepackage{tabu}
\usepackage{booktabs}
\usepackage{changepage}
\usepackage{array}
\usepackage{pdfpages}
\usepackage{microtype} 
 \usepackage[margin=1in]{geometry}

\usepackage{algorithm}
\usepackage{algorithmicx}
\usepackage{algpseudocode}
\usepackage[parfill]{parskip} 
\usepackage[title,titletoc]{appendix}
\allowdisplaybreaks
\usepackage{tikz, amsthm, enumitem, tikz-cd, bm, xfrac, mathtools}
\usepackage{csquotes}

\usepackage{etoolbox}

\makeatletter

\DeclareMathOperator{\diag}{diag}
\DeclareMathOperator*{\argmin}{argmin}
\def\utilde#1{\mathord{\vtop{\ialign{##\crcr
$\hfil\displaystyle{#1}\hfil$\crcr\noalign{\kern1.5pt\nointerlineskip}
$\hfil\tilde{}\hfil$\crcr\noalign{\kern1.5pt}}}}}

\usepackage{textcomp}

\usepackage{setspace}

\newcommand{\RR}{\mathbb{R}}
\newcommand{\normal}{\mathcal{N}}
\newcommand{\NN}{\mathbb{N}}
\newcommand{\lik}{\mathcal{L}}

\newcommand{\Exp}{\mathbb{E}}
\newcommand{\sighat}{\hat{\Sigma}_s}
\newcommand{\sighatone}{\hat{\Sigma}_1}
\newcommand{\sighatS}{\hat{\Sigma}_S}
\newcommand{\rank}{\text{rank}}
\newcommand{\Var}{\text{Var}}

\usepackage[hidelinks]{hyperref}
\hypersetup{
    colorlinks=true,
    citecolor = black,
    linkcolor = black, 
    urlcolor = black,
    }
\usepackage{cleveref}
\crefname{equation}{equation}{equations}
\Crefname{equation}{equation}{equations}

\crefname{table}{Table}{Tables}
\crefname{figure}{Figure}{Figures}

\begin{document}


\title{Cross-study analyses of microbial abundance using generalized common factor methods}

\date{\vspace{-5ex}}
\author{Molly G. Hayes$^1$, Morgan G. I. Langille$^{2,3}$,  Hong Gu$^1$ }

\maketitle
$^1$Department of Mathematics \& Statistics, Dalhousie University, Halifax NS, Canada\\                    
$^2$Department of Microbiology \& Immunology, Dalhousie University, Halifax NS, Canada\\
$^3$Department of Pharmacology, Dalhousie University, Halifax NS, Canada

\begin{abstract}
By creating networks of biochemical pathways, communities of micro-organisms are able to modulate the properties of their environment and even the metabolic processes within their hosts. Next-generation high-throughput sequencing has led to a new frontier in microbial ecology, promising the ability to leverage the microbiome to make crucial advancements in the environmental and biomedical sciences. However, this is challenging, as genomic data are high-dimensional, sparse, and noisy. Much of this noise reflects the exact conditions under which sequencing took place, and is so significant that it limits consensus-based validation of study results. We propose an ensemble approach for cross-study exploratory analyses of microbial abundance data in which we first estimate the variance-covariance matrix of the underlying abundances from each dataset on the log scale assuming Poisson sampling, and subsequently model these covariances jointly so as to find a shared low-dimensional subspace of the feature space. By viewing the projection of the latent true abundances onto this common structure, the variation is pared down to that which is shared among all datasets, and is likely to reflect more generalizable biological signal than can be inferred from individual datasets. We investigate several ways of achieving this, and demonstrate that they work well on simulated and real metagenomic data in terms of signal retention and interpretability. 


\end{abstract}



\section*{Background}

Communities of microbes inhabit all areas of the environment, including the body cavities and exterior surfaces of larger organisms. A microbiome can be thought of as a community of shared genes and metabolic pathways that acts as a complex system in ways defined in part by their composition, or which microbial taxa are present and in what abundance \citep{boon}. These communities vary widely in composition, and certain patterns of colonization seem characteristic across individuals---and analogously across environmental sites---given similar conditions. In fact, mounting evidence suggests that micro-organisms comprising the gut microbiome interact with host systems in myriad ways, and that understanding these associations could have profound implications for our ability to predict, diagnose, or treat pathologies \citep{blaser,nieuwdorp,thaiss}. Similarly, the environmental microbiota, such as the communities characterizing a particular stratum of the soil or a particular water column, have enormous influence on local physiochemical conditions with far-reaching implications for ecology, agriculture, fisheries, and biotechnology. Moreover, the study of micro-organisms, their community dynamics, and their interactions with the cellular systems of their hosts continues to help researchers elucidate the origin of life on Earth. 

Two of the most common ways to characterize the composition of the microbiome involve the use of high-throughput next-generation sequencing technologies, followed by bioinformatic algorithms that allow researchers to ultimately obtain counts of the observed representatives of each microbial taxon in each sample. With 16S targeted gene (amplicon) sequencing, only the counts of the bacterial constituents of the sample are available, while shotgun metagenomics allows us to infer the full taxonomic profile and also permits analysis of functional pathways in the community. Just like in macro-ecology, the distribution of functions, or niches of particular taxa, can be highly specific to the conditions of an individual community, and things are further complicated in the microbiome by the fact that genes are also laterally transferred between micro-organisms. While taxonomy may not be able to tell the whole story, in this paper we restrict our attention to analyzing the microbiome by taxonomic rather than functional profiling. While taxonomy-based studies have some drawbacks, they also have benefits significant enough that they are likely to remain popular for the foreseeable future \citep{langille}. Also, note that because taxonomic abundance data from 16S sequencing (be it in the form of operational taxonomic units or amplicon sequence variants) and metagenomic sequencing each pose similar obstacles to classical statistical analyses, the exploratory methods we propose later in this paper apply to both.

Regardless of sequencing protocol, the output comprises sequences of base pairs from a random sample of the total collection of genes in the community. This means that we only observe a given count of each taxon, and not its true abundance in the community. In addition, the ``depth" of sequencing, or the average number of reads in a sample that align to a known reference, varies significantly by sample and is thus a source of multiplicative error on the counts. Since many short fragments of a sequence have to be read and aligned with each other in order for that sequence to be recognizable, samples with lower sequencing depth have lower observed counts and more uncertainty. Furthermore, thousands of microbial taxa can be present in a single sample, many of which are present in extremely low numbers, while the number of samples is---as in any experiment---limited by practical constraints such as cost and participation. As a result, each dataset has far more features than samples \citep{kurtz}, and in a given sample there will be zero instances of many taxa. The high dimension and sparsity of these data invalidate most existing methods of inferring factors associated with large variation among conditions \citep{sill}. There is also evidence that the ``compositional effects" \citep*{aitchison} arising from sampling can cause spurious correlations when distributions of taxa are unbalanced. Despite the promise held by microbial abundance data, analysis is so statistically challenging \citep{tsilimigras} that scope for application is currently limited.

As it happens, further complications arise when genomic samples from different studies are compared with one another. In order to sequence DNA, it first has to be isolated from a sample, fragmented, and potentially amplified. Each of these processes requires a number of laboratory techniques and reagents, and procedures vary substantially between labs. Sequencing platforms also differ, and presumably there are also machine calibration differences between two sequencers of the same model. The result is that when two studies of similar design look at samples of similar origin, or even when identical samples are sent to two different labs for sequencing, the signal patterns are very different due to dominating ``batch effects". This noise persists even under highly controlled conditions \citep{oytam} and can obscure the signal of interest: for example, machine learning classifiers enjoying good within-study performance may become grossly inaccurate when applied cross-study \citep{sze}. Batch effects impair our ability to determine whether results generalize to other cohorts, and preclude meaningful validation and meta-analysis \citep{buhule,leek2,miller}.

During statistical analysis, microbial abundances have often been modelled as if they were continuous by computing proportions of observed counts to the read depth of the sample (we will call this ``relative abundance" data). Some workflows instead rarefy counts, which sacrifices observed data in order to equalize read depth. Hence, two generally acknowledged (but not necessarily enacted) recommendations for better statistical treatment of abundance data are 1) that an appropriate discrete generating distribution be used to model the sampling of counts, such as Poisson, negative binomial, or multinomial, and 2) that sequencing depth error be treated within a statistical framework \citep{mcmurdie}. Additionally, it is well supported in all fields of ecology that the logarithmic scale can be useful when modeling populations of organisms in a community \citep{olesen}. However, the issue of batch effects has not been as thoroughly investigated. 

In the RNA microarray literature, several approaches to correcting batch effects have been proposed. The most popular of these, ComBat \citep{johnson}, performs gene-wise Bayesian location-scale adjustment. Several methods that combine regression and singular value decomposition have also been proposed, such as surrogate variable analysis \citep*{leek} and RUV-4 \citep{gagnon}, aiming to project away noise, which is identified as such based on gene expression signatures gleaned from regression. However, microarray data are very different from microbial abundance data; critically, we have no equivalent of ``housekeeping genes" with which to base inferences about signal source. With the goal of pooling data across case-control microbial abundance studies, \citet{gibbons} proposed a within-study non-parametric normalization technique in which abundance of taxa in case samples are converted to percentiles of the abundance of equivalent taxa in control samples. However, their results are based on naive relative abundance models, which were run on a subset of taxa chosen in an ad hoc fashion (i.e., those that occurred in at least one third of case samples or one third of control samples). Multi-study factor analysis \citep[MSFA;][]{deveet} and Bayesian MSFA \citep{deveet2} extend classical factor analysis to multiple groups to decompose features into factors that reflect shared vs. group-specific variability. However, like classical factor analysis, these methods decompose the naive sample variances of the observed data, which cannot capture the covariance structure of microbial abundances. \citet{argelaguet2} proposed MOFA+, a multi-group multi-omics Bayesian factor analysis method that can consider multiple data types simultaneously in addition to multiple studies or sample groups, but MOFA+ \citep[like its predecessor, MOFA;][]{argelaguet} is unsuitable for microbial abundance data due to the use of an inappropriate transformation in the Poisson case. Most recently, \citet{liu} performed multi-group decomposition of correlations estimated by latent Gaussian copula models, but again this method lacks the machinery to address count data with multiplicative error.

This brings us to our purpose, which is to address the broadly meta-analytic difficulties presented by batch effects or technical variation in high-throughput microbial abundance data. We will operate under the assumption that signal shared among $S$ different datasets on the same $p$ variables is likely to represent biological variability of interest, whereas disparate signal likely reflects variability attributable only to irrelevant differences in experimental conditions. Accordingly, we propose that if the variances $\Sigma_1,\dots,\Sigma_S$ are estimated reasonably well from individual datasets using an existing method, then an existing multi-group method can then be applied to the variance estimates $\sighatone,\dots,\sighatS$ simultaneously to remove unshared variation. We assume Poisson sampling, estimate quantities on the logarithmic scale, and account for sequencing depth. We deal with sparsity and the high-dimensional feature space by analyzing abundance at the genus level, which is a trade-off we make in order to obtain high-quality variance estimates.

For this two-step ensemble method, we will consider two possible approaches to the variance estimation step, each of which assumes that the observations are conditionally Poisson-distributed. Since our application is the analysis of microbiome composition, we treat the latent means $\Lambda_{is}$, $i=1,\dots,n_s$, $s=1,\dots,S$ on the logarithmic scale, which arises naturally in the first method via the canonical link function, and in the second method can be achieved by a transformation. The first approach is to leverage the Poisson log-normal PCA \citep[PLNPCA;][]{chiquet}, a fully parametric model in which the log Poisson mean is a function of a latent variable and follows a multivariate normal distribution, and the observed counts are---given the log means---independently Poisson-distributed. Sequencing depth can be treated as an offset in this model. The second approach is to use Poisson measurement error corrected PCA \citep*[PoissonPCA;][]{kenney} to sidestep a complex likelihood-based model in favor of assuming only that the observed counts are---given the latent Poisson means---independently Poisson-distributed, potentially including sequencing depth as a nuisance random variable. The authors derived an unbiased variance estimator for any non-linear transformation of the latent means, and we make use of the log-transformed case. See the Methods for details on PLNPCA and PoissonPCA.

After obtaining estimated variance-covariance matrices for each dataset, the second step of the ensemble method is to simultaneously decompose them to find a $q$-dimensional basis that is common to all groups, and we compare two general approaches by which to perform this step. One possible approach is a multi-group extension of PCA called common principal components analysis (CPCA), first described by \citet{flury}, which assumes that there exists an orthogonal matrix that can approximately diagonalize the covariance matrices of all $S$ groups simultaneously. We test two algorithms for CPCA: Flury's CPCA (FCPCA) and stepwise CPCA \citep[SCPCA;][]{trendafilov}. The other model under consideration is the previously mentioned MSFA \citep{deveet}, which is an extension of classical factor analysis. Both CPCA and MSFA require Wishart or multivariate normal assumptions, which our variance estimates satisfy because we treat abundances on the log scale. See Methods for more details on FCPCA, SCPCA, and MSFA. After estimating the common loadings, the final step of our proposed ensemble method is to express the underlying abundances in each multivariate observation with respect to our new common basis, which we will do by projecting the estimated latent Poisson means from each group into a common subspace spanned by the loadings using a quasi-likelihood procedure that \citet{kenney} developed for the single-group case. The remainder of the paper will demonstrate how this ensemble method can eliminate most technical noise while retaining shared biological signal, facilitating novel exploratory findings.




\section*{Results}

\subsection{Simulation Study}
\begin{figure}
    \centering
    \includegraphics[scale=0.2]{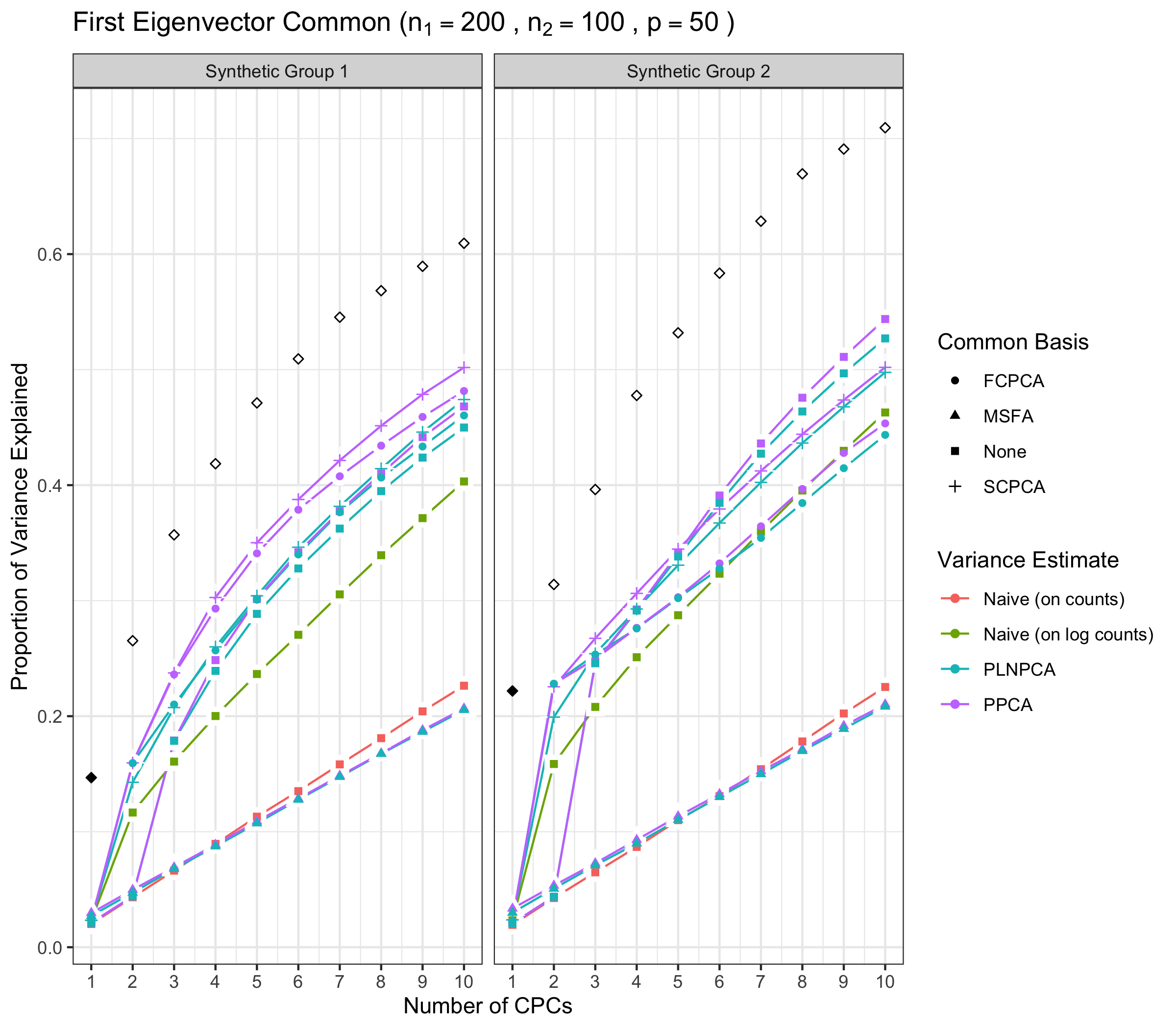}
    \caption{Simulation results for decreasing eigenvalues and one common eigenvector, with no sequencing depth correction; $p$=50, $n_1=200$, $n_2 = 100$. "None" as a common basis label means that Group 1 and Group 2 data were concatenated prior to variance estimation.}
 \label{fig:CnnNoseq1}
\end{figure}

\begin{figure}
    \centering
    \includegraphics[scale=0.2]{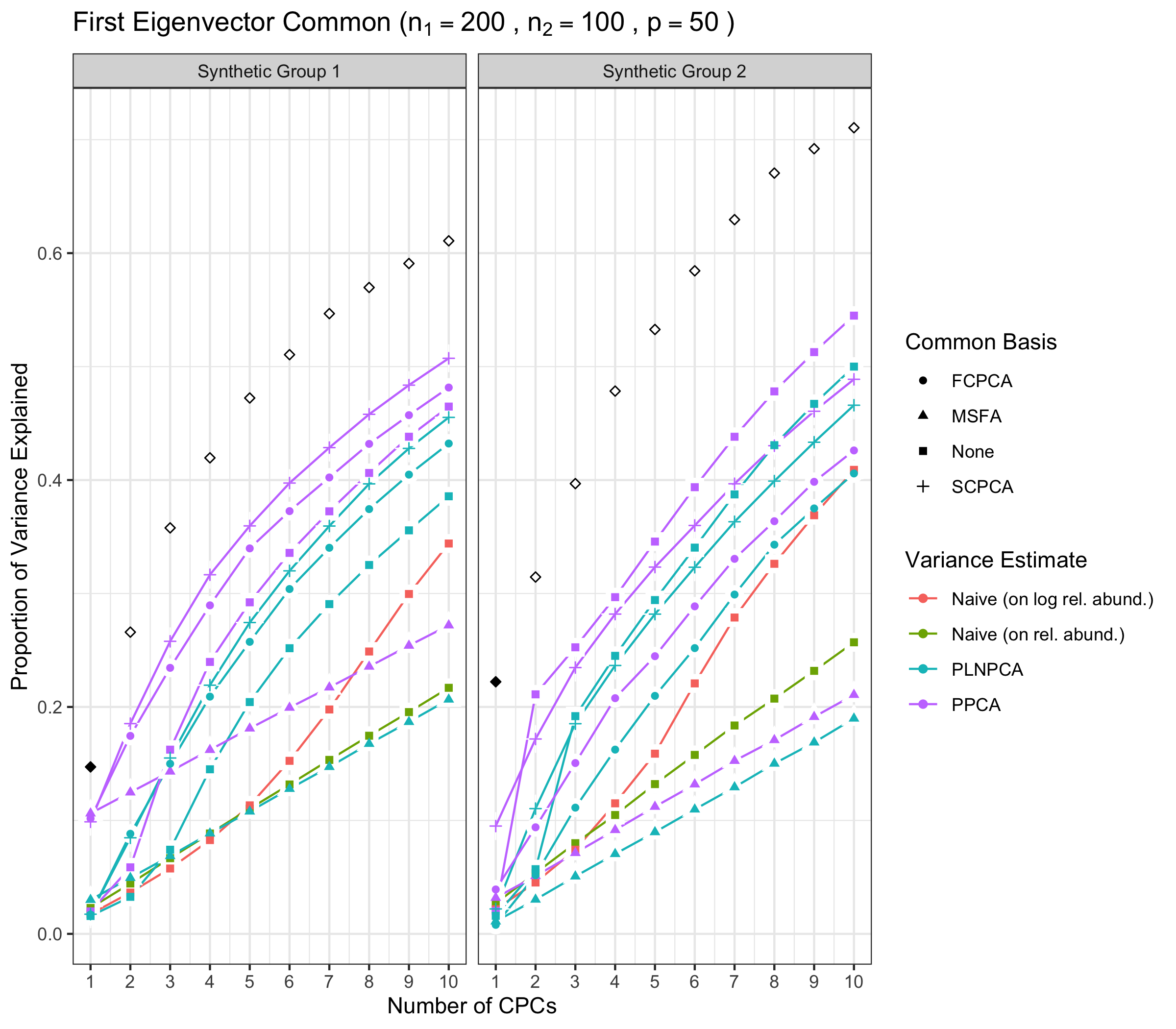}
    \caption{Simulation results for decreasing eigenvalues and one common eigenvector, with sequencing depth correction; $p$=50, $n_1=200$, $n_2 = 100$. "None" as a common basis label means that Group 1 and Group 2 data were concatenated prior to variance estimation.}
    \label{fig:CnnSeq1}
\end{figure}

\begin{figure}
    \centering
    \includegraphics[scale=0.2]{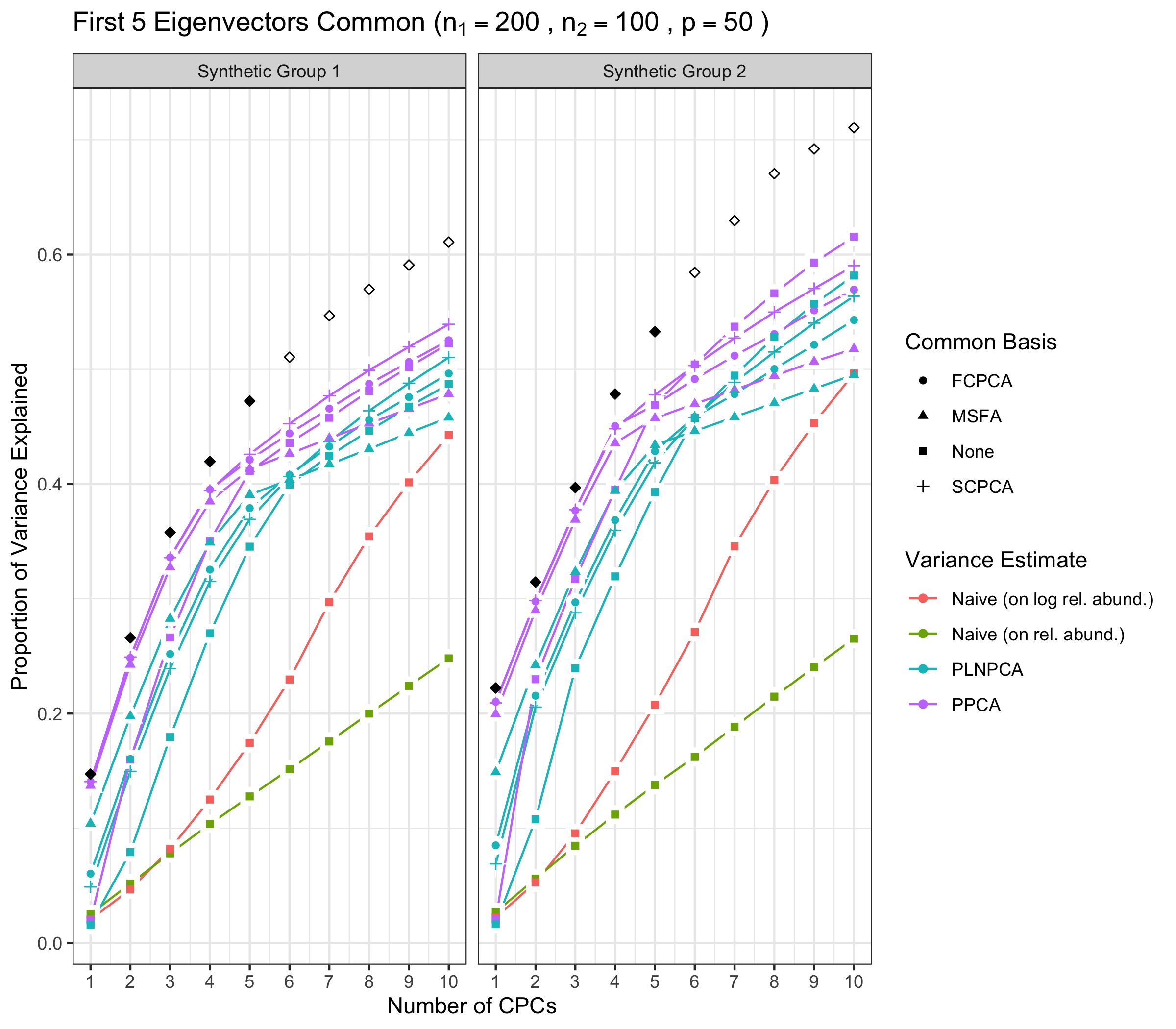}
    \caption{Simulation results for decreasing eigenvalues and five common eigenvectors, with sequencing depth correction; $p$=50, $n_1=200$, $n_2 = 100$. "None" as a common basis label means that Group 1 and Group 2 data were concatenated prior to variance estimation.}
 \label{fig:CnnSeq5}
\end{figure}

\begin{figure}
    \centering
    \includegraphics[scale=0.2]{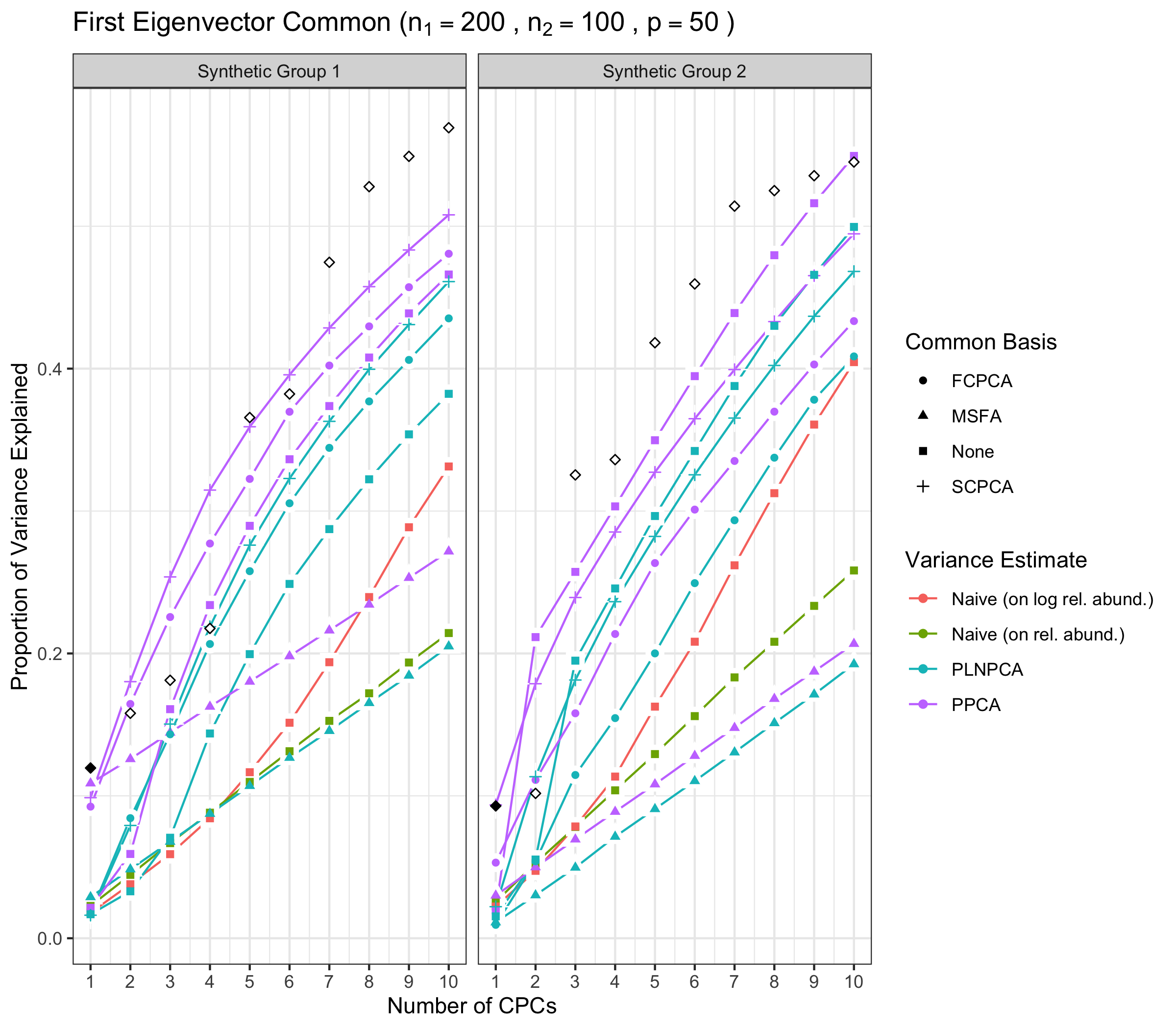}
    \caption{Simulation results for non-decreasing eigenvalues and one common eigenvector, with sequencing depth correction; $p$=50, $n_1=200$, $n_2 = 100$. "None" as a common basis label means that Group 1 and Group 2 data were concatenated prior to variance estimation.}
 \label{fig:DnnSeq1}
\end{figure}

\begin{figure}
    \centering
    \includegraphics[scale=0.2]{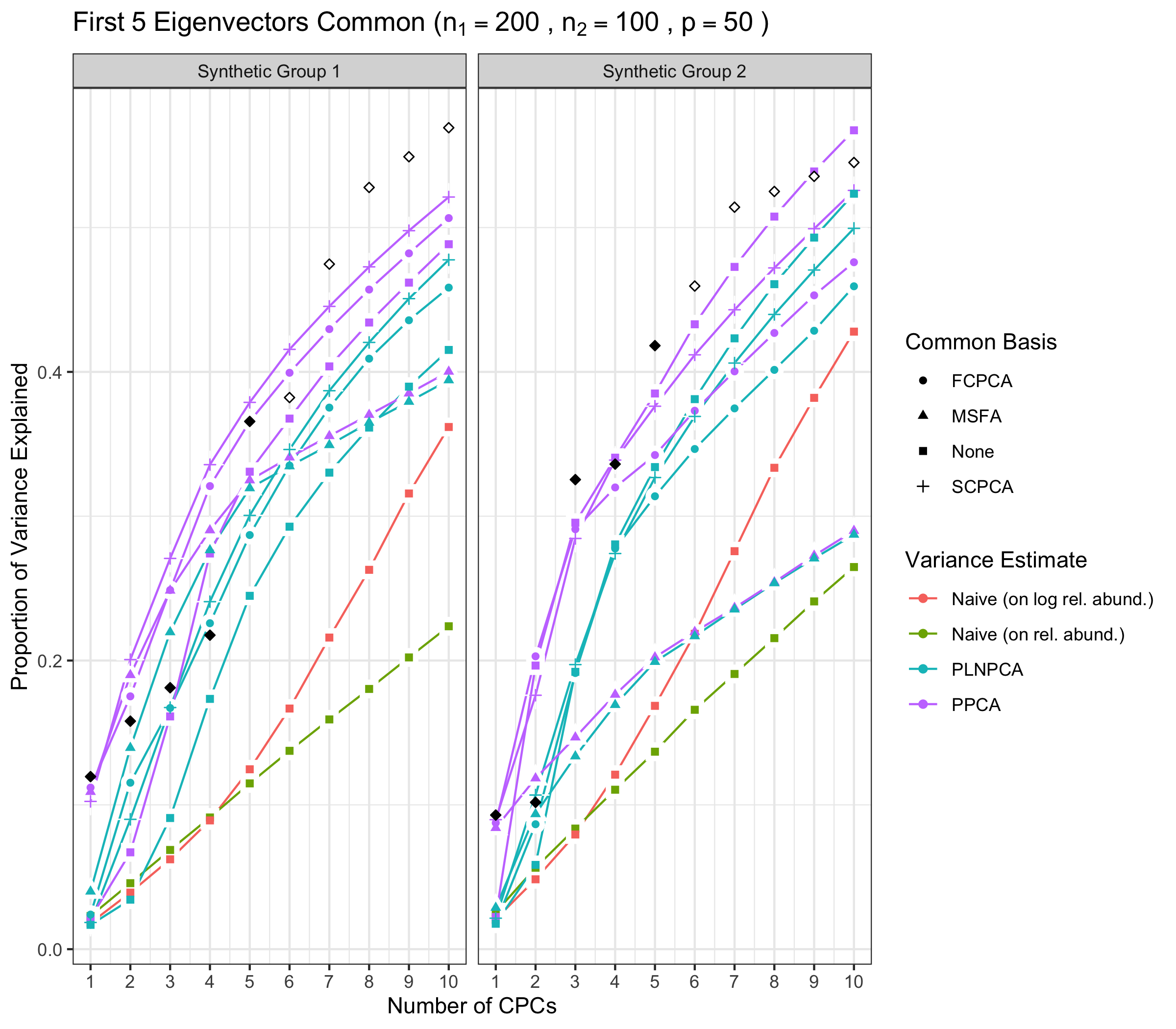}
    \caption{Simulation results for non-decreasing eigenvalues and five common eigenvectors, with sequencing depth correction; $p$=50, $n_1=200$, $n_2 = 100$. "None" as a common basis label means that Group 1 and Group 2 data were concatenated prior to variance estimation.}
 \label{fig:DnnSeq5}
\end{figure}

We performed simulation studies of two synthetic groups of multivariate Poisson log-normal observations across several scenarios. These scenarios differed on the true signal (including the number---$q$---of eigenvectors that were common to both groups' variance-covariance matrices, and whether the eigenvalues of the variance-covariance matrices were simultaneously decreasing), the sample sizes $n_1$ and $n_2$, whether or not the sample sizes were balanced, and whether or not sequencing depth correction was performed in the variance estimation stage. See the Methods section for more details of the simulation design. We now describe the results of the candidate ensemble methods listed in \cref{tab:methods} and some single-group alternatives, which we ran on the simulated data. The single-group methods were all run on the data from the two groups concatenated together, and these methods comprised PoissonPCA, PLNPCA, naive PCA on counts, naive PCA on log counts, naive PCA on relative abundances, and naive PCA on log relative abundances. Representative results of the simulations are given by \cref{fig:CnnNoseq1} through \cref{fig:DnnSeq5}, where in each plot the true eigenvalues are given in black. We show here results for small, unbalanced sample sizes ($n_1=200$, $n_2=100$) only, which have the most relevance for real studies, while some results for larger sample sizes and balanced sample sizes are provided in the Appendix. The abbreviation ``SDC" in tables and figures henceforth refers to ``sequencing depth correction".

\begin{table}[t]
\begin{centering}
\begin{tabular}{@{}cll@{}}
\toprule
\multicolumn{1}{l}{} & \multicolumn{2}{c}{\textbf{Ensemble Methods}} \\ \midrule
\textbf{}            & PoissonPCA + SCPCA \ \     & PLNPCA + SCPCA     \\
\textbf{No SDC} \ \ \      & PoissonPCA + FCPCA      & PLNPCA + FCPCA     \\
\textbf{}            & PoissonPCA + MSFA       & PLNPCA + MSFA      \\ \midrule
\textbf{}            & PoissonPCA + SCPCA      & PLNPCA + SCPCA     \\
\textbf{SDC}        & PoissonPCA + FCPCA      & PLNPCA + FCPCA     \\
\textbf{}            & PoissonPCA + MSFA       & PLNPCA + MSFA      \\ \bottomrule
\end{tabular}
\caption{The twelve candidate ensemble methods.}
\label{tab:methods}
\end{centering}
\end{table}

 We first show the results for all methods under simulation conditions in which the variance-covariance matrices of Group 1 and Group 2 share only the first eigenvector and their eigenvalues are decreasing. In \cref{fig:CnnNoseq1}, the methods were run with no sequencing depth corrections applied: since all results without sequencing depth correction show similar patterns, we report this case only. In \cref{fig:CnnSeq1}, all methods were run under the same simulation conditions as the aforementioned but with sequencing depth corrections applied: that is, PoissonPCA with its compositional correction, PLNPCA with observed read counts as offsets, and the synthetic counts were transformed to synthetic relative abundances before application of the naive PCA methods. \cref{fig:CnnSeq5} shows the results of the analogous scenario except that $\Sigma_1$ and $\Sigma_2$ shared the first five principal eigenvectors instead of only the first.

Then, \cref{fig:DnnSeq1} shows the results of the methods applied with sequencing depth correction to synthetic data for which the variance-covariance matrices of Group 1 and Group 2 shared one eigenvector, but with non-decreasing eigenvalues (i.e., the eigenvalues on the shared eigenvector are not the largest eigenvalues). \cref{fig:DnnSeq5} shows the results for the analogous case except with $\Sigma_1$ and $\Sigma_2$ sharing five common eigenvectors instead of one. In each of these figures, the true variances (the eigenvalues used to simulate the data) associated with the $q$ shared eigenvectors were plotted with solid black points, and subsequently the true variances associated with their unique eigenvectors were plotted with outline-only points.

The simulations showed generally good performance by the ensemble methods based on PoissonPCA or PLNPCA, and by PoissonPCA or PLNPCA alone. This is especially clear in comparison to the naive PCA methods on relative abundances or raw counts, which were not able to find good rotations for the data. As a result, these methods showed very slow increases in their cumulative explained variance, with no improvement even as the number of shared eigenvectors increased. This suggests that methods tailored to deal with Poisson measurement error should be the first choice among the different options. While both variance estimation methods seem to do fairly well, PoissonPCA is seen here to have consistently outperformed PLNPCA in terms of the reconstruction of the dominant signal in each group, despite the fact that the data were simulated under PLNPCA's generative model. One possible explanation for this is that PoissonPCA's moment-based variance estimator may be less sensitive than PLNPCA's variance estimator to the outliers that arise from the Poisson log-normal generating process. Interestingly, the difference in performance between PoissonPCA and PLNPCA is negligible without sequencing depth correction, which suggests that PoissonPCA's compositional correction to the variance estimate may be superior to PLNPCA's use of observed read count as a model offset. Incidentally, PoissonPCA was also an order of magnitude faster to run in our simulations. However, regardless of which method was used to estimate the variance (or which simulation scenario), when no sequencing depth correction was applied, the first CPC explained very little variance: instead that axis captured the unique variation from sequencing depth. The second CPC then usually explained a large proportion of common variance, and thereafter the points typically tracked the true signal.

As for the estimation of the low-dimensional basis, the two CPCA methods are very similar and both did well. SCPCA appeared to outperform FCPCA when the signal was more difficult to resolve, such as when there were very few shared eigenvectors and the true cumulative variance increased slowly (e.g., Group 2 with one shared eigenvector). While technically SCPCA, FCPCA, and MSFA are all designed for positive-definite symmetric matrices, in practice SCPCA performed well even when the covariance matrix was indefinite, which occasionally occurred in the PoissonPCA case (see Methods section). MSFA, which assumes that only $q$ axes are common among the $S$ groups (unlike CPCA, which assumes that all $p$ axes are common), performed well in some cases, with the caveat that we were only able to specify the true $q$ because this was a simulation. However, when common signal was very low ($q$=1), MSFA struggled to capture any variation on the first axis for Group 2 compared to SCPCA. Moreover, the MSFA optimization routine sometimes failed, and the simulation results for MSFA had to be averaged over the successful replicates (typically around 90\%, depending on the simulation scenario). In addition, MSFA had by far the slowest run-time, and so from these simulations it would appear that the CPCA approaches have more practical utility.

PoissonPCA or PLNPCA alone on the concatenated data generally performed well. In each case, the second axis showed good recovery, while the first axis tended to capture a smaller amount of common signal even when a sequencing depth correction was applied. The fact that these methods alone can often adequately reconstruct the signal of the two groups while naive PCA methods consistently fail to do so suggests that misspecification of models with respect to sampling can sometimes pose a larger obstacle to cross-study microbial abundance analyses than study-specific sources of variation. However, the multi-group methods dramatically outperform PoissonPCA or PLNPCA alone when some of the large eigenvalues of $\Sigma_s$ are not associated with the shared eigenvectors, especially on the estimation of the first eigenvector: when the variance estimates from PoissonPCA or PLNPCA are decomposed using SCPCA, FCPCA, or MSFA, we observe highly desirable dimension reduction behavior. This provides evidence that our ensemble method holds value for extracting signal and biological insight from collections of noisy metagenomic and 16S datasets, which may well possess extensive unique variation.

    \begin{figure}[t]
    \centering
    \includegraphics[scale=0.55]{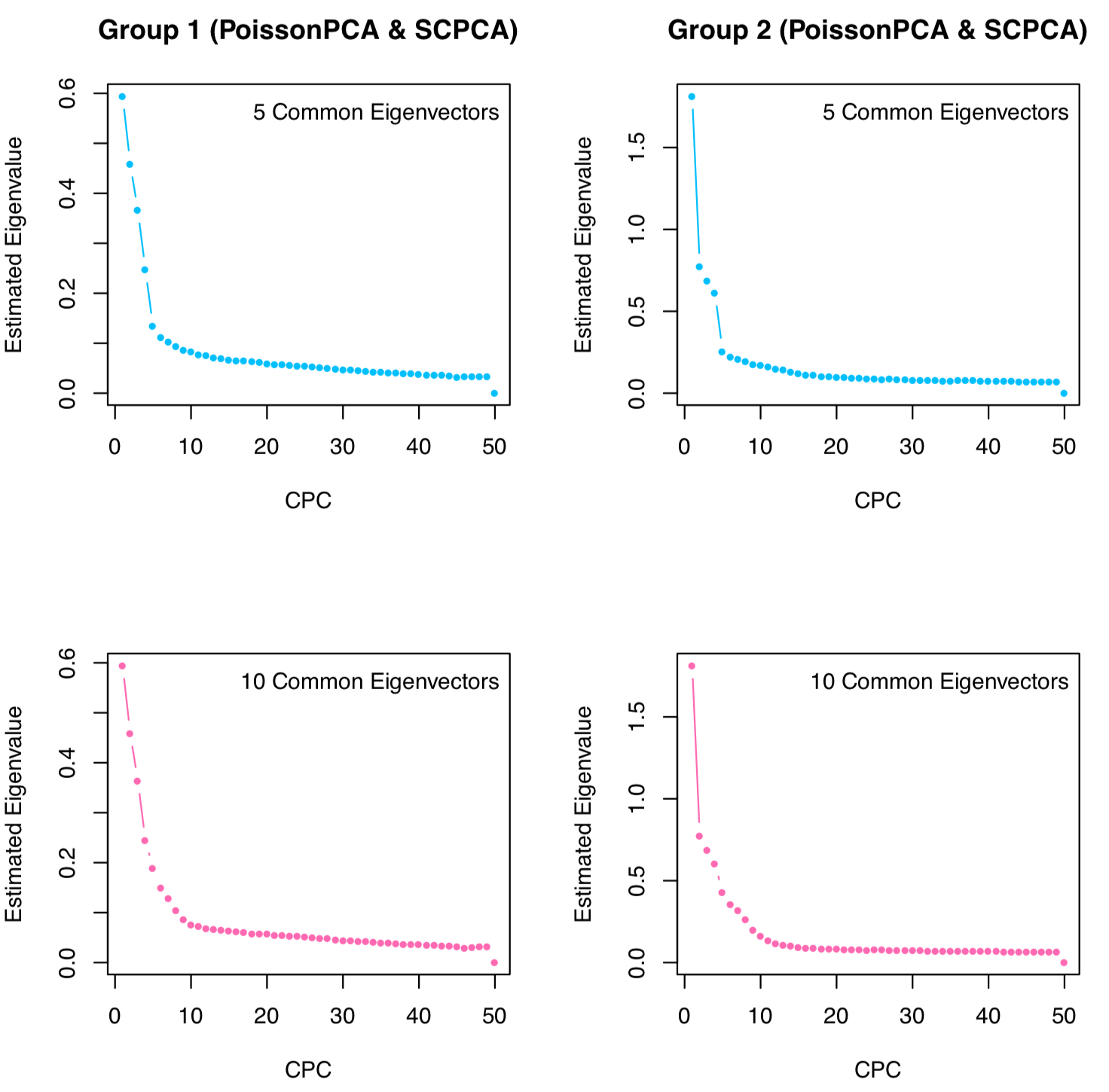}
    \caption{Scree plots of estimated eigenvalues from PoissonPCA \& SCPCA for each group with 5 or 10 shared eigenvectors.}
    \label{fig:simscree}
\end{figure}

Finally, \cref{fig:simscree} contains scree plots for the estimated eigenvalues from PoissonPCA followed by SCPCA for each group, where $\Sigma_1$ and $\Sigma_2$ share either 5 or 10 common eigenvectors and the true eigenvalues are simultaneously decreasing. These plots show that the differences between successive estimated eigenvalues drop to near zero when the number of CPCs is larger than the true number of shared eigenvectors, which provides evidence that we will be able to make a good choice of $q$ when applying the method to real data for which we cannot know the true number of shared eigenvectors.

\subsection{Colorectal Cancer Data Analysis}
It is not surprising that many studies have found links between the gut microbiota and cancer of the colon. To investigate the performance of our candidate ensemble methods on real data, we re-analyzed metagenomic datasets from \citet{feng} and \citet{zeller}, each consisting of fecal samples from participants diagnosed with colorectal carcinoma (CRC) or non-malignant colorectal adenoma and from controls (study and participant characteristics are summarized in \cref{tab:fz}). 
\begin{table}[]
\begin{centering}
\begin{tabular}{@{}lll@{}}
\toprule
                                   & \textbf{\citet{zeller}}               & \textbf{\citet{feng}} \\ \midrule
\textbf{Number of samples}         & 199                                         & 154                         \\
\textbf{Country of origin}         & France, Germany                             & Austria                     \\
\textbf{Sequencing technology}     & Illumina HiSeq                              & Illumina HiSeq              \\
\textbf{Number of CRC samples}     & 91                                          & 46                          \\
\textbf{Number of adenoma samples} & 42                                          & 47                          \\ \bottomrule
\end{tabular}
\caption{Comparison of \citet{zeller} and \citet{feng}.}
\label{tab:fz}
\end{centering}
\end{table}

\begin{figure}
    \includegraphics[scale=0.8]{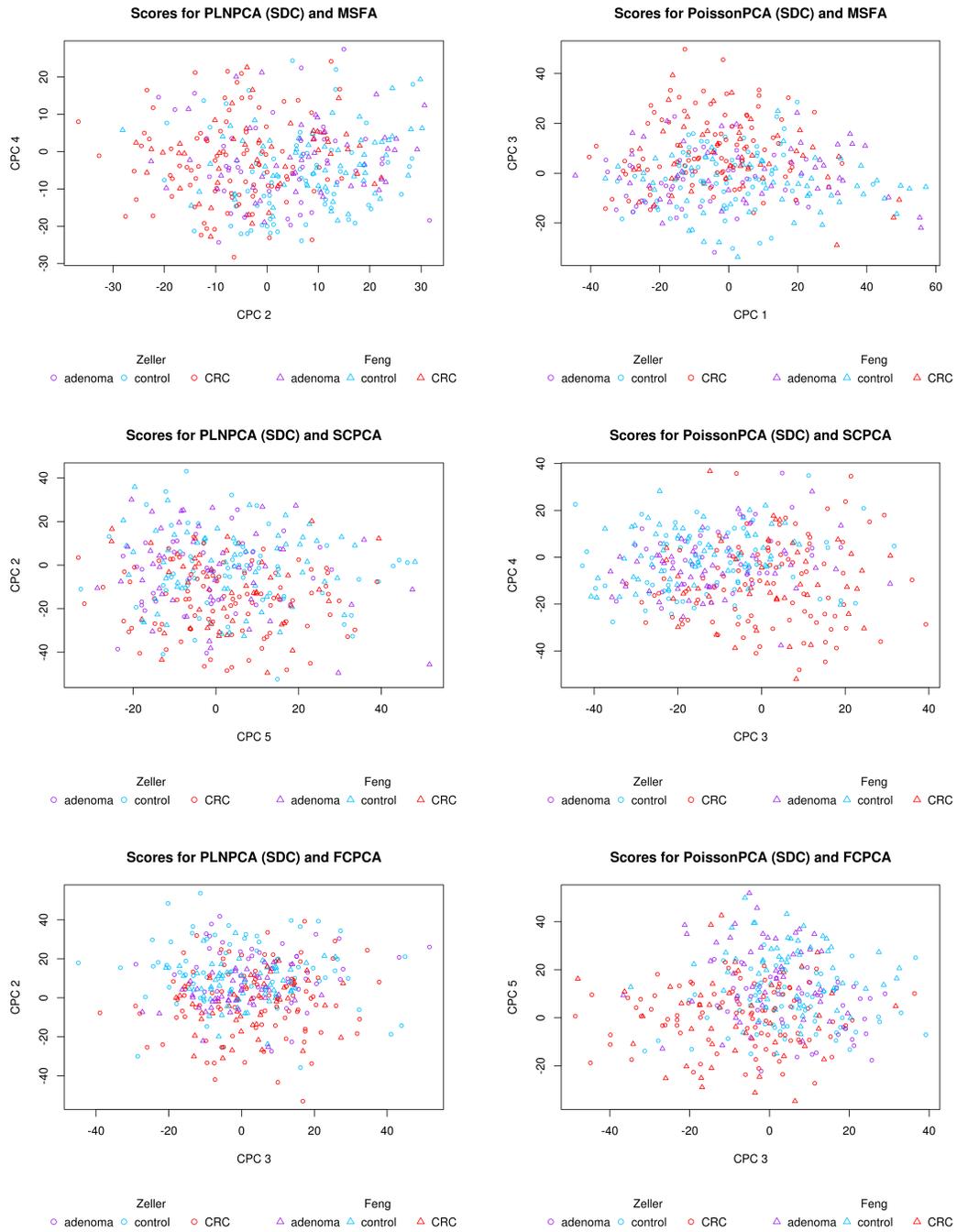}
    \caption{Scores from ensemble methods with SDC by disease state.}
    \label{fig:manyscores}
\end{figure}

\begin{figure}
    \centering
    \includegraphics[scale=0.75]{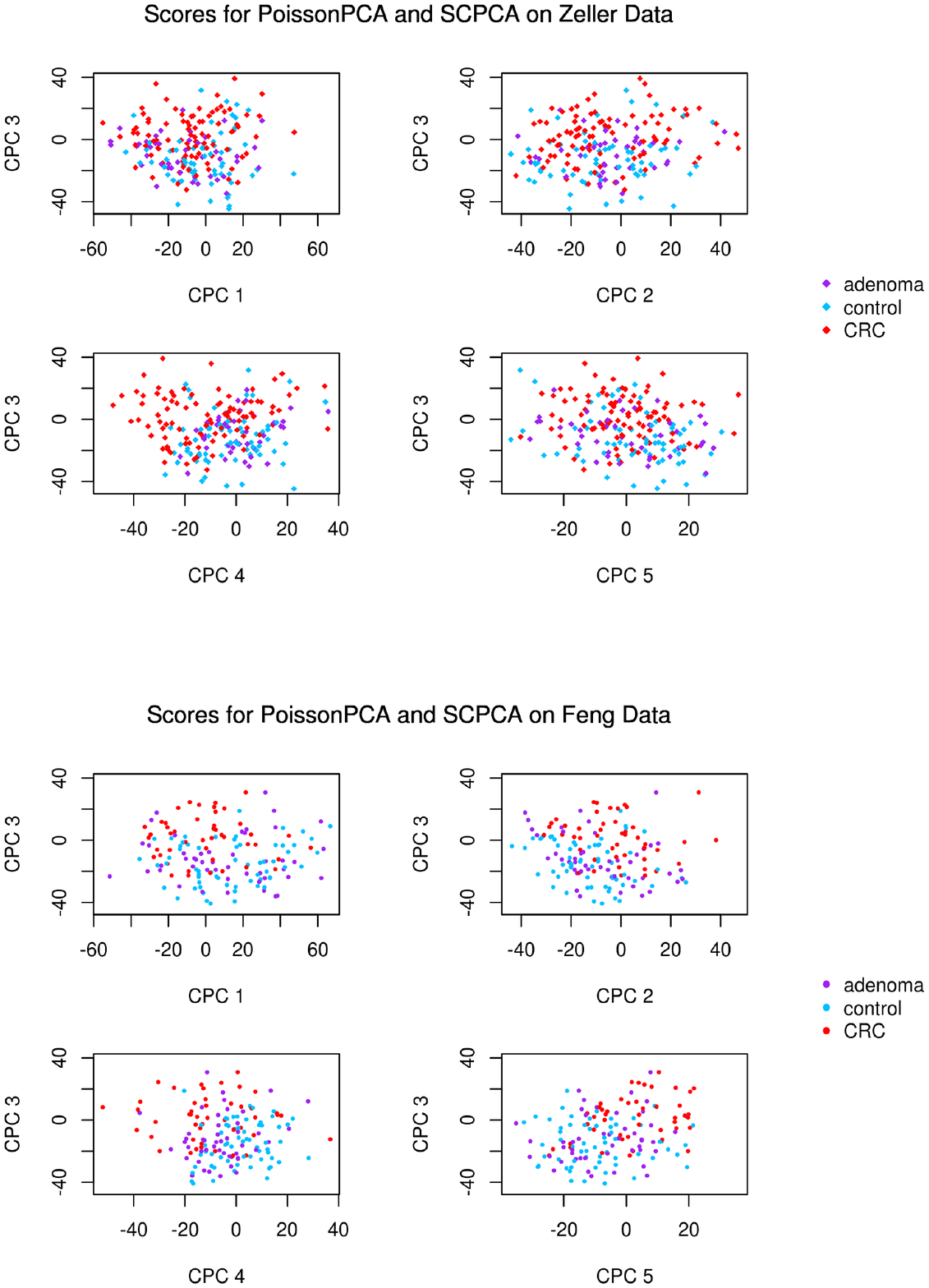}
    \caption{Score plots by disease state (PoissonPCA with SDC \& SCPCA).}
 \label{fig:ppstepscore}
\end{figure}

\begin{figure}
	    \includegraphics[scale=0.85]{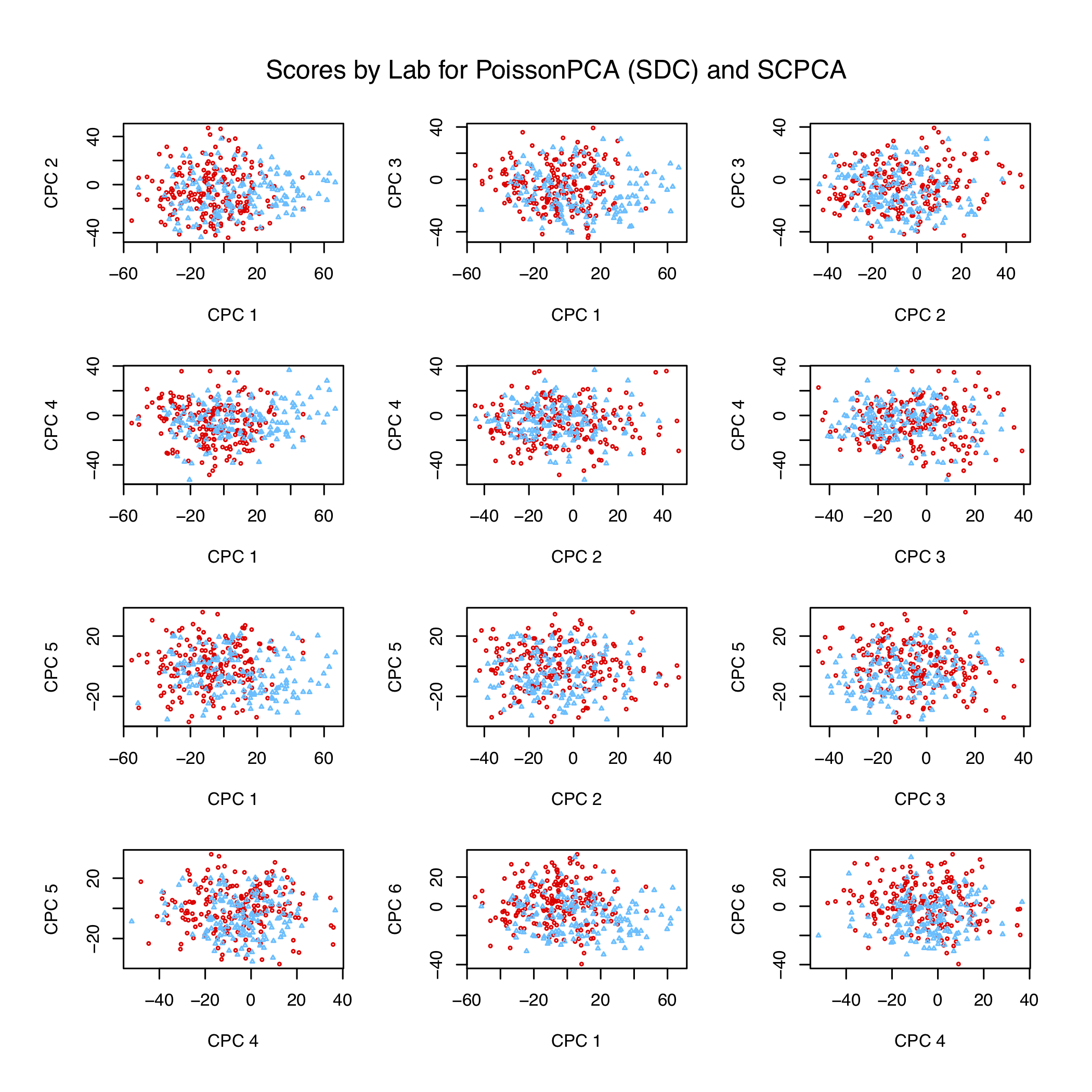} 
    \caption{Scores by study of origin (PoissonPCA with SDC and SCPCA).}
    \label{fig:bylab3}
\end{figure}

\begin{figure}
	    \includegraphics[scale=0.85]{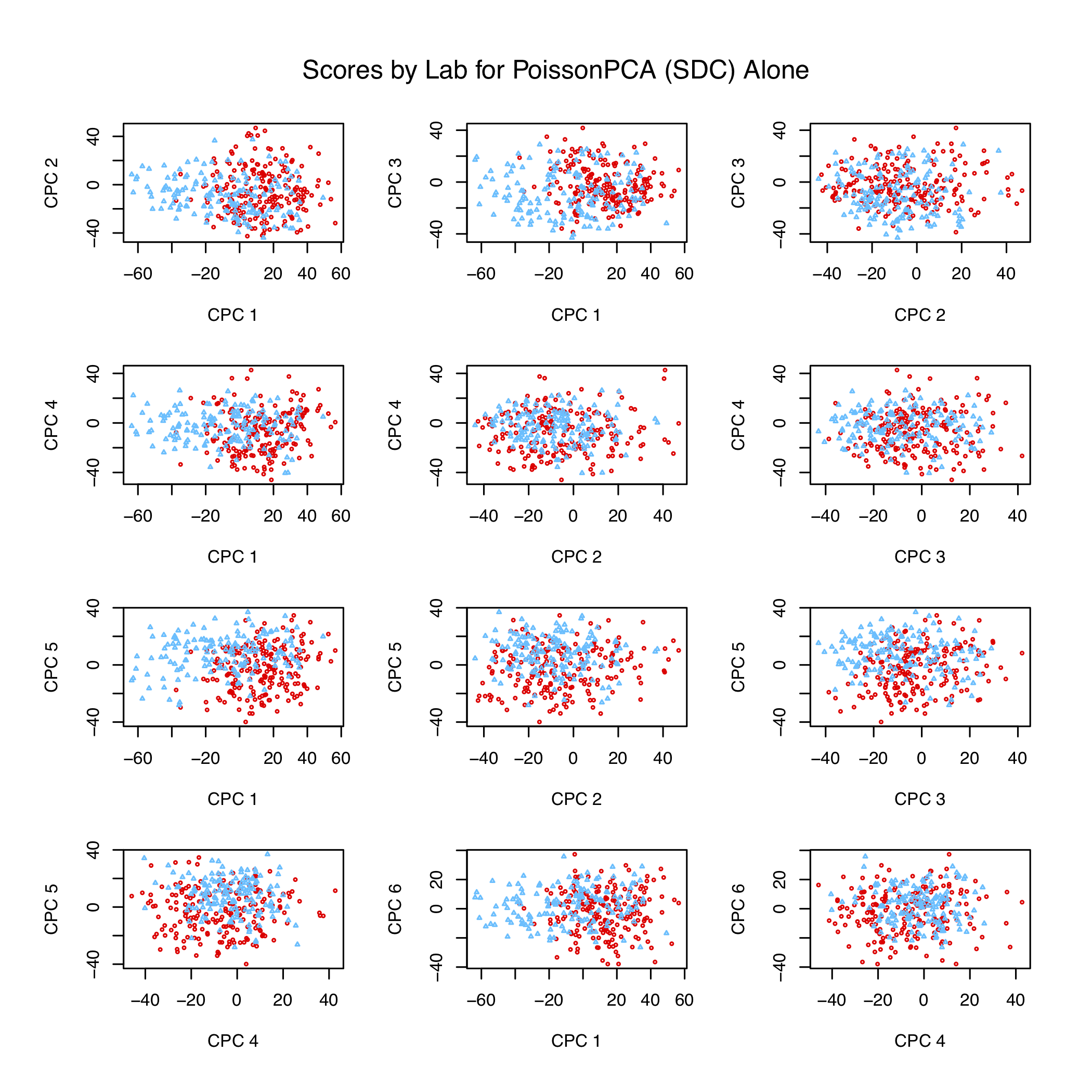} 
    \caption{Scores from PoissonPCA alone with SDC by study of origin.}
    \label{fig:bylabpp}	
\end{figure}

\begin{figure}
   \centering
    \includegraphics[scale=0.7]{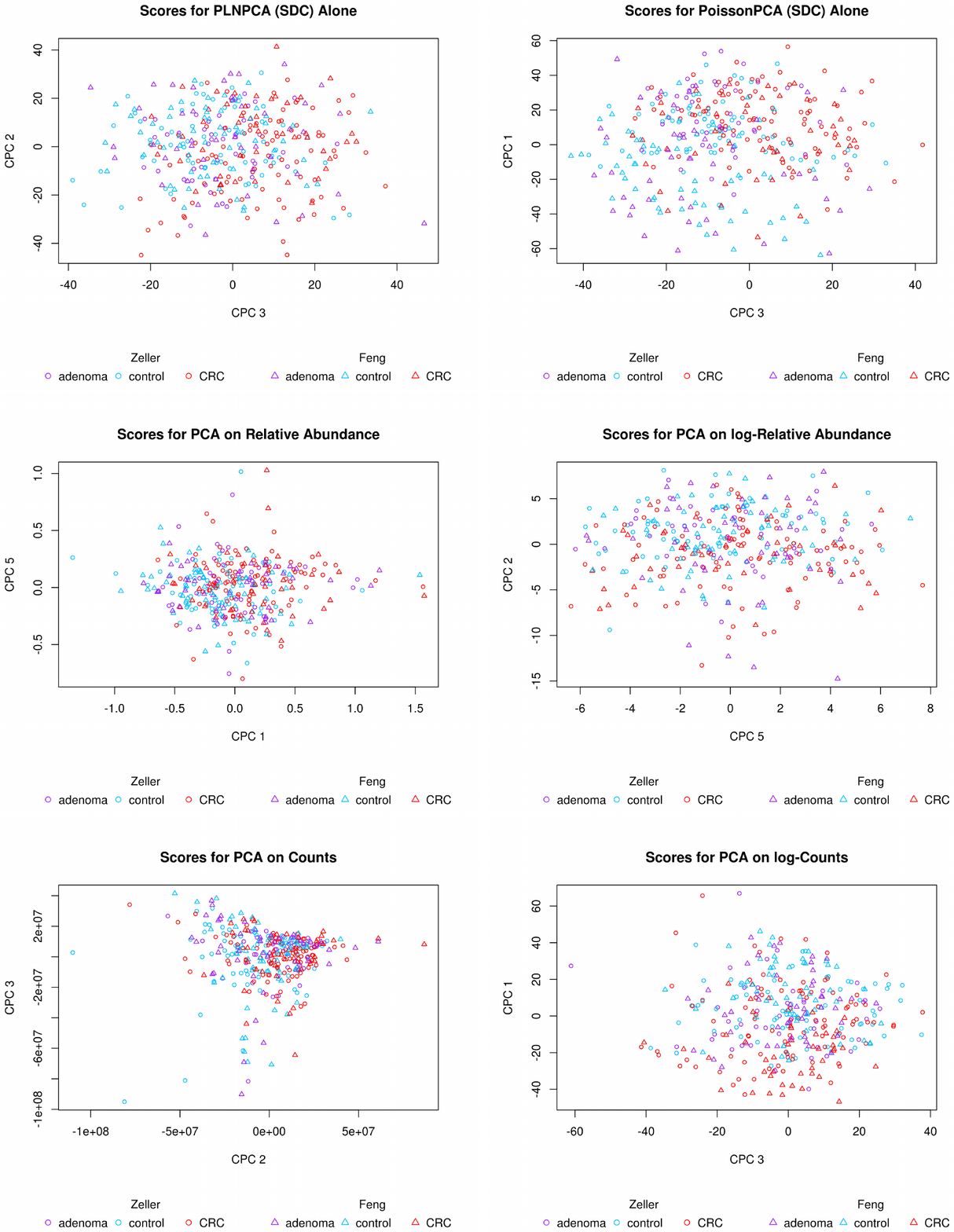}
    \caption{Scores from single-group and naive methods (with SDC for PoissonPCA/PLNPCA) by disease state.}
    \label{fig:naive}
\end{figure}

\cref{fig:manyscores} depicts selected score plots for each candidate ensemble method with sequencing depth corrections applied (see Supplemental \cref{app:fig:noseqs} for the equivalent plots with no sequencing depth correction), which show participants with CRC clustering distinctly from participants without CRC on the common axes. From \cref{fig:ppstepscore}, which shows several axes on each dataset separately for PoissonPCA \& SCPCA with sequencing depth correction, we can see that the \citet{feng} data seem to be better behaved than \citet{zeller} in terms of clustering. In terms of lab-specific signal, Supplemental \cref{app:fig:bylab1} through Supplemental \cref{app:fig:bylab10} show score plots for each ensemble method colored by study of origin, with Supplemental \cref{app:fig:bylabpln} through Supplemental \cref{app:fig:bylabcountlog} containing those for the single-group and naive methods. While the sequencing depth correction does appear to create some clustering by study on some axes, the CPCs that best discriminate disease state do not show this. For example, for PoissonPCA with sequencing depth correction followed by SCPCA, the best clustering of control vs. CRC samples is shown by CPC 4 and CPC 3, and \cref{fig:bylab3} shows that lab-driven signal only appears in plots of CPC 1 or CPC 6. This suggests that these axes indeed correspond to common, generalizable, CRC-related biological signal. In contrast, although PoissonPCA with sequencing depth correction does show clustering by disease state on the selected PCs in \cref{fig:naive}, the score plots in \cref{fig:bylabpp} show clustering by study on these and most other PCs. Of the naive PCA methods, each one shows some clustering by disease state (\cref{fig:naive}). Supplemental \cref{app:fig:bylabpln} through Supplemental \cref{app:fig:bylabcountlog} do not suggest pervasive clustering by study of origin, although this is expected since the data were individually mean-centered by study. 

We then investigated the predictive ability of these scores. Since we, like \citet{zeller}, found that samples positive for colorectal adenoma tended to cluster with control samples, we collapsed the two groups together. As a benchmark of the discriminating signal in the data, we trained random forest models on 80\% of the full genus-level concatenated data (see Methods for details). We used the final model to predict disease state on the test set, and consider the performance of this nonlinear classifier to represent the extent to which the signal in these data can be used to discriminate CRC samples. We then trained logistic regressions on the first five or ten common scores estimated by each method. The results are summarized in \cref{tab:pred}; area under the receiver operating curve (AUC) entries greater than 0.85 and test accuracy entries greater than 0.80 are in bold.

\begin{table}
\begin{centering}
\begin{tabular}{lllllllll}
\toprule
                                &                                   & \multicolumn{2}{l}{\textbf{First 5 CPCs}}          &  & \multicolumn{3}{l}{\textbf{First 10 CPCs}} &  \\ \cmidrule{3-4} \cmidrule{6-7}
                                & \textbf{Scores}                   & \textbf{AUC}   & \multicolumn{1}{c}{\textbf{Acc.}} &  & \textbf{AUC}      & \textbf{Acc.}     &    &  \\ \midrule
\multirow{6}{*}{\textbf{No SDC}}& PoissonPCA + SCPCA                & 0.835          & 0.757                             &  & 0.839    &    0.786    &    &  \\
                                & PLNPCA + SCPCA                    & \textbf{0.877}          & 0.800                             &  & \textbf{0.885}    & \textbf{0.814}             &    &  \\
                                & PoissonPCA + FCPCA                & 0.826          & 0.771                             &  & 0.832             & 0.786             &    &  \\
                                & PLNPCA + FCPCA                    & \textbf{0.868}     & 0.786                        &  & \textbf{0.860}             & 0.800             &    &  \\
                                & PoissonPCA + MSFA, $q=4$          & 0.804          & 0.729                             &  & \textbf{0.860}             & 0.771             &    &  \\
                                & PLNPCA + MSFA, $q=3$              & 0.811          & 0.786                             &  & \textbf{0.858}             & 0.757             &    &  \\ \midrule
\multirow{6}{*}{\textbf{SDC}}   & PoissonPCA + SCPCA                & \textbf{0.860} & \textbf{0.814}                    &  & 0.848    & 0.771             &    &  \\
                                & PLNPCA + SCPCA                    & 0.766          & 0.671                             &  & 0.710             & 0.700             &    &  \\
                                & PoissonPCA + FCPCA                & \textbf{0.863}          & \textbf{0.814}                   &  & 0.849             & 0.800             &    &  \\
                                & PLNPCA + FCPCA                    & 0.780          & 0.771                             &  & 0.779             & 0.743             &    &  \\
                                & PoissonPCA + MSFA, $q=4$          & 0.843          & 0.771                             &  & \textbf{0.860}             & 0.786            &    &  \\
                                & PLNPCA + MSFA, $q=5$              & 0.752         & 0.743                             &  & 0.826            & 0.800    &    &  \\ 
                                & PoissonPCA alone                  & 0.818          & 0.771                             &  & \textbf{0.853}    & 0.771            &    &  \\    
                                & PLNPCA alone                  & 0.744          & 0.629                            &  & 0.756    & 0.714              &    &  \\ \midrule
\multirow{2}{*}{\textbf{Naive}} & PCA of Proportion                 & 0.669          & 0.657                             &  & 0.645             & 0.614             &    &  \\
                                & PCA of Log-Proportion             & 0.677          & 0.671                             &  & 0.738             & 0.743             &    &  \\
                                & PCA of Count                & 0.637          & 0.643                             &  & 0.628             & 0.614             &    &  \\
                                & PCA of Log-Count             & 0.822          & 0.743                              &  & 0.816             & 0.771             &    &  \\ \midrule
\textbf{}                       & \textbf{}                         & \textbf{AUC}   & \textbf{Acc.}                     &  &                   &                   &    &  \\
                                & \textbf{All Genus-Level Features} & 0.828          & 0.771                             &  &                   &                   &    &  \\ \bottomrule
\end{tabular}
\caption{AUC and test set accuracy for classification of CRC samples.}
\label{tab:pred}
\end{centering}
\end{table}

In general, the ensemble methods performed well with many obtaining high accuracy using only five CPCs, insinuating that the biological signal characterizing CRC samples across the two groups was captured in very few CPCs. In fact, the linear classifier using the scores as predictors often performs better than the nonlinear classifier using all genus-level taxa, suggesting that uninformative noise was reduced. Sequencing depth correction does not appear to increase the predictive ability of scores estimated by CPCA, which may suggest that the sequencing depth noise is not complex enough to throw the classifier off the scent of the CRC-related signal, which provides further evidence that correcting estimates for Poisson measurement error is a critical factor in these analyses. 

PoissonPCA on the concatenated data performed well with no common factor estimation, as did PCA of the individually mean-centered log-count data. This is not entirely surprising given their performances in simulation. Also, \cref{fig:ppstepscore} suggests that the ability to discriminate disease state in Zeller samples may be the main determinant of classification performance, and so the success of PoissonPCA and log-count PCA on the concatenated data may be owed to some unique biological signal that is relevant to CRC status in the Zeller data but missing or undetectable in the Feng data, which would naturally give the single-group procedures an edge over the ensemble methods for prediction. In general, unique signal could be helpful or unhelpful for machine learning prediction of a given response, but in either case it could potentially obstruct meaningful interpretation of the predictors. The results in \cref{tab:pred} support our hypothesis that the proposed ensemble methods are able to find a very low-dimensional representation of the data that retains virtually all of the discriminating biological signal that is shared among groups, which is our main interest.

    \begin{figure}
    \centering
    \includegraphics[scale=0.7]{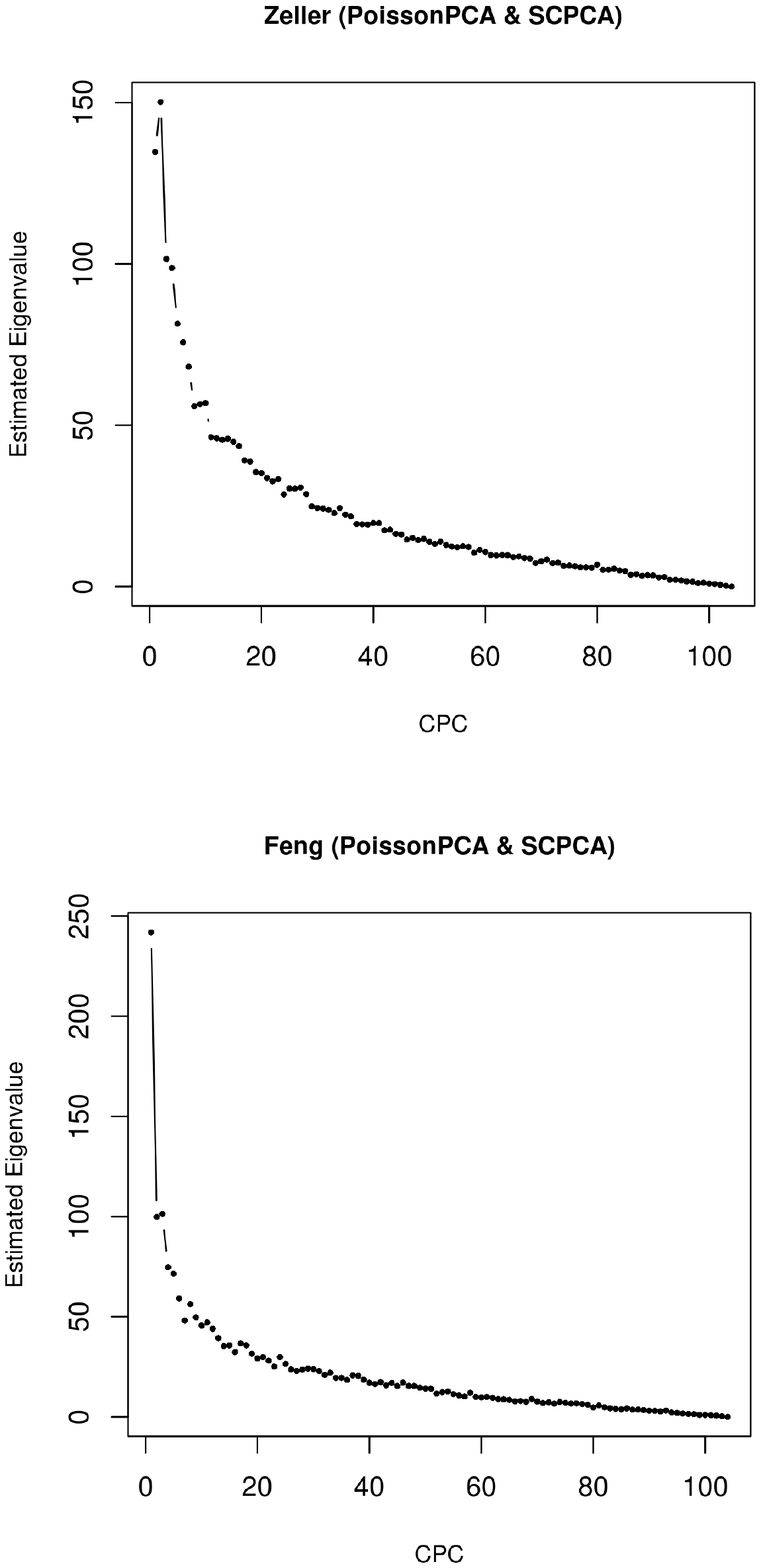}
    \caption{Scree plots of PoissonPCA followed by SCPCA on the Zeller and Feng data.}
    \label{fig:fzscree}
\end{figure}
	
	    \begin{figure}
    \includegraphics[scale=0.7]{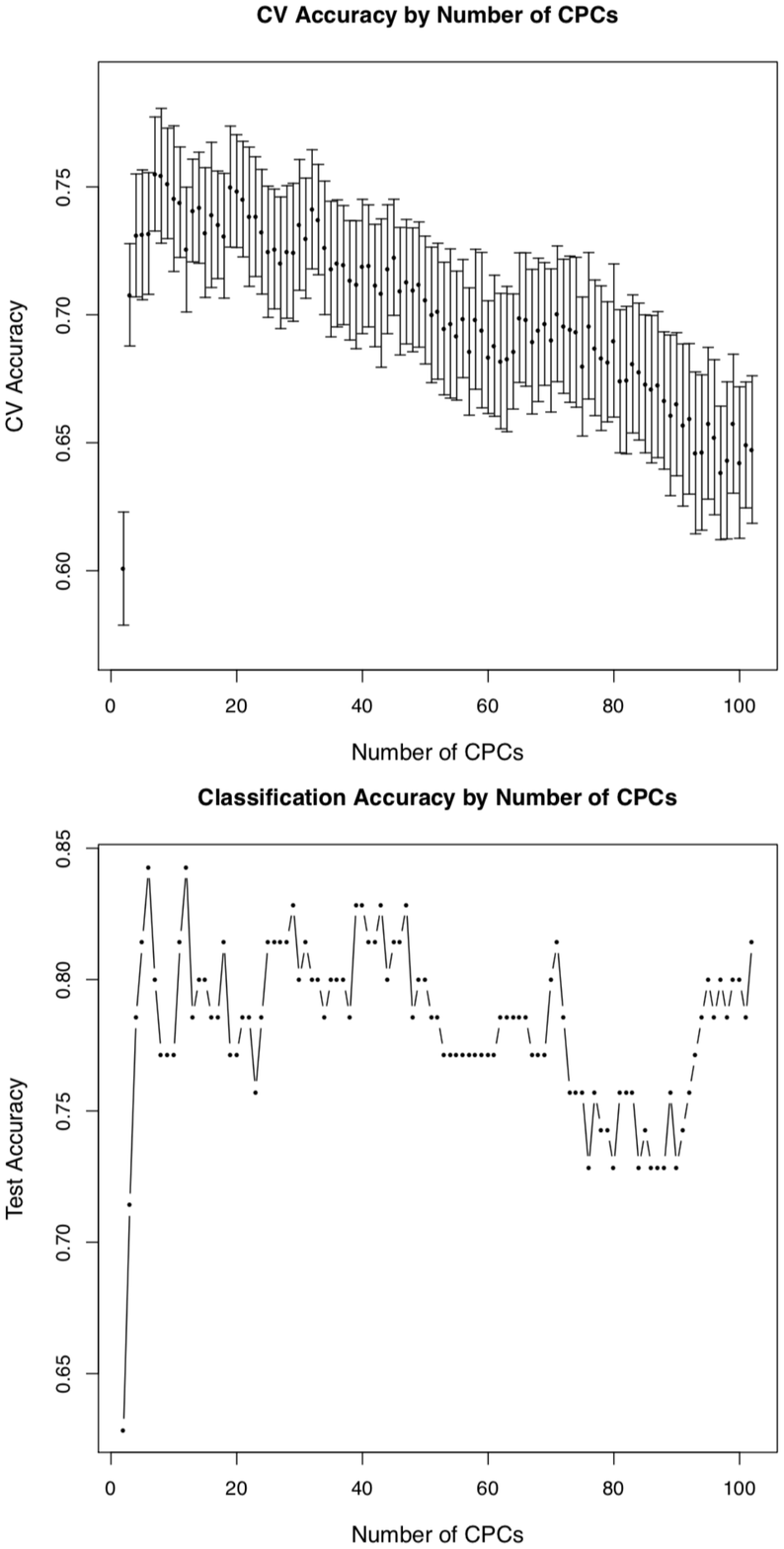}
    \caption{Cross-validation accuracy and test set accuracy of logistic regression models trained on the scores from PoissonPCA and SCPCA on the Zeller and Feng data.}
    \label{fig:fzerror}
\end{figure}
	
To choose the optimal number of common principal components to describe the common signal on CRC status, we can choose $q$ using scree plots of the estimated eigenvalues or by prediction such that the test set misclassification error is lowest. For example, using PoissonPCA followed by SCPCA, scree plots in \cref{fig:fzscree} suggest that the Feng data have a pronounced elbow, while the eigenvalues level off more slowly in the Zeller data. However, in both datasets, the last large difference between successive eigenvalues occurs by about $q=11$. In \cref{fig:fzerror}, which shows the cross-validation and test classification accuracies, it is apparent that after 6 CPCs the misclassification error tends to increase with additional CPCs and so we choose $q=6$. The prediction results in \cref{tab:pred} corroborate that in general, such parsimonious models perform well.

Finally, we turn to biological interpretation of the common loadings, wherein lies the strength of the ensemble method. Even assuming that single-group methods were able to resolve some common underlying variation patterns, the resultant PCs reflecting this would be indistinguishable from those capturing within-study variation or contrasts between studies. However, we have been assured by our simulation results and assessment of the CPC scores that the ensemble methods can yield $q<<p$ common axes that load on fully shared factors with strong signal. Furthermore, since our estimated loadings are related to the decomposition of variance on the log scale, we can interpret the values given by these loadings as log ratios of geometric means between groups of taxa with large-magnitude values in opposite directions. The following was observed from the common loadings estimated from SCPCA on the PoissonPCA sequencing depth-corrected variance estimate. 
 
Within our two CRC datasets, we found that on the two estimated common loading vectors that appeared to differentiate between samples with and without CRC, taxa with loading coefficient magnitudes in the $90^{\text{th}}$ or greater percentile were often described in other studies of CRC-associated microbiota. Peptostreptococcus, Butyrivibrio, Phascolarctobacterium, Fusobacterium, and Porphorymonas genera were identified by our common loadings as well as by the analyses of \citet{zeller}, while Acidaminococcus, Parvimonas, Gemella, and Peptostreptococcus were found in common with \citet{feng}. Genera we identified that have been previously associated with CRC in additional studies include Fusobacterium \citep{purcell,chenw}, Porphorymonas \citep{purcell,chenw}, Acidaminococcus \citep{azcarate}, Phascolarctobacterium \citep{weir}, Enterococcus \citep{kong}, Gemella \citep{kong}, Klebsiella \citep{chenw}, Prevotella \citep{purcell}, Solobacterium \citep{yu}, and the Siphoviridae viruses \citep*{handley}. Of those highlighted by our analyses, only Dialister \citep{weir}, Butyrivibrio \citep{zeller}, and Flavonifractor \citep{ai} have been reported to be protective. Lastly, our loadings suggest that Adlercreutzia contributes to distinguishing CRC samples, but this genus has not yet been implicated in the literature, although it has been observed to participate in the metabolism of flavonoids \citep{liy}, which are antioxidants generally known to be anti-inflammatory and anti-tumorigenic. Moreover, on the loading in which Butyrivibrio (a genus of butyrate producers) and Adlercreutzia have large negative coefficients, all other large magnitude coefficients are positive. Flavonoids, along with dietary fibre, are linked to butyrate production, all of which have been linked to metabolic health and decreased risk of cancer \citep{jakobsdottir,wu,ohkawara}, although little is known about how flavonoids are metabolized. This loading vector may then represent a contrast between these genera---which may be involved in some protective pathway related to the metabolic intermediates of flavonoids---and the other genera dominating the loading. This is an insight that neither \citet{zeller} nor \citet{feng} were able to resolve from their single-group analyses.


\section*{Conclusion}
We have demonstrated that by appropriate modeling of microbial abundance count datasets so as to estimate their respective variances, followed by estimation of a basis for a common low-dimensional subspace for the true underlying abundances, we can remove unwanted variation while retaining shared variation patterns. Our ensemble method performs well in simulation and appears to have utility for real data analysis. 

While we have demonstrated the method on data from two studies, in theory it can be applied to any number of datasets. PoissonPCA and SCPCA, the constituent methods showing the best performance, computation speed, and flexibility, can scale up to handle estimation of a common basis for many groups. Hence, our next step is to find a larger pool of similar datasets to determine whether strong common signal can still be resolved, since the relative amount of common signal will presumably drop with each additional dataset. We can also validate our findings by classification prediction or graphically by applying leave-one-out cross-validation across the groups, and comparing the projections of the estimated abundances for the held-out data and the rest of the data onto the subspace common to all except the held-out set. If we find that the ensemble method can indeed find common generalizable signal across many groups, this will lead us closer to impactful meta-analysis and the translation of research to application.

Similarly, we have shown the performance of our ensemble method on independent genomic samples from different experiments with very similar characteristics, but it remains to be investigated what range of common-signal-to-unique-noise ratios the method can handle. In this article we assumed that the variation of interest is shared across samples, which is true at least in the sense that the common metadata collected in all samples reflects what is of interest to us in a given exploratory analysis. However, the common signal may be harder to differentiate from strong unique biological signatures than it is from technical noise, and in our real data analysis we avoided this issue by choosing datasets from two clinical studies with very similar designs. That is, certain types of samples may be presumed to have more constraints on their similarity than other types: two independent fecal samples from different individuals are more similar than would be, for instance, two independent water samples from different lakes, because we would expect that the conditions within the human intestinal lumen are more constrained across individuals (by physiology) than the conditions of lake water are constrained across lakes. Because of the abundance of biomedical and clinical data available from studies of human and animal gut microbiota that tend to investigate similar themes, our method already has a solid basis of applicability based on our results. The next step in our research is to investigate whether our proposed method can resolve the common axes among different microbial abundance data even when there is presumably large unique biological variation among groups, in addition to unique technical noise and common biological variation. If so, the scope of our method could be extended to exploring commonalities in community structure that are conserved across a diverse range of communities.

In other future work, we may develop a multi-group extension of PLNPCA, which could make direct use of the dimension reduction capabilities of modeling counts as conditional on linear combinations of common and unique factors lying in separate low-dimensional subspaces, potentially removing the need for an ensemble method. However, we would have to consider carefully how best to fit this model, as PLNPCA and especially MSFA are already slow to estimate parameters. We would also need to consider whether the restrictive parameterization that these models share might be inappropriate in general for exploratory analyses of metagenomic and 16S data, and whether in fact an ensemble method using semi-parametric approaches is preferable, as was found in our analyses. Yet another option would be to build an equivalent hierarchical Bayesian factor analysis model, although MCMC run times scale poorly with dimension compared to likelihood methods. Even so, given the difficulty that MSFA's ECM algorithm had in achieving convergence in our simulations and data analysis, it is possible that a Bayesian approach would be easier to implement.

In conclusion, the study of microbiota poses a number of challenges, including heterogeneous signal-obstructing noise. In this paper, we addressed the lack of methods for cross-study generalization in microbial abundance data, and proposed a framework to ``remove" some of the observed within-study variation by seeking the latent structure that provides a scaffold for the common variance. We found that our two-step ensemble approach---first estimating the variance of the underlying abundances on the log scale for each group, and then decomposing the variances with a multi-group method---faithfully captured shared signal in our simulation scenarios, even when the groups had significant unique signal. The analysis of the Feng and Zeller colorectal cancer datasets demonstrated the excellent dimension reduction and signal retention capabilities of the ensemble methods, and moreover the ease with which biological interpretations can be drawn from the common loadings.

\section*{Methods}

\subsection{Estimation of Variance}
Before we can estimate the common signal among several datasets, we first have to deal with the Poisson error in each dataset individually. In this work we will consider two approaches, each of which assumes that the observations are conditionally Poisson-distributed. Since our application is the analysis of microbiome composition, we are primarily interested in treating the means on the logarithmic scale, which arises naturally in the first method via the canonical link function, and in the second method can be achieved by a transformation.

 The first approach is to leverage the Poisson log-normal PCA \citep[PLNPCA;][]{chiquet}, a fully parametric model that extends the probabilistic PCA of \citet*{tipping} such that the emission layer is Poisson (or any natural exponential family distribution) rather than normal. That is, the log of the Poisson mean is a function of a latent variable and follows a multivariate normal distribution of lower dimension than the feature space, and the observed counts are---given the log Poisson means---independently Poisson-distributed. Using this framework one has the option of introducing row-wise sums as an offset to treat sequencing depth as observed sampling effort, and can obtain an estimate of the variance-covariance matrix of the log means.
 
 The second approach is to use Poisson PCA \citep*[PoissonPCA;][]{kenney} to sidestep a complex likelihood-based model in favor of assuming only that the observed counts are---given the latent Poisson means---independently Poisson-distributed, potentially including sequencing depth as a nuisance random variable. The authors derived an unbiased variance estimate for any non-linear transformation of the latent means, as well as a semi-parametric method for estimating the scores of the transformed means, which we adapt for use in our ensemble method.


\subsubsection{Poisson Log-Normal PCA}
Let $X_i \in \NN^p$, \ $i=1,\dots,n$ be a random vector of which we have observed $n$ realizations. We briefly review the results from \citet*{chiquet} for PLNPCA, which provides an estimator for $\Sigma=\text{Var}(\log\Lambda)$ (where $\log$ is assumed to be applied element-wise whether we speak of $\log \Lambda_i \in \RR^p$ or the $n \times p$ matrix $\log\Lambda$) using variational inference.

Because the original method is based on probabilistic PCA \citep{tipping}, PLNPCA also provides the opportunity to estimate a low-rank covariance matrix depending on the dimension chosen for the latent space, but for our ensemble method we wish only to reduce dimension based on the common variation across all $S$ datasets, so we will be using the rank-$p$ estimate. Also, we wish to avoid disrupting signal in a supervised fashion so we will not include any covariate effects in the model. We denote sample-wise offsets by $\xi_i\in\RR^p$ (i.e., $\xi_{i1}=\xi_{i2}=\dots=\xi_{ip}$). In the microbiome setting, we choose these offsets to correspond to log-total read count, which is the closest we have to an observed value for sequencing depth. Should we wish not to apply a sequencing depth correction, the following results hold as written for $\xi_i=0_p$. 

PLNPCA assumes that the $X_i$ are conditionally independent given latent variables $w_{i}$ iid $\mathcal{N}(0_{\ell},I_{\ell})$. In the parameter space dwells $\log\Lambda_i\in\RR^p$, where
	\begin{align*}
	\log \Lambda_{i} = \xi_i + \mu + \beta w_{i}, \ \ \ i=1,\dots,n
	\end{align*}
	or equivalently, $\log\Lambda_i$ is multivariate normal with mean $\xi_i + \mu$ and variance $\Sigma=\beta \beta^T\in\RR^{p \times p}$. No uniqueness constraints are put on $\beta$. The final layer allows observations $X_{ij}$ to be generated according to the Poisson distribution with mean $\Lambda_{ij}$:
	\begin{equation*}
	p(X_{ij}|\log \Lambda_{ij})=\frac{1}{X_{ij}!}\exp(X_{ij}\log \Lambda_{ij}-\exp(\log \Lambda_{ij}))
	\end{equation*}
	such that, for the canonical log link, $\log(\mathbb{E}(X_{ij}|\log \Lambda_{ij}))=\log \Lambda_{ij}$. 
    
 	Because the marginal likelihood isn't analytic for the Poisson case, the authors integrate out $w$ under a variational approximation of $p(w|X)$, and maximize the variational lower bound for the marginal log-density of $X$ instead of maximizing the marginal log-likelihood. Details of this process and the resultant variance estimator can be found in their paper.

PLNPCA is implemented in the R package \textbf{PLNmodels} \citep*{chiquet}.

	\subsubsection{PoissonPCA}
Let $X_i \in \NN^p$, \ $i=1,\dots,n$ be a random vector of which we have observed $n$ realizations. PoissonPCA \citep*{kenney} assumes that conditional on the latent Poisson means $\Lambda_{ij}$, the $X_{ij}$ are independently distributed as Poisson with parameter $\Lambda_{ij}$. In contrast to PLNPCA, the distribution of $\Lambda$ itself is not parameterized. The authors use a method of moments procedure for estimating $\Sigma=\text{Var}(\log\Lambda)$. 

Finding an estimator for the variance of the log-transformed latent variables is contingent on finding a corresponding (element-wise) transformation $f$ for the data. The PoissonPCA model specifies that conditionally on $\Lambda$, the mean of $X$ is $\Lambda$, which implies that we are looking for $f(X)$ such that, conditionally on $\Lambda$, the mean of $f(X)$ is $\log \Lambda$. Accordingly, substituting $\log \Lambda$ for $\Exp[f(X)|\Lambda]$ in the total variance $\Var(f(X))=\Exp[\Var(f(X)|\Lambda)]+\Var(\Exp[f(X)|\Lambda])$, we get the formula 
\begin{align*}
    \Sigma=\Var(\log\Lambda)=\Var(f(X))-\Exp[\Var(f(X)|\Lambda)] 
\end{align*}
To find $f(X)$, PoissonPCA approximates $\log\Lambda_{ij}$ for small values with a Taylor series about a specified point and truncated at a specified value, whereas for larger $X$ it sets $f(X)=\log X$ since $\Exp[\log X|\Lambda]\approx \log \Lambda$. Estimating the conditional variance is achieved in PoissonPCA by finding a function $k(X)$ such that the average conditional mean of $k(X)$ is approximately $\Var(\log X|\Lambda)$. 

PoissonPCA accounts for sequencing depth by introducing a random variable $\xi_i\in \RR^p$ such that $X_{ij}|(\Lambda_{ij},\xi_{ij}) \sim \text{Poisson}(\xi_{ij}\Lambda_{ij})$, where $X_{i1},\dots,X_{ip}$ are independent given $\Lambda_i$ and $\xi_i$. Unlike in PLNPCA, $\xi_i$ is not considered to be observable, so under this model we end up estimating $\Sigma=\Var(\log\Lambda)=\Var(\log(\xi \circ \Lambda^*))$, where $\circ$ is the element-wise product, when in fact what we want is $\Sigma^*=\Var(\log\Lambda^*)$. Thus, in order to account for sequencing depth error, \citet{kenney} add constraints to the variance estimator given above in order to get the desired estimator $\Sigma^*$. They suggest that the best method of sequencing depth correction is to imbue it with the characteristic properties of a variance-covariance matrix of a compositional random vector, which is to say that $\Sigma^*$ should be symmetric and contained in the orthogonal complement of the vector $1$. Obviously the aim is to preserve the map defined by $\Sigma$, and so $\Sigma^*$ should have the same bilinear form as $\Sigma$ in the orthogonal complement of the vector $1$. With these constraints, the authors show that the sequencing depth-corrected variance estimate is 
\begin{align*}
\Sigma^*=\Sigma-(pI_p+1_p1_p^T)^{-1} \Sigma 1_p1_p^T - 1_p1_p^T\Sigma(pI_p+1_p1_p^T)^{-1}. 
\end{align*}
See their paper for details on the PoissonPCA variance estimator and its sequencing depth correction. In addition, the construction of these estimators makes no guarantee about the definiteness of the matrices, and in practice there can be several negative terminal eigenvalues. Since we found that the multi-group methods FCPCA and MSFA (to be described in the following sections) were sensitive to whether or not input variance-covariance matrices were positive-definite, if any $\sighat$, $s=1,\dots,S$ computed from PoissonPCA in our analyses was indefinite, we eigen-decomposed it, replaced the negative eigenvalues with small decreasing positive values, and then used the eigenvectors to reconstruct the variance before running FCPCA or MSFA.

PoissonPCA is implemented in the R package \textbf{PoissonPCA} \citep*{kenney}.


\subsection{Multi-Group Analysis}
We are ultimately interested in estimating the loadings that are common to all groups, and projecting the estimated latent Poisson means from each group into a common space spanned by the loadings. To achieve this, the natural choice would be to apply multi-group extensions of PCA or of factor analysis to the Poisson-corrected estimates from each group, since by dealing with abundance on the log scale we are able to decompose our variance estimates under Wishart/multivariate normal assumptions. Both PCA-based and factor analysis-based approaches allow us to find a common space of low dimension, but the latter is prescriptive in this sense while the former inherits the exploratory nature of PCA. 

Perhaps the most direct multi-group generalization of PCA is CPCA \citet{flury}, which assumes that there exists an orthogonal matrix that can approximately diagonalize the covariance matrices of all $S$ groups simultaneously. Using a generalized PCA approach, the dimension of the common space can be chosen after estimating the full loadings matrix, based on which $q$ loading vectors are associated with the largest variances for all groups simultaneously. Unlike the usual PCA, CPCA assumes that the sample covariance matrices follow a Wishart distribution, and so the common loadings matrix is estimated in a maximum likelihood framework. More recently Flury's algorithm was revisited by \citet{trendafilov}, who criticized the fact that FCPCA does not constrain the eigenvalues of each group to be simultaneously decreasing, which in some cases could disallow its use as a dimension reduction strategy. Instead, \citet{trendafilov} suggested the stepwise algorithm, which finds the common loadings sequentially in order of variance explained. In the present study, we compared the performance of FCPCA and SCPCA for simultaneously decomposing the $S$ covariance matrices that have been estimated using either PoissonPCA or PLNPCA.

The second model under consideration is MSFA \citep{deveet}, which is an extension of classical factor analysis and likewise assumes that the latent variables and measurement error are multivariate normal and hence that the observations have a multivariate normal marginal distribution. However, in MSFA, there are $q$ latent common factors and $\ell_1,\dots,\ell_S$ latent unique factors, and so $S+1$ loadings matrices have to be estimated by maximum likelihood, which is a much more difficult optimization problem than that posed by classical factor analysis. Also, as with any factor analytic approach, dimensions of the latent subspaces must be considered a hyperparameter of the generative model. 

Finally, after estimating the common loadings with either SCPCA, FCPCA, or MSFA, we want to express the underlying abundances in each sample with respect to our new common basis, and we will refer to these quantities as the scores.


\subsubsection{Common Principal Components Analysis}
Let $\Sigma_s \in \mathbb{R}^{p \times p}$ be symmetric and positive definite and assume that each $(n_s-1) \hat{\Sigma}_s$ is independently distributed as $\mathrm{W}_p(n_s-1, \ \Sigma_s)$, \ $s=1,\dots S$, where $\mathrm{W}_p$ is the $p$-variate Wishart distribution. If there is a rotation matrix $V \in O(p)$ (where $O(p)$ denotes the set of orthogonal $p \times p$ matrices) for which
\begin{align}
   V^T \Sigma_s V = D_s  \label{eq:cpc}
\end{align}
for all $s$, where $D_s=\diag(d_{s1},\dots,d_{sp})$, then the subspace spanned by the columns of $V$ is common to all groups; the assumption in CPCA is that $V$ exists as such. Hence, CPCA comes down to simultaneous diagonalization (or an approximation thereof) of the $S$ sample covariance matrices. Fortunately, for $\sighatone,\dots,\sighatS$ positive-definite, their quadratic forms are strictly convex, and so the optimization problem posed in CPCA is highly tractable. The log-likelihood is given by
\begin{align*}
    \lik(\Sigma_1,\dots,\Sigma_S)&= \log \prod_{s=1}^S p(\sighat;\Sigma_1,\dots,\Sigma_S)  \\
    &\propto \sum_{s=1}^S \big(-\frac{n_s-1}{2} \log |\Sigma_s|-\frac{n_s-1}{2} \mathrm{tr} (\Sigma_s^{-1}\hat{\Sigma}_s)\big),
\end{align*}
which easily leads to the minimization problem
\begin{align}
\hat{V}&=\argmin_{\substack{v_h^Tv_j = 0, \ h\ne j \\ v_h^Tv_j = 1, \ h=j}}\sum_{s=1}^S (n_s-1) \big(\sum^p_{j=1}(\log d_{sj}+\frac{v^T_j \hat{\Sigma}_s v_j}{d_{sj}}) \big). \label{eq:objfn}
   \end{align}
From this, \citet{flury} obtains $D_s$, the diagonal $p \times p$ matrix whose entries $d_{s1},\dots,d_{sp}$ are the optimal unique eigenvalues and the diagonal entries of $V^T\sighat V$. The minimum also satisfies
\begin{align}
  v_h^T\big(\sum^S_{s=1}(n_s-1) \frac{d_{sj}-d_{sh}}{d_{sh}d_{sj}}\sighat \big) v_j=0 \label{eq:flury}
\end{align}
for $h,j=1,\dots,p$ and $h \neq j$, that is, that $\sum^S_{s=1}(n_s-1)(D_s^{-1}V^T \sighat V)$ is symmetric.

FCPCA uses the FG algorithm \citep*{fluryandgaut} to compute $\hat{V}$, in which all pairs of vectors of $V$ are rotated in each iteration to satisfy \cref{eq:flury}. Using FCPCA we can reduce the basis to $v_1,\dots,v_q$ only provided that the last $p-q$ eigenvalues are small for all $S$ groups. 

On the other hand, SCPCA \citep{trendafilov} starts with the same objective function, but performs minimization to find the optimal axes sequentially based on the fact that if covariance matrices $C_1,\dots, C_S$ share a common eigenvector $e$, then $e$ is also an eigenvector of the average of $C_1,\dots,C_S$ weighted by their unique eigenvalues associated with $e$. Using the results of \citet{flury}, \citet{trendafilov} writes the same CPCA minimization problem given by \cref{eq:objfn} more compactly as
\begin{align}
    \hat{V}=\argmin_{\substack{v_h^Tv_j = 0, \ h\ne j \\ v_h^Tv_j = 1, \ h=j}}\sum_{s=1}^S (n_s-1) \big(\sum^p_{j=1}\log (v_j^T\sighat v_j) \big). \label{eq:step}
\end{align}
If the eigenvalues estimated by FCPCA for each estimated covariance matrix are all simultaneously decreasing, then it can be shown that one can solve a sequence of minimization problems that gives the $v_j$ in ascending order of the minima, such that obtaining the $j^\text{th}$ eigenvector associated with the $j^\text{th}$ largest eigenvalue yields the same result as FCPCA. However, if the FCPCA eigenvalues are not simultaneously decreasing in all $S$ groups, then the stepwise approach will not solve \cref{eq:step}. \citet{trendafilov} argues that despite this, the stepwise solution is useful because it can always be used to find a set of $q \leq p$ common principal component vectors forming a basis for $\RR^q$, such that their variances $d_{s1},\dots, d_{sq}$ are approximately simultaneously decreasing for all $s=1,\dots,S$, and all $d_{1j},\dots,d_{Sj}$ are as similar as possible for a given $j$, $j=1,\dots,q$.

FCPCA and SCPCA were implemented using code adapted and modified from source code in the R packages \textbf{multigroup} and \textbf{cpca} respectively.


\subsubsection{Multi-Study Factor Analysis}
Let $X_{is}\in \RR^p$, $i=1,\dots,n_s$, $s=1,\dots,S$ be a random vector with $n_s$ realizations in the $s^\text{th}$ group. MSFA assumes that there exist iid latent variables $f_{is} \in \RR^q$ and $w_{is} \in \RR^{\ell_s}$, which generate $X_{is}$ by
\begin{align*}
    X_{is}=\Phi f_{is}+\beta_s w_{is}+\mu_s+\epsilon_{is}, \ \ \text{with} \ \ f_{is}\sim\mathcal{N}(0_q,\text{I}_q), \ w_{is}\sim \mathcal{N}(0_{\ell_s},\text{I}_{\ell_s})
\end{align*}
for $i = 1, \dots , n_s$, \ $s=1, \dots, S$, where $\mu_s\in\RR^p$ is the mean vector, $\epsilon_{is}$ are iid with $\epsilon_{is}\sim \normal(0_p,\Psi_s)$, where $\Psi_s=\diag(\psi_{s1},\dots,\psi_{sp})$, and $\epsilon_{is}$ is independent from the latent factors $f_{is}$ and $w_{is}$. $\Phi$ is a $p \times q$ matrix of common loadings, and $\beta_s$ is a $p \times \ell_s$ matrix of group-specific loadings, which provide structure to the latent factors. Alternatively, we can say that each $X_{is}$ is conditionally independent given the latent variables $f_{is}$ and $w_{is}$ as follows: \\

    $X_{is}|f_{is},w_{is} \sim \mathcal{N}(\Phi f_{is}+\beta_s w_{is} + \mu_s, \Psi_s) \ \ \text{with} \ \ f_{is}\sim\mathcal{N}(0_q,\text{I}_q), \ w_{is}\sim \mathcal{N}(0_{\ell_s},\text{I}_{\ell_s}).$ \\
    
Since there is a multivariate normal distribution for $X_{is}$ conditional on the multivariate normal latent variables, $X_{is}$ has a multivariate normal marginal distribution $X_{is} \sim \normal(\mu_s,\Sigma_s=\Phi \Phi^T + \beta_s \beta_s^T + \Psi_s)$. Hence, the log-likelihood is given by 
\begin{align*}
    \lik(\Phi, \beta_s, \Psi_s)&=\log \prod_{s=1}^S \prod_{i=1}^{n_s} p(X_{is}|\Phi, \beta_s,\Psi_s)\\
    &\propto \sum_{s=1}^S \big(-\frac{n_s}{2} \log |\Sigma_s|-\frac{n_s}{2} \mathrm{tr} (\Sigma_s^{-1}\hat{\Sigma}_s)\big).
\end{align*}

From this we see that even though MSFA is designed for normally distributed data, we can use it as the multi-group step in our ensemble method because its likelihood depends only on the estimated variance-covariance matrices for each group. 

As identifiability constraints, the authors impose that $\Phi$, $\beta_1, \dots, \beta_S$ all be lower triangular matrices, which is typical of classical factor analysis and forces the first loading to correspond only to the first axis of the factor space, the second loading to the first and second axes, and so on and so forth. The authors further note that an additional constraint---that $\rank \big([\Phi \ \beta_1 \dots \beta_S]\big)=q+\sum^S_{s=1}\ell_s$---is needed to ensure uniqueness of the solution, since we have to estimate $S$ group-specific loadings plus the common loadings from the $S$ covariance matrices. $\hat{\Phi}$, $\hat{\beta}_1,\dots,\hat{\beta}_S$, and $\hat{\Psi}_s$ are estimated by expectation-conditional maximization (ECM). Of course, we are interested only in $\hat{\Phi}$, the matrix that describes how the latent factors characterizing the common signal are weighted to generate the observed data.

MSFA was implemented using the source code from the R package \textbf{msfa} \citep{deveet}, modified so as to use $\sighatone,\dots,\sighatS$ estimated from PoissonPCA/PLNPCA instead of from the standard sample covariance matrices. Additionally, for our method we require a subspace spanned by a set of orthonormal vectors in order of the variance on the common factors, and so what we seek are actually the rotated loadings $v_1,\dots,v_q$, or the first $q$ columns of $V \in O(p)$ computed from $\hat{\Phi}\hat{\Phi}^T$ by the following spectral decomposition:
\begin{align}
    \hat{\Phi}\hat{\Phi}^T=V A V^T,  \label{eq:phiphi}
\end{align}
where $A$ is the diagonal matrix of eigenvalues.


\subsubsection{Computing Scores}
Since, in all cases, for any given multivariate observation we are interested in the scores of the unobserved $\log \Lambda_{is}$ rather than the scores of the observed $X_{is}$, we adopt the following procedure from \citet*{kenney}. To apply the classical PCA criterion of minimizing the squared reconstruction error between the original points and their projections onto principal component space, we minimize the squared reconstruction error between $\log \Lambda_{is}$ and its projection $P_{is}^q$ onto the $q$-dimensional common subspace spanned by the orthonormal vectors $v_1,\dots,v_q$ that we estimated using CPCA or MSFA. Since $\Sigma_s$ is the variance of $\log\Lambda_s$, this error is given by the squared distance
\begin{align}
    D^2=(\log\Lambda_{is}-\mu_s-P_{is}^q)^T\Sigma_s^{-1}(\log\Lambda_{is}-\mu_s-P_{is}^q), \label{eq:squar}
\end{align}
where, because the $v_j$'s are orthonormal, the projection is $P_{is}^q=\sum^q_{j=1}v_jv_j^T(\log\Lambda_{is}-\mu_s)$.
At the same time, since $\log \Lambda_{is}$ is unobserved, we should still seek to maximize the likelihood of the observed data. Whether we use PoissonPCA or PLNPCA to estimate $\Sigma_s=\Var{(\log\Lambda)}$, the underlying assumption is always that $X_{ijs}$ was generated by a Poisson distribution with mean $\Lambda_{ijs}$. So as in the single-group case in \citet*{kenney}, by combining the Poisson log-likelihood and \cref{eq:squar}, we arrive at an objective function of the form
\begin{equation*}
    L(\Lambda_{is})=\sum^{n_s}_{i=1}\big(X_{is}^T\log \Lambda_{is}-1^T\Lambda_{is}-(\log\Lambda_{is}- \mu_s-P_{is}^q)^T\Sigma_s^{-1}(\log\Lambda_{is}-\mu_s-P_{is}^q)\big).
\end{equation*}
This is optimized by Newton-Raphson iteration in the \textbf{PoissonPCA} R package, and we implement the procedure using adapted portions of this code.

\subsection{Ensemble Method}
We have reviewed two very different ways of estimating the full- (or near full-) rank variance-covariance matrix from a set of conditionally independent realizations of Poisson sampling in which the Poisson means are subject to additional multiplicative noise: PoissonPCA and PLNPCA, each of which can either be performed with a sequencing depth correction or without. These methods are applied to each data set $X_{1},\dots,X_{S}$ individually. We went on to explore two distinct methods that can take a set of $S$ estimated variance-covariance matrices and estimate a set of $q<p$ common vectors forming a shared orthogonal basis for $\RR^q$: CPCA (for which we have a choice of two algorithms, SCPCA and FCPCA) and MSFA. The twelve possible combinations of variance estimation and common factor extraction techniques are given in \cref{tab:methods}

\subsection{Simulation Study}
We performed simulation studies of two synthetic groups of multivariate Poisson log-normal observations across several scenarios. These scenarios differed on the true signal (including the number of eigenvectors that were common to both groups' variance-covariance matrices, and whether the eigenvalues of the variance-covariance matrices were simultaneously decreasing), the sample sizes $n_1$ and $n_2$, whether or not the sample sizes were balanced, and whether or not sequencing depth correction was performed in the variance estimation stage. For each simulation experiment, the process of simulating Poisson log-normal data and applying each method was performed 100 times. For each of the 100 replicates, synthetic data for two ``groups" were simulated as follows. Synthetic eigenvectors $E_1$ were constructed for Group 1 by spectral decomposition of a synthetic covariance matrix $FF^T$ where each column $F_j$, $j=1,\dots,p$ was a normalized length-$p$ vector of standard normal variates. We then constructed the eigenvectors $E_2$ for Group 2 so as to share the first $q$ columns of $E_1$ for several values of $q$, while the remaining columns $e_{2,q+1},\dots,e_{2,p}$ were replaced by vectors of standard normal variates regressed on the preceding $q$ columns and normalized. These eigenvectors, along with a pre-determined set of eigenvalues for each group, were used to construct variance-covariance matrices $\Sigma_1$ and $\Sigma_2$. Note that with decreasing eigenvalues, for $q<<p$, by this construction each shared eigenvector will be a principal eigenvector of the two covariance matrices, whereas with non-decreasing eigenvalues some of the large variances will not be associated with the shared axes: we performed simulation experiments for both these scenarios. Next, these covariance matrices $\Sigma_1$ and $\Sigma_2$ in turn were used to simulate the transformed latent Poisson means $\log \Lambda_1$ and $\log \Lambda_2$ using the multivariate normal, with mean vectors for each group consisting of $p$ normal random variates (see \cref{tab:simtable}) such that the means differed between simulation replicates only. We then performed scalar multiplication of $\Lambda_1$ and $\Lambda_2$ respectively by length-$n_1$ and length-$n_2$ vectors of gamma random variates to simulate sequencing depth error, and finally these means were used to generate $n_1 \times p$ and $n_2 \times p$ synthetic data matrices of Poisson random variates for each group respectively (see \cref{tab:simtable}). 
\begin{table}[]
\begin{centering}
\begin{tabular}{@{}lll@{}}
\toprule
                                                                           & \textbf{Group 1}                                  & \textbf{Group 2}                     \\ \midrule
$\log \Lambda_{is}\sim\normal(\mu_{is},\Sigma_s)$ \ \ \ \ \ \ \ \ \ \ \ \ \ & $\mu_{ij1}\sim\normal(4, 3)$                      & $\mu_{ij2}\sim\normal(3,2)$          \\
$X_{ijs}\sim\text{Poisson}(\gamma_{is}\Lambda_{ijs})$                      & $\gamma_{i1}\sim\text{Gamma}(7, 1)$ \ \ \ \ \ \ \ & $\gamma_{i2}\sim\text{Gamma}(10, 1)$ \\ \bottomrule
\end{tabular}
\caption{Distributions used to simulate Poisson log-normal data.}
\label{tab:simtable}
\end{centering}
\end{table}

Then, the candidate ensemble methods listed in \cref{tab:methods} and some single-group alternatives were performed on the synthetic data. The single-group methods were all run on the data from the two groups concatenated together, and these methods comprised PoissonPCA, PLNPCA, naive PCA on counts, naive PCA on log counts, naive PCA on relative abundances, and naive PCA on log relative abundances. Before log-transforming count or relative abundance data for naive PCA, zero values were first imputed with 0.001.

In the case of SCPCA and FCPCA, the explained variances (eigenvalues) for each estimated orthogonal loading vector were computed by 
\begin{align}
    \hat{d}_{sj}= \hat{v_j}^T \Sigma_s \hat{v_j}, \ j=1,\dots,p, \ s=1,2 \label{eq:simexplain}
\end{align}
where $\Sigma_s$ is the true variance-covariance matrix of $\log\Lambda_s$. Cumulative sums of $\hat{d}_{11},\dots,\hat{d}_{1p}$ and $\hat{d}_{21},\dots,\hat{d}_{2p}$ were divided by the true eigenvalues $\sum^p_{j=1}d_{1j}$ and $\sum^p_{j=1}d_{2j}$ respectively to find the proportion of the true variance explained by each method.

In the case of MSFA, as the common loadings are not constrained to orthogonality, the variance $\hat{\Phi}\hat{\Phi}^T$ of the common factors was computed and then eigen-decomposed as described in \cref{eq:phiphi}. The resultant $\hat{v}_1,\dots,\hat{v}_p$ (of which only the first $q$ contain signal, but all $p$ are retained in this case for ease of visibility in plots) were used to compute the estimated eigenvalues and the proportion of true variance explained as according to \cref{eq:simexplain}.

\subsection*{Real Data Analysis}
\citet{zeller} and \citet{feng} collected fecal samples and processed them as described in their respective publications. Although each team had their own bioinformatic pipeline to process the raw reads, the taxonomy tables used for our data analysis were obtained from the R package \textbf{curatedMetagenomicData} \citep{pasolli}, whose authors applied a standard pipeline for assembly, gene prediction, and taxonomic assignment to the raw files from each study. 

The taxonomic abundance tables for each dataset included strain-level taxa from all three domains of life, as well as viruses. Since our candidate methods do not involve any regularization, we separately collapsed the data to the genus level and removed features with near-zero variance to reduce the number of taxa to $p=104 < \min(n_Z=199,n_F=154)$. We then subsetted the features to include only those common to the two datasets, and ran the twelve candidate combinations of methods listed in \cref{tab:methods} just as we did for the simulation experiments. We then computed the scores by projecting the latent means into the common space.

For disease state prediction, we collapsed colorectal adenoma labels together with control labels. We used the R package \textbf{caret} for data splitting and model training. As a benchmark of the discriminating signal in the data, we trained random forest models (using the method implemented in the R package \textbf{randomForest}) on 80\% of the full genus-level concatenated data (in which the training/testing split was determined by disease state) using five-times-repeated five-fold cross validation. We fit 5000 trees and tuned the hyperparameter \textit{mtry} based on classification accuracy. We used the final model to predict disease state on the test set, and computed AUC and accuracy. We then trained logistic regressions on the first 5 or 10 common scores for each method and similarly computed AUC and accuracies on the test sets. 

\newpage


\section*{Funding}
MGH was supported by scholarships from the Natural Sciences and Engineering Research Council of Canada (NSERC) and the Nova Scotia Health Research Foundation. HG was supported by NSERC RGPIN/05108-2017.

\section*{Competing interests}
The authors declare that they have no competing interests.

\section*{Authors' contributions}
MGH performed the analyses and wrote the manuscript. MGIL provided biological and practical insight. HG designed the statistical method.

\newpage
\begin{appendices}
\section*{Supplementary Results}
\renewcommand{\thefigure}{S.\arabic{figure}}
\setcounter{figure}{0}
	
\begin{figure}
    \centering
    \includegraphics[scale=0.2]{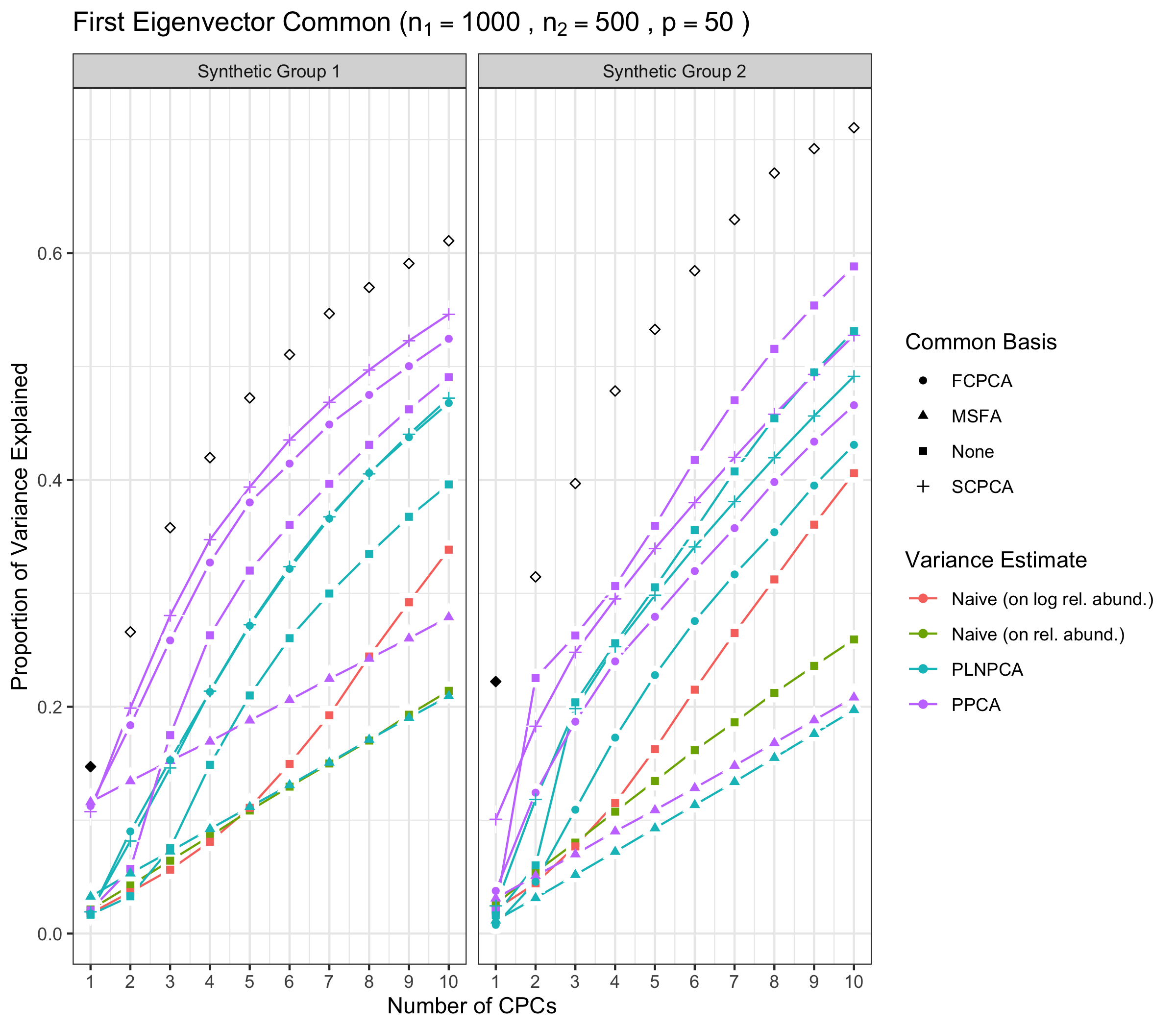}
    \caption{Simulation results for decreasing eigenvalues and one common eigenvector, with sequencing depth correction; $p$=50, $n_1=1000$, $n_2 = 500$. "None" as a common basis label means that Group 1 and Group 2 data were concatenated prior to variance estimation.}
 \label{app:fig:ag1}
\end{figure}

\begin{figure}
    \centering
    \includegraphics[scale=0.2]{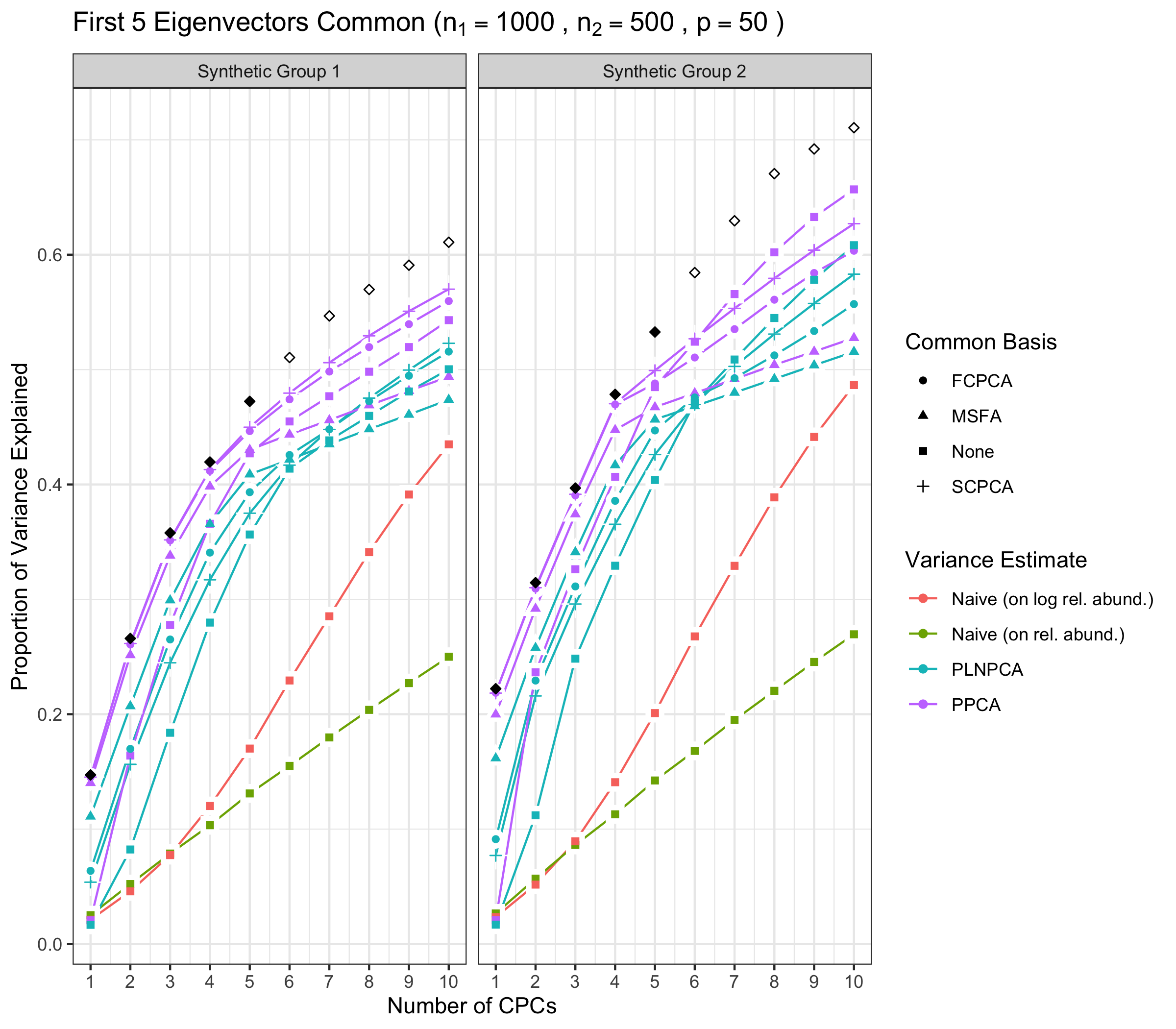}
    \caption{Simulation results for decreasing eigenvalues and five common eigenvectors, with sequencing depth correction; $p$=50, $n_1=1000$, $n_2 = 500$. "None" as a common basis label means that Group 1 and Group 2 data were concatenated prior to variance estimation.}
 \label{app:fig:ag15}
\end{figure}


\begin{figure}
    \centering
    \includegraphics[scale=0.2]{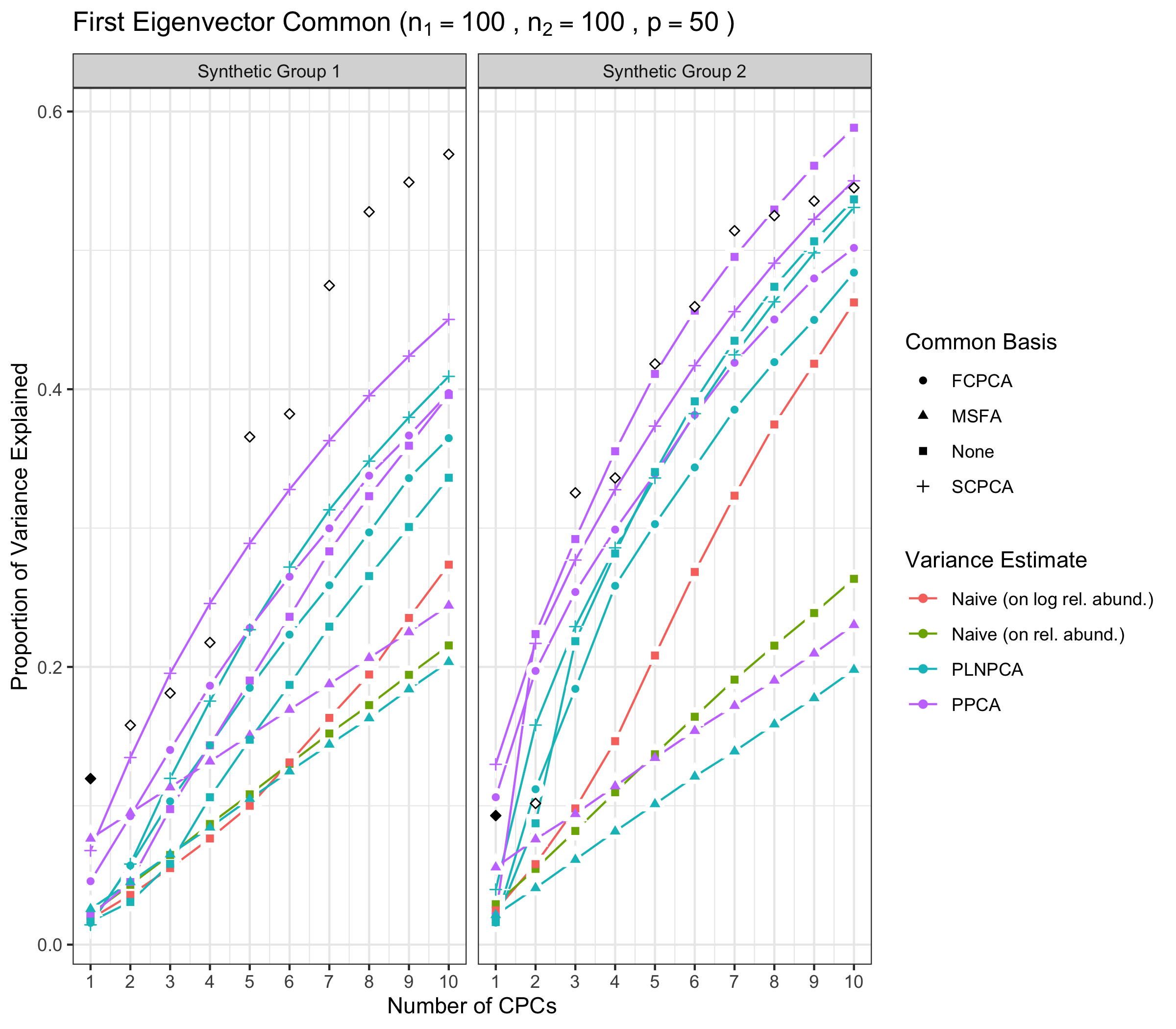}
    \caption{Simulation results for non-decreasing eigenvalues and one common eigenvector, with sequencing depth correction; $p$=50, $n_1=100$, $n_2 = 100$. "None" as a common basis label means that Group 1 and Group 2 data were concatenated prior to variance estimation.}
 \label{app:fig:ag2}
\end{figure}

\begin{figure}
    \centering
    \includegraphics[scale=0.2]{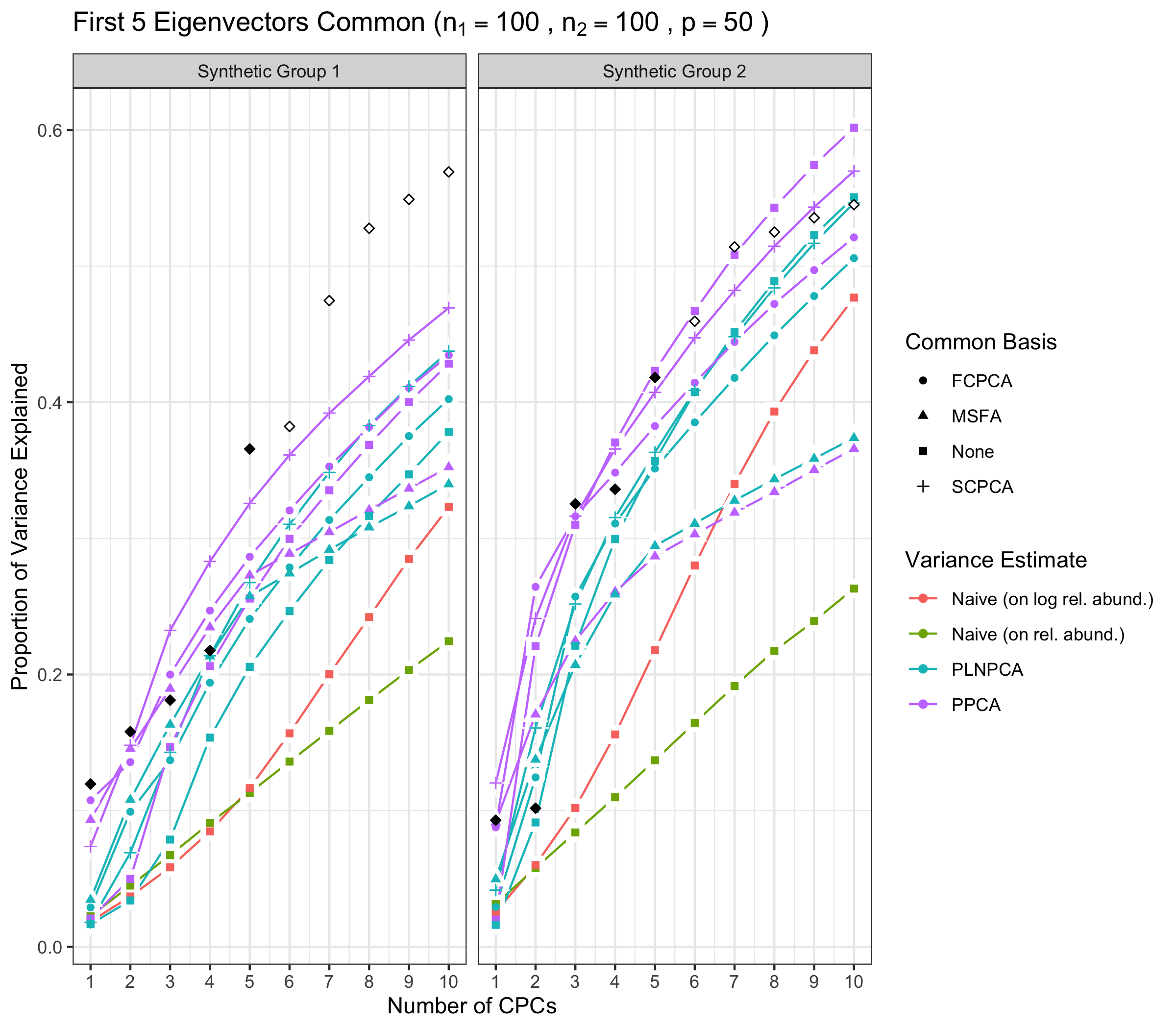}
    \caption{Simulation results for non-decreasing eigenvalues and five common eigenvectors, with sequencing depth correction; $p$=50, $n_1=100$, $n_2 = 100$. "None" as a common basis label means that Group 1 and Group 2 data were concatenated prior to variance estimation.}
 \label{app:fig:ag25}
\end{figure}


\begin{figure}
    \centering
    \includegraphics[scale=0.2]{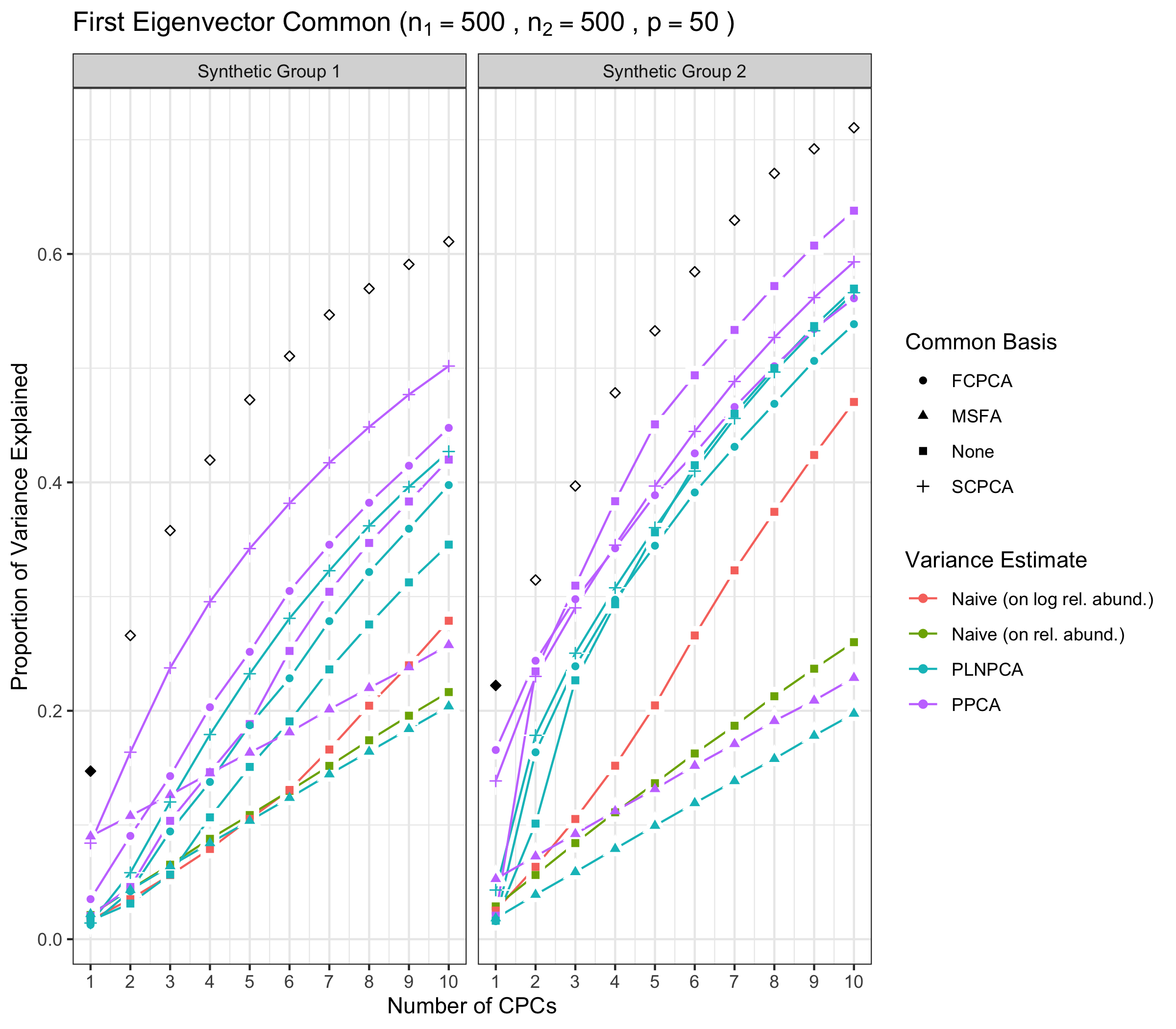}
    \caption{Simulation results for decreasing eigenvalues and one common eigenvector, with sequencing depth correction; $p$=50, $n_1=500$, $n_2 = 500$. "None" as a common basis label means that Group 1 and Group 2 data were concatenated prior to variance estimation.}
 \label{app:fig:ag1big}
\end{figure}

\begin{figure}
    \centering
    \includegraphics[scale=0.2]{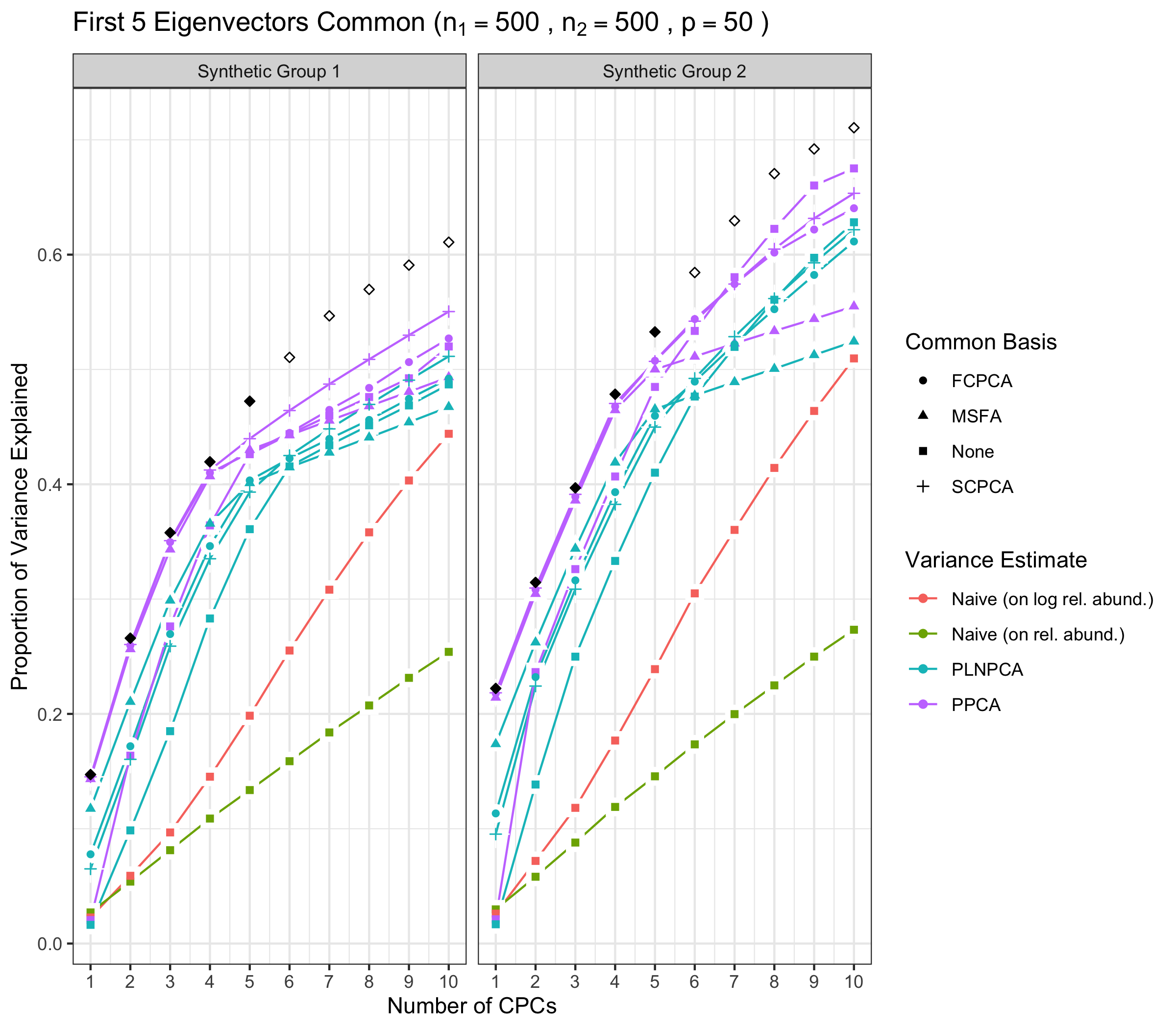}
    \caption{Simulation results for decreasing eigenvalues and five common eigenvectors, with sequencing depth correction; $p$=50, $n_1=200$, $n_2 = 500$. "None" as a common basis label means that Group 1 and Group 2 data were concatenated prior to variance estimation.}
 \label{app:fig:ag15big}
\end{figure}


\begin{figure}
    \centering
    \includegraphics[scale=0.2]{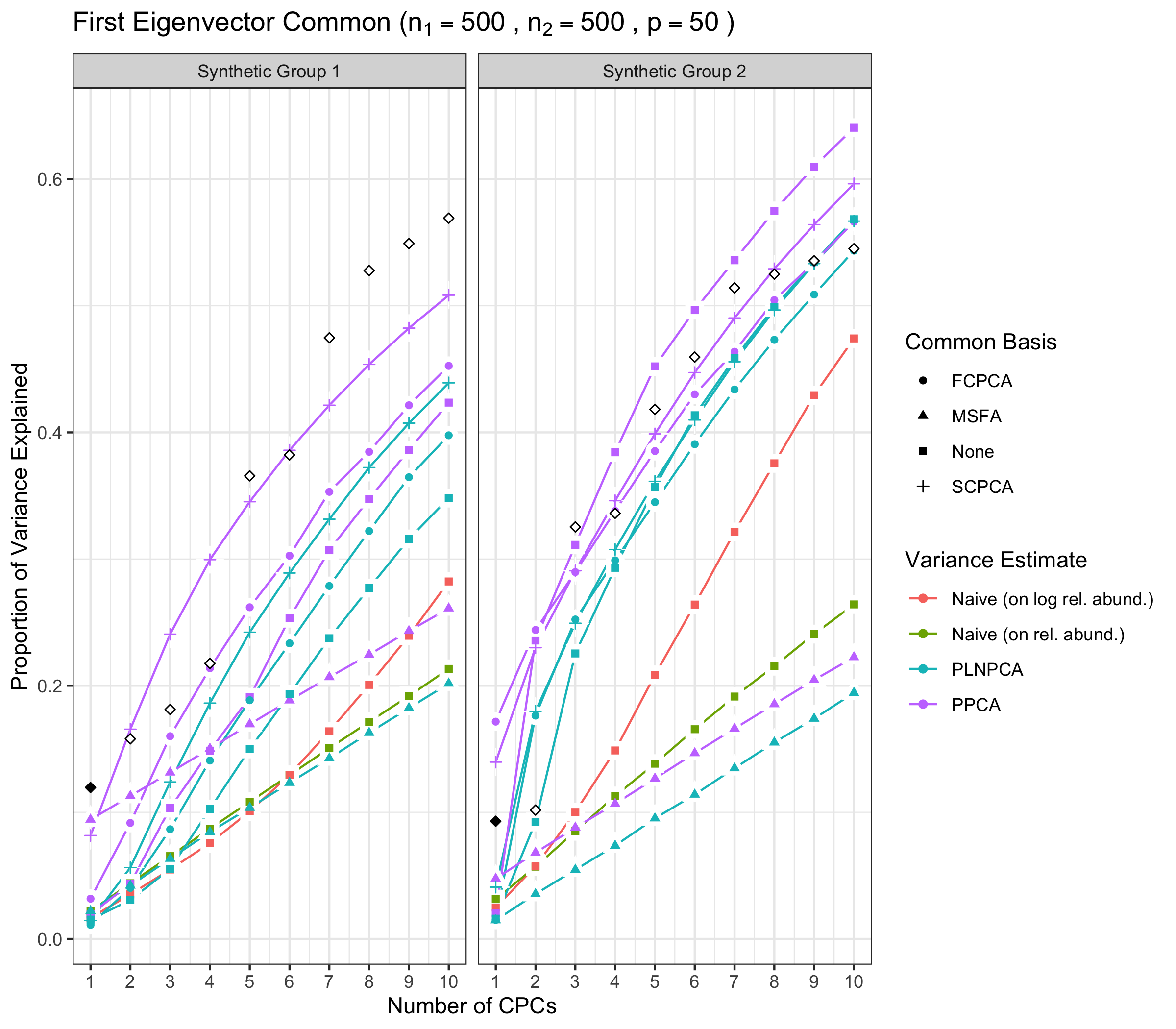}
    \caption{Simulation results for non-decreasing eigenvalues and one common eigenvector, with SDC; $p$=50, $n_1=500$, $n_2 = 500$. "None" as a common basis label means that Group 1 and Group 2 data were concatenated prior to variance estimation.}
 \label{app:fig:ag2big}
\end{figure}

\begin{figure}
    \centering
    \includegraphics[scale=0.2]{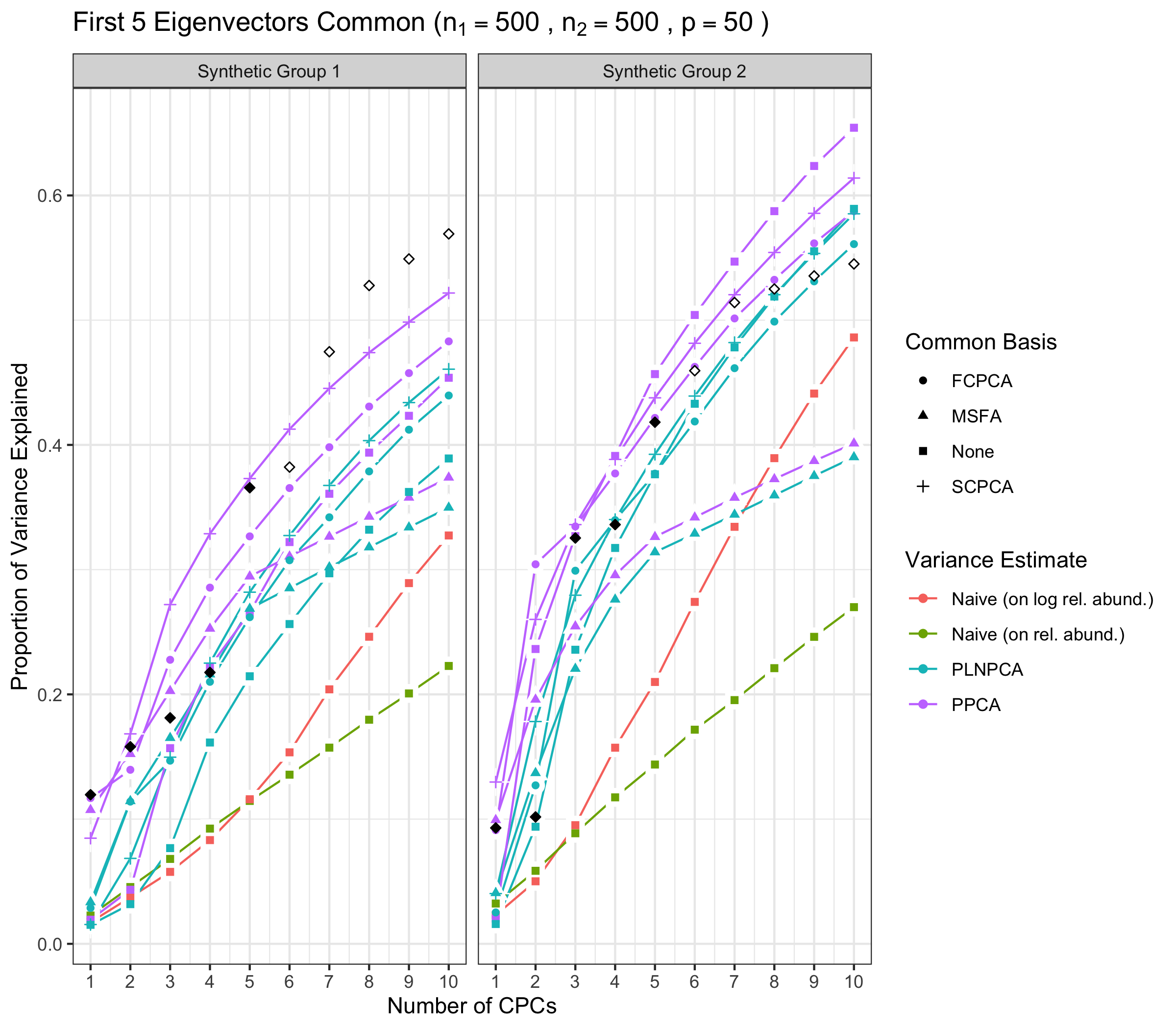}
    \caption{Simulation results for non-decreasing eigenvalues and five common eigenvectors, with SDC; $p$=50, $n_1=500$, $n_2 = 500$. "None" as a common basis label means that Group 1 and Group 2 data were concatenated prior to variance estimation.}
 \label{app:fig:ag25big}
\end{figure}



	    \begin{figure}
	    \centering
	  \includegraphics[scale=0.7]{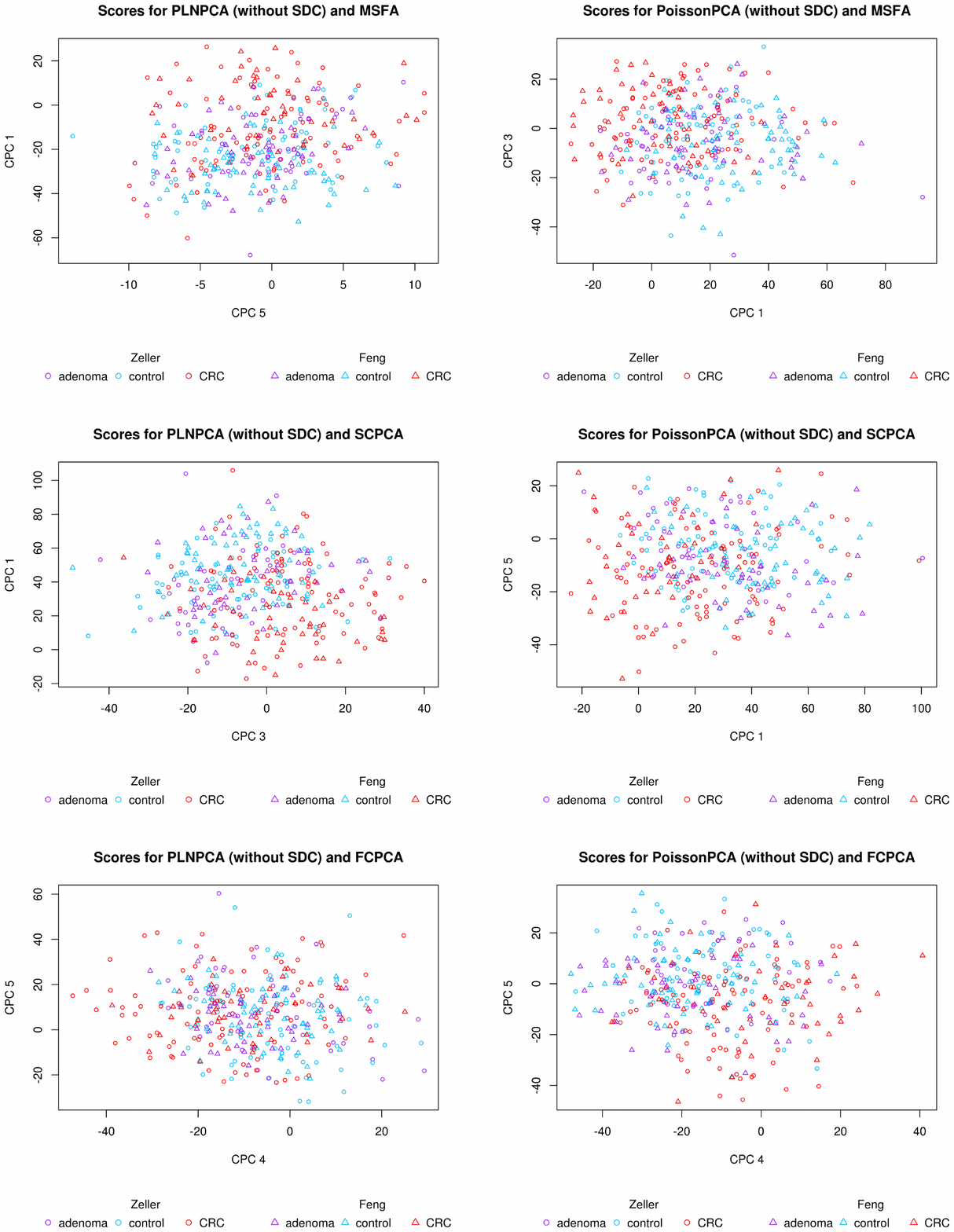}
    \caption{Scores from ensemble methods without SDC by disease state.}
    \label{app:fig:noseqs}
\end{figure}
	
		 	    \begin{figure}
	    \includegraphics[scale=0.85]{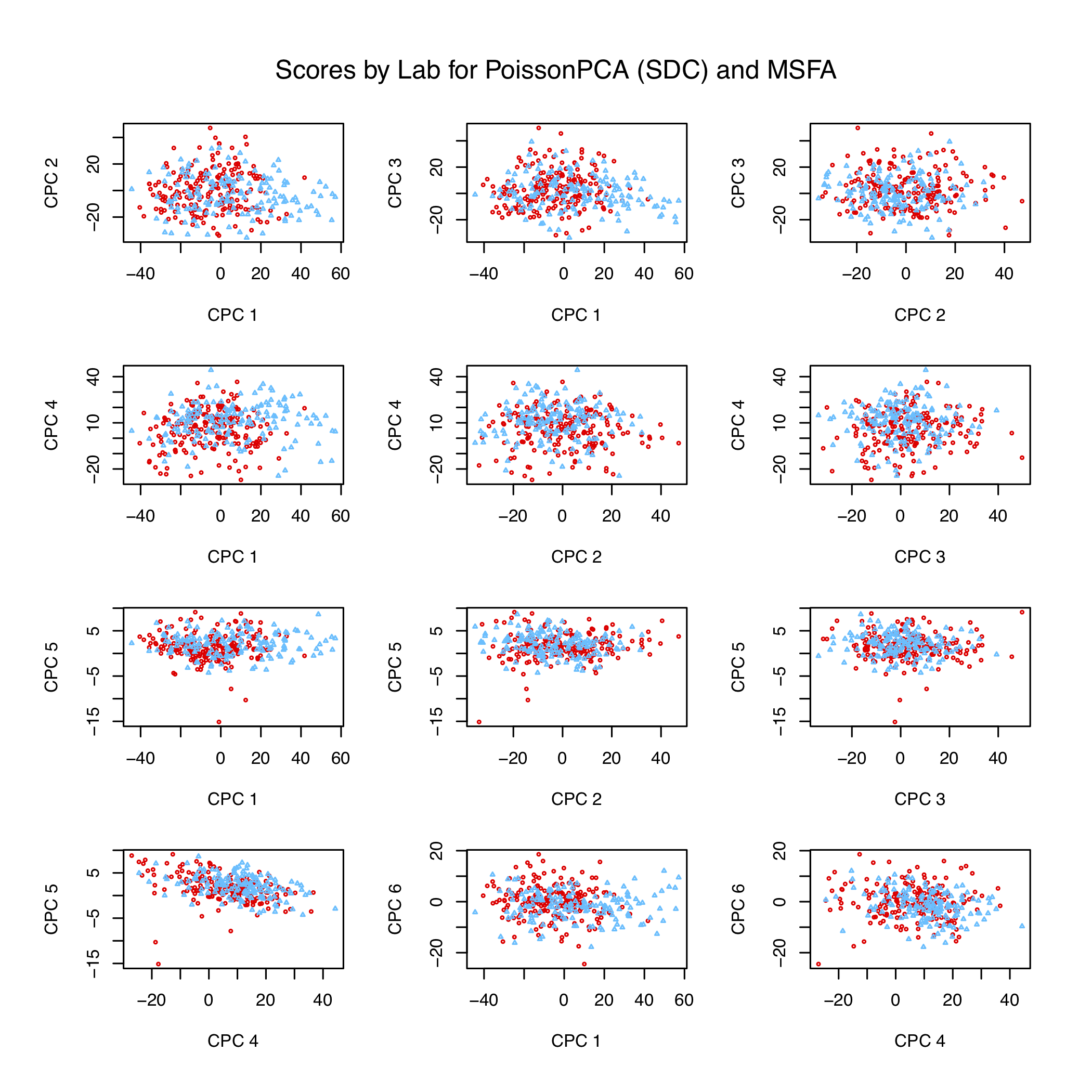} 
    \caption{Scores from PoissonPCA (SDC) and MSFA by study of origin.}
    \label{app:fig:bylab1}
\end{figure}
	          
	    \begin{figure}
    \includegraphics[scale=0.85]{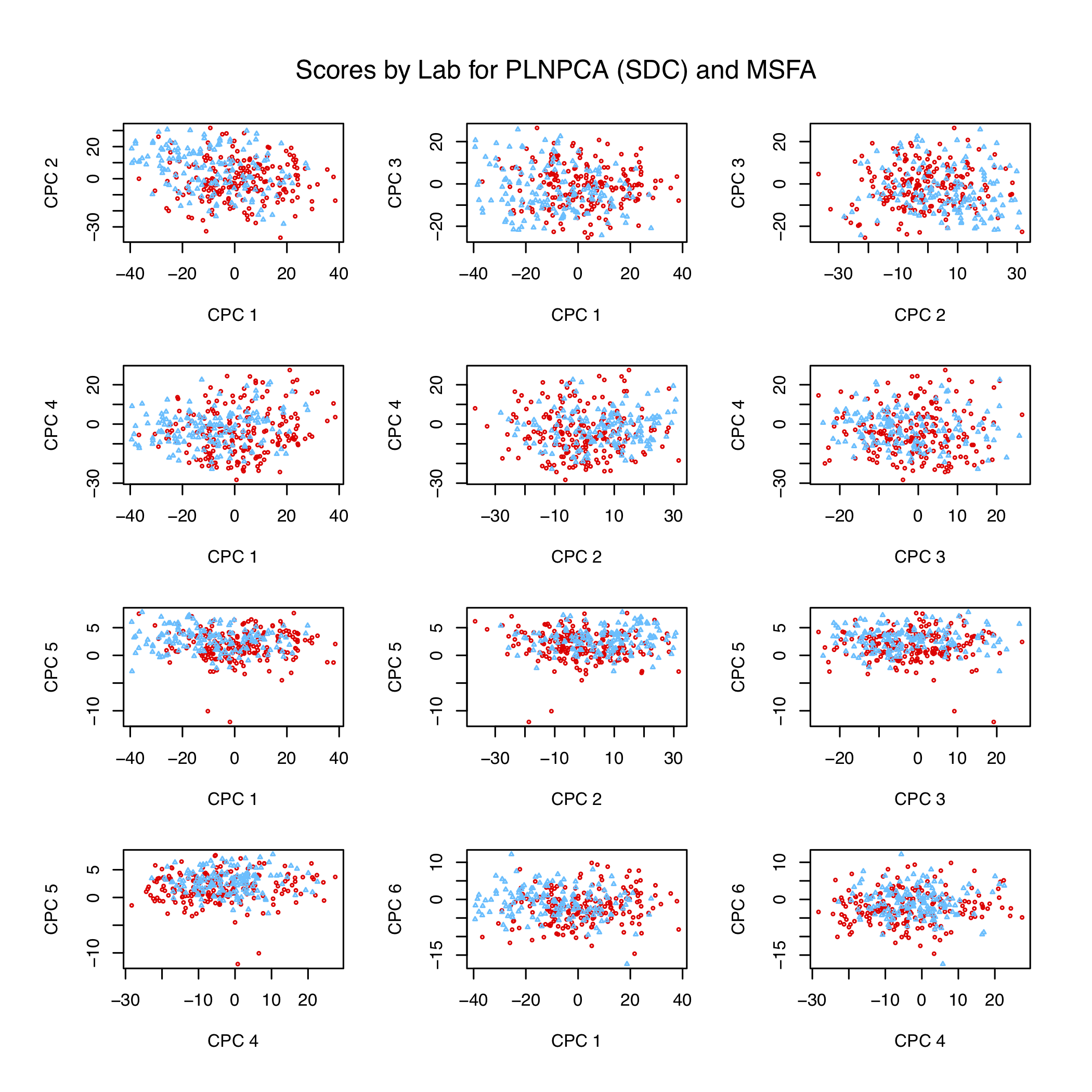} 
    \caption{Scores from PLNPCA (SDC) and MSFA by study of origin.}
    \label{app:fig:bylab2}
\end{figure}
	
	    \begin{figure}
    \includegraphics[scale=0.85]{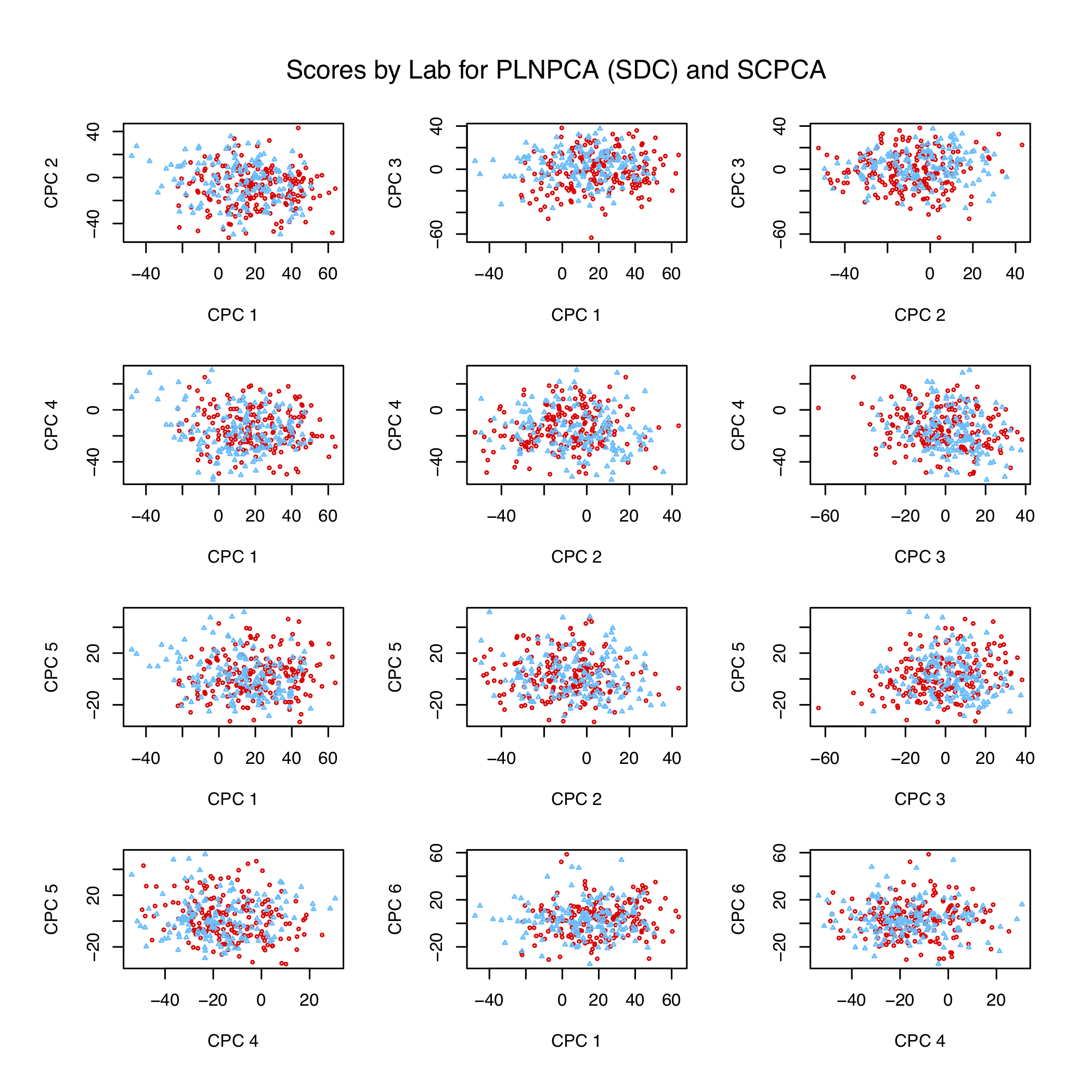} 
    \caption{Scores from PLNPCA (SDC)  and SCPCA by study of origin.}
    \label{app:fig:bylab4}
\end{figure}
	
		  	    \begin{figure}
    \includegraphics[scale=0.85]{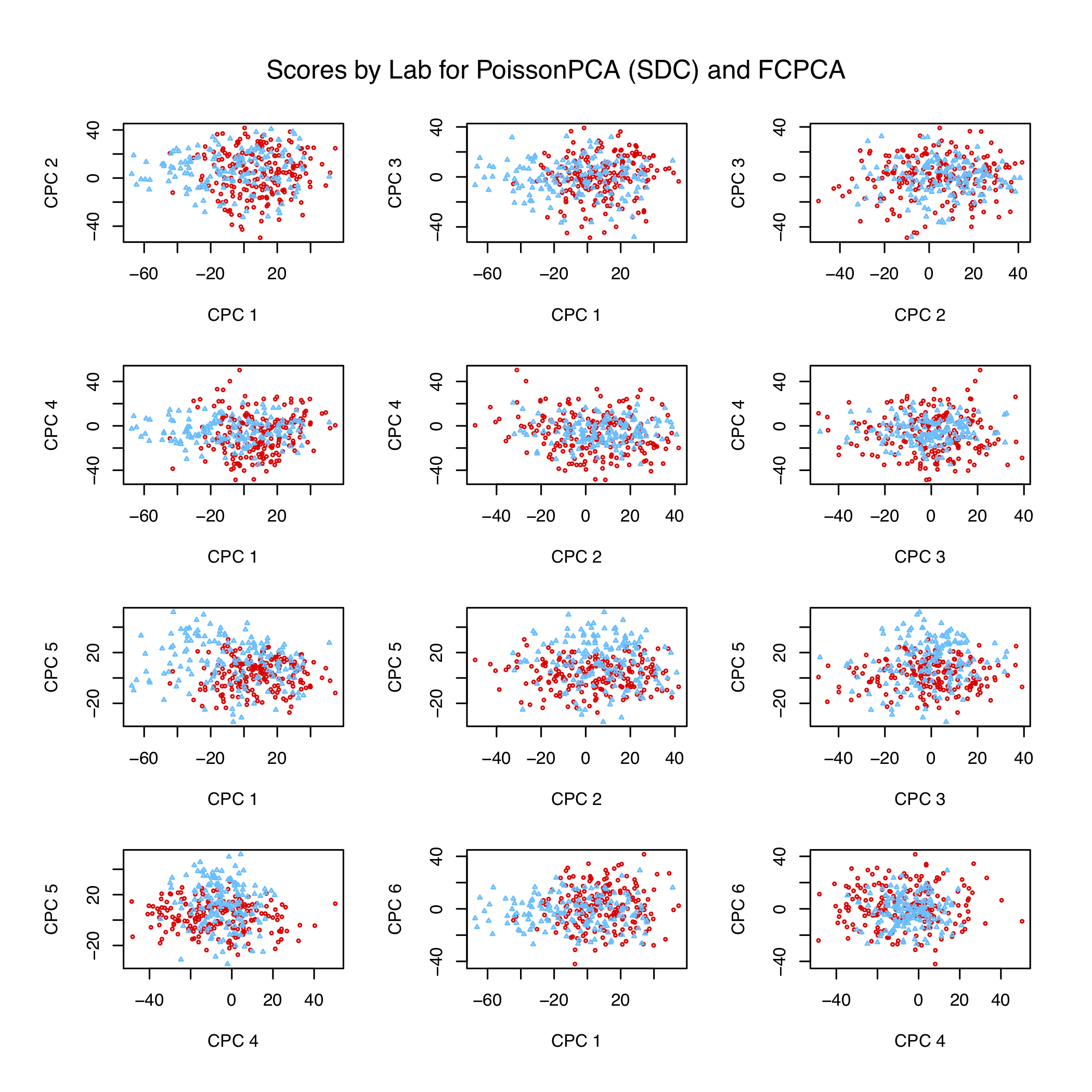}
    \caption{Scores from PoissonPCA (SDC) \& FCPCA by study of origin.}
    \label{app:fig:bylab5}
\end{figure}
	          
	  	    \begin{figure}
    \includegraphics[scale=0.85]{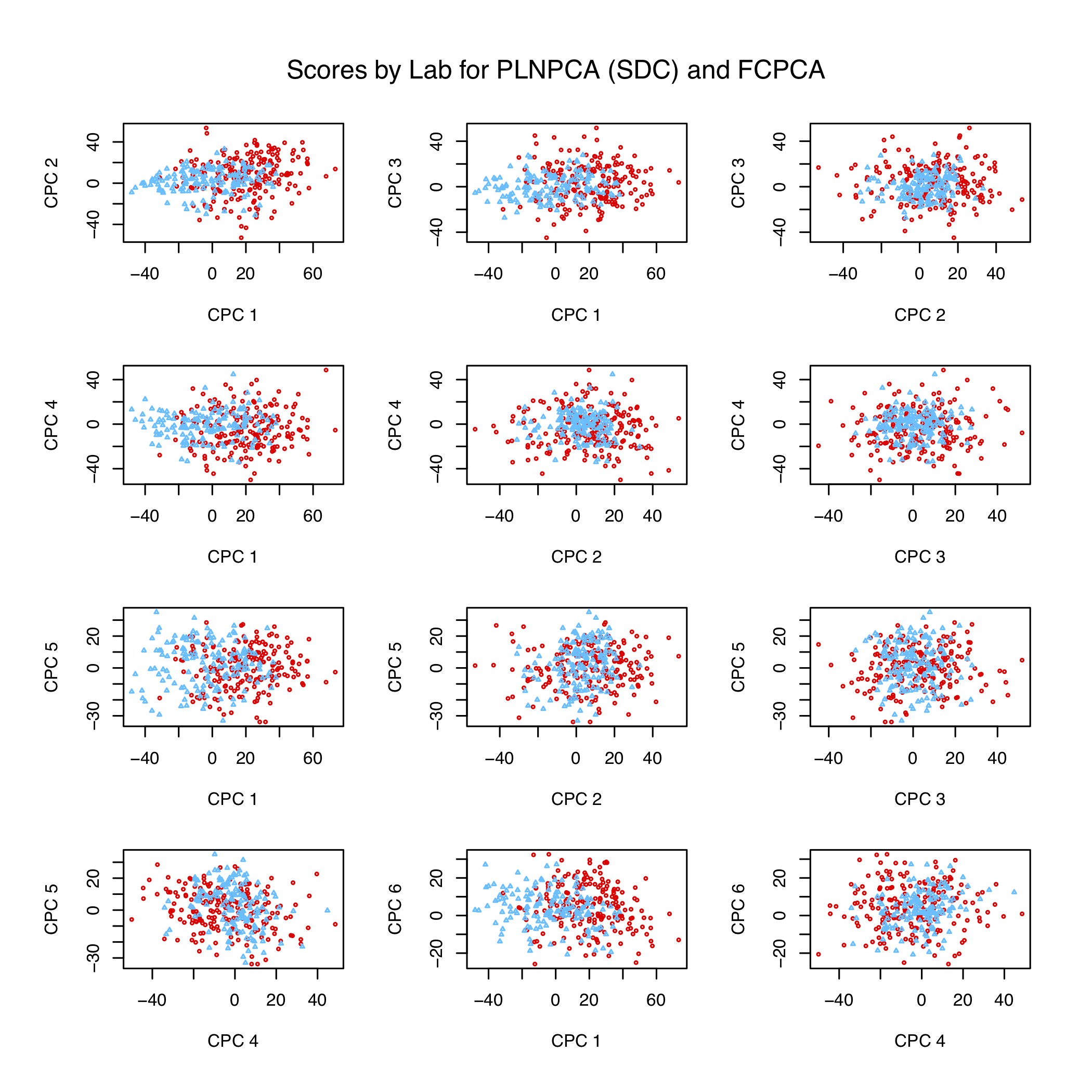}
    \caption{Scores from PLNPCA (SDC) and FCPCA by study of origin.}
    \label{app:fig:bylab6}
\end{figure}
	          
			 	    \begin{figure}
	    \includegraphics[scale=0.85]{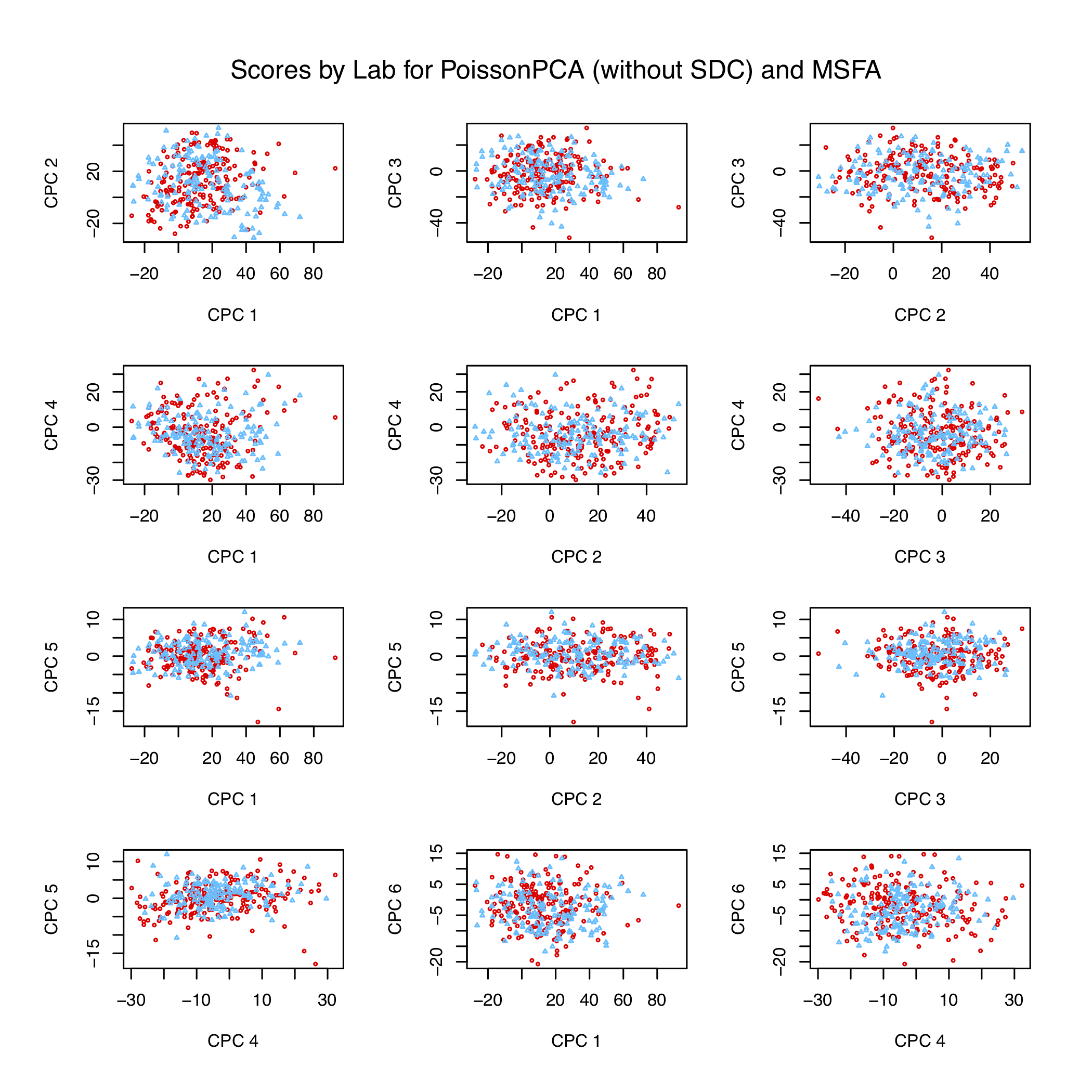} 
    \caption{Scores from PoissonPCA (no SDC) \& MSFA by study of origin.}
    \label{app:fig:bylab1no}
\end{figure}
	          
	    \begin{figure}
    \includegraphics[scale=0.85]{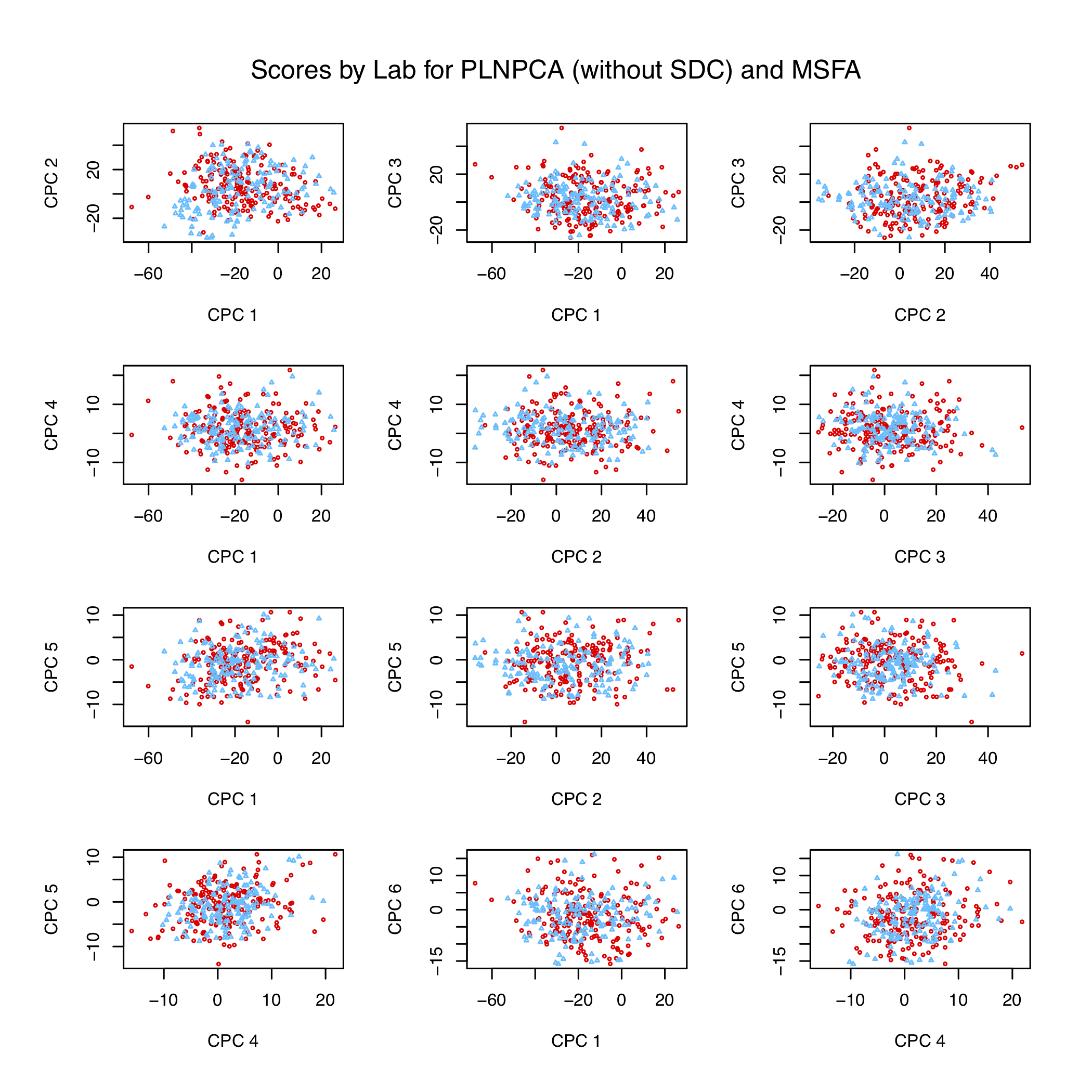} 
    \caption{Scores from PLNPCA (no SDC) and MSFA by study of origin.}
    \label{app:fig:bylab2no}
\end{figure}

  	    \begin{figure}
	    \includegraphics[scale=0.85]{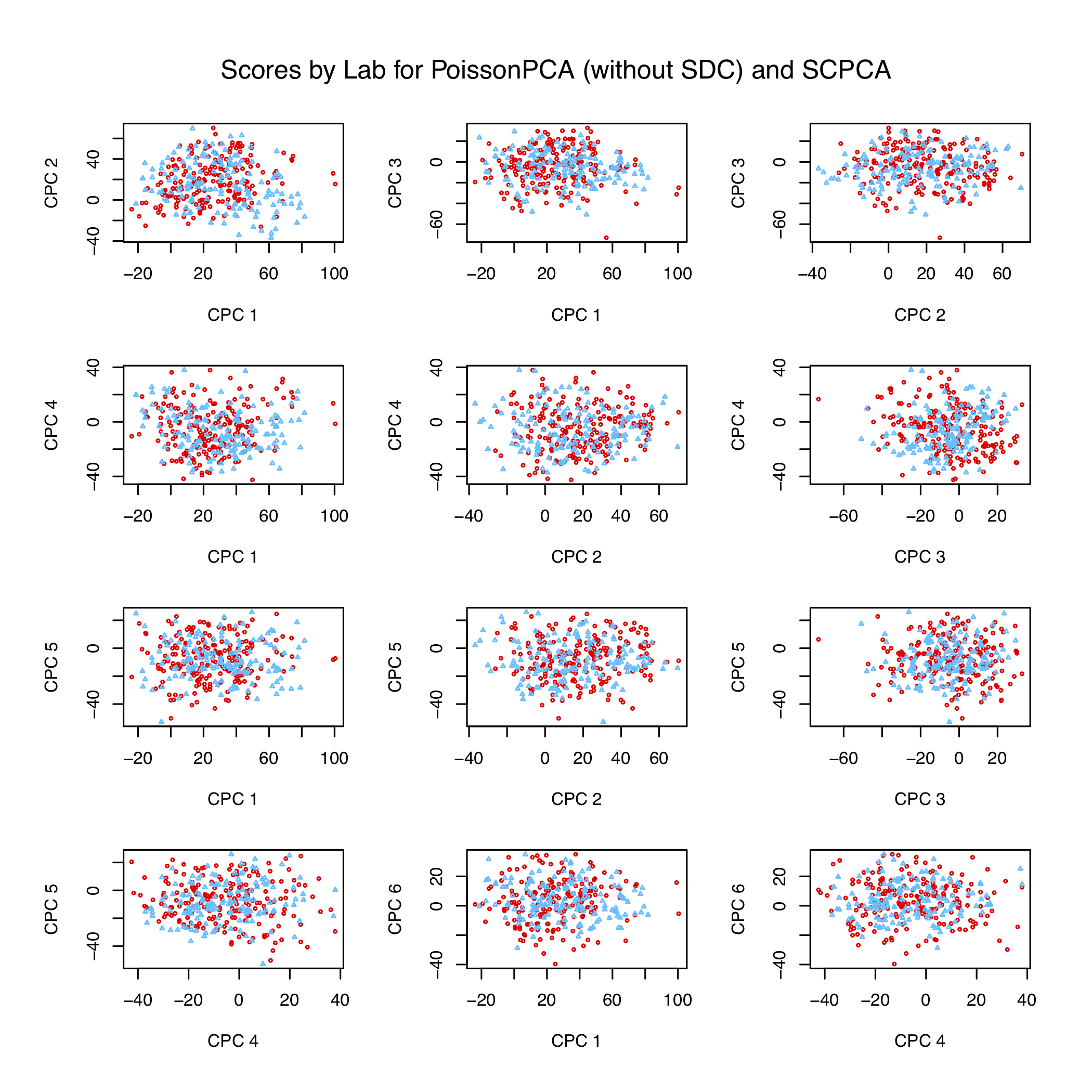} 
    \caption{Scores from PoissonPCA (no SDC) \& SCPCA by study of origin.}
    \label{app:fig:bylab7}
\end{figure}
	          
	    \begin{figure}
    \includegraphics[scale=0.85]{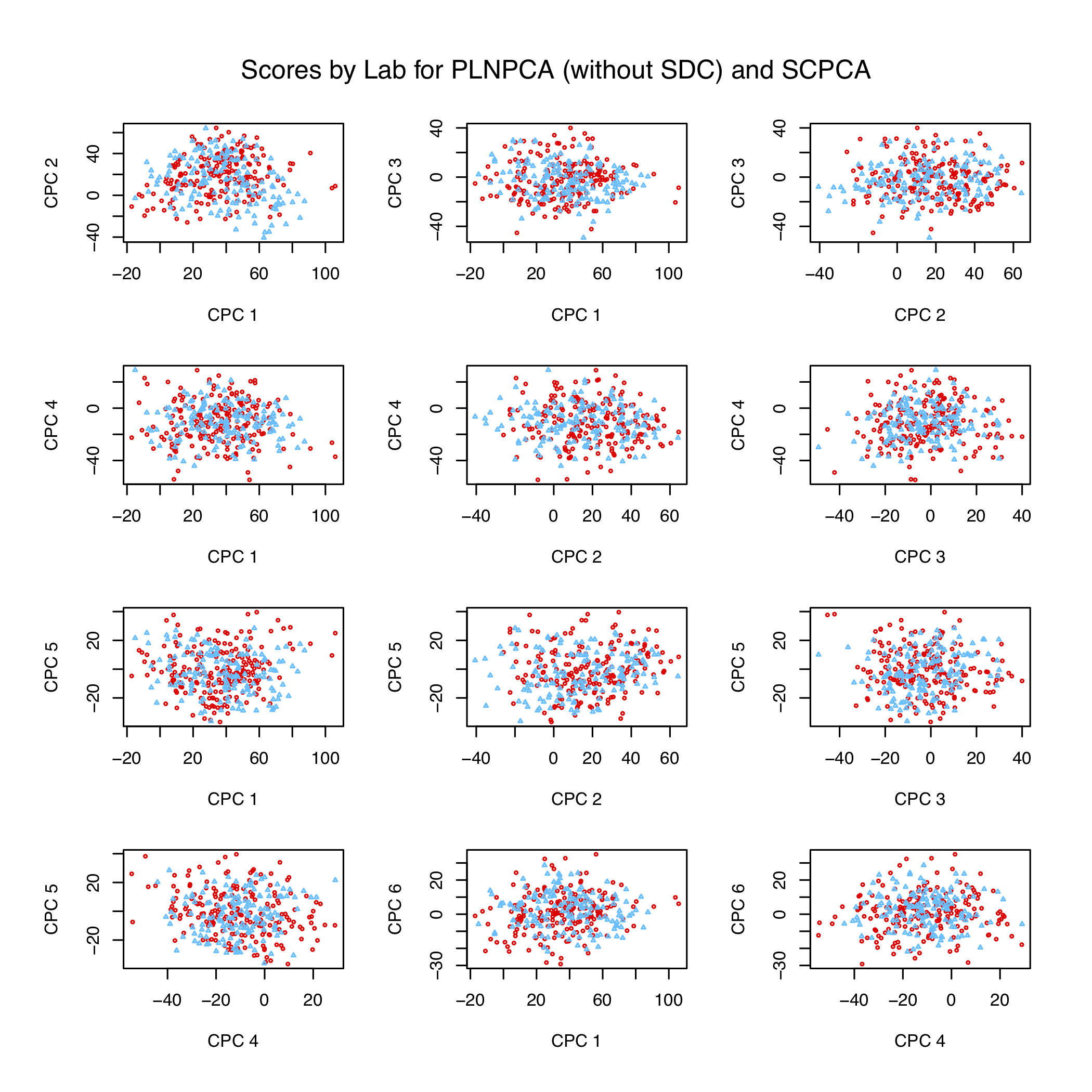} 
    \caption{Scores from PLNPCA (no SDC) \& SCPCA by study of origin.}
    \label{app:fig:bylab8}
\end{figure}
	
		  	    \begin{figure}
    \includegraphics[scale=0.85]{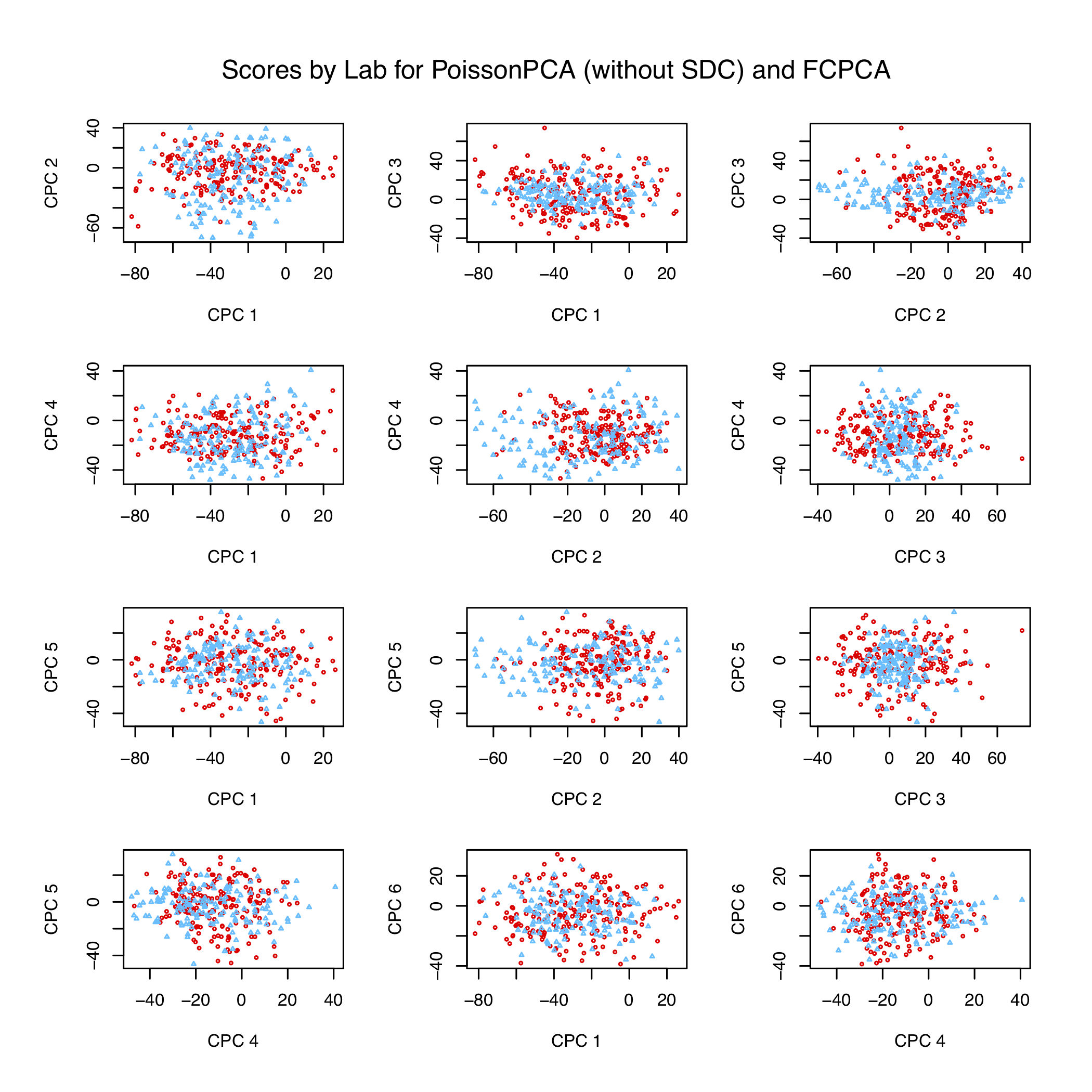}
    \caption{Scores from PoissonPCA (no SDC) \& FCPCA by study of origin.}
    \label{app:fig:bylab9}
\end{figure}
	          
	  	    \begin{figure}
    \includegraphics[scale=0.85]{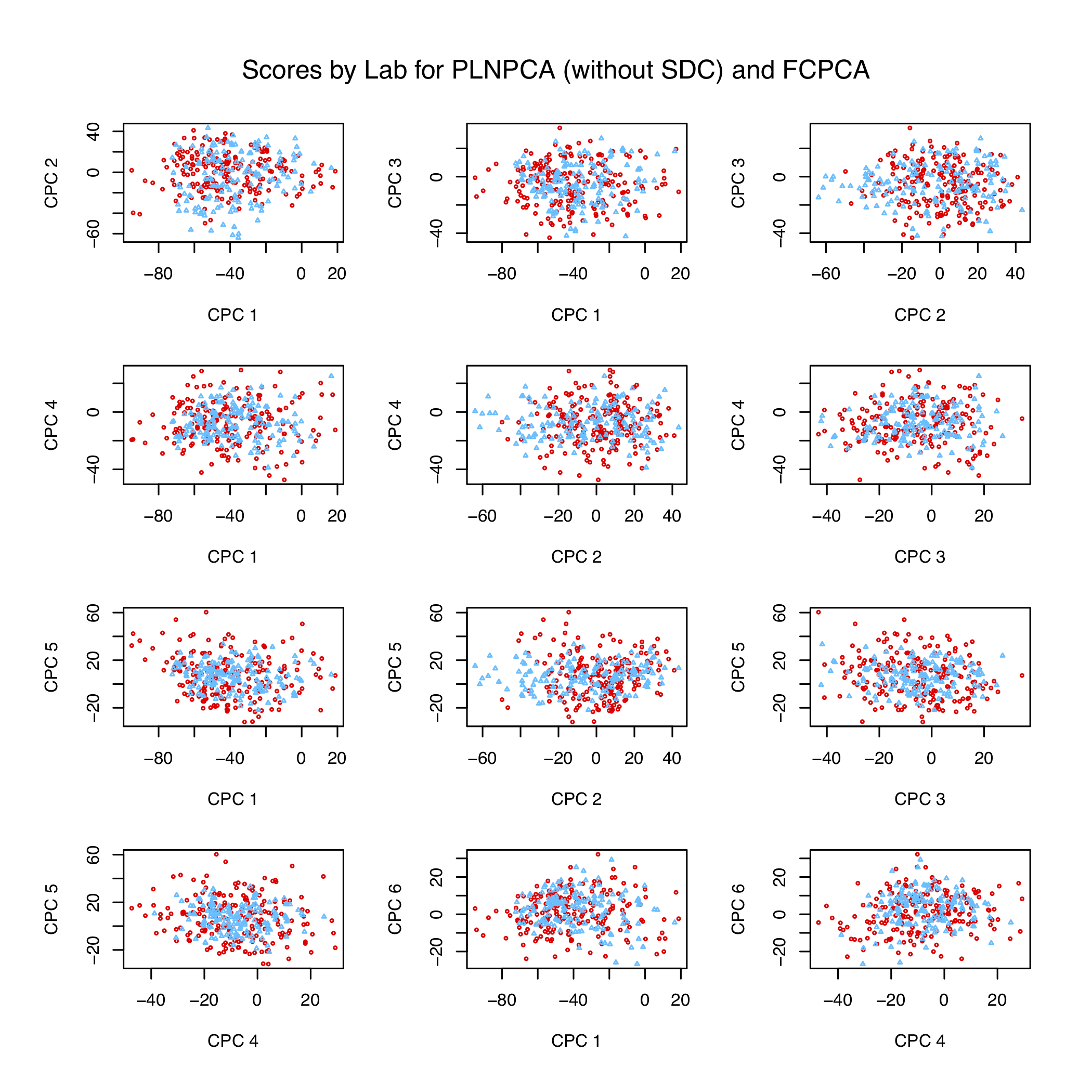}
    \caption{Scores from PLNPCA (no SDC) \& FCPCA by study of origin.}
    \label{app:fig:bylab10}
\end{figure}
	          
	          	 	    \begin{figure}
	    \includegraphics[scale=0.85]{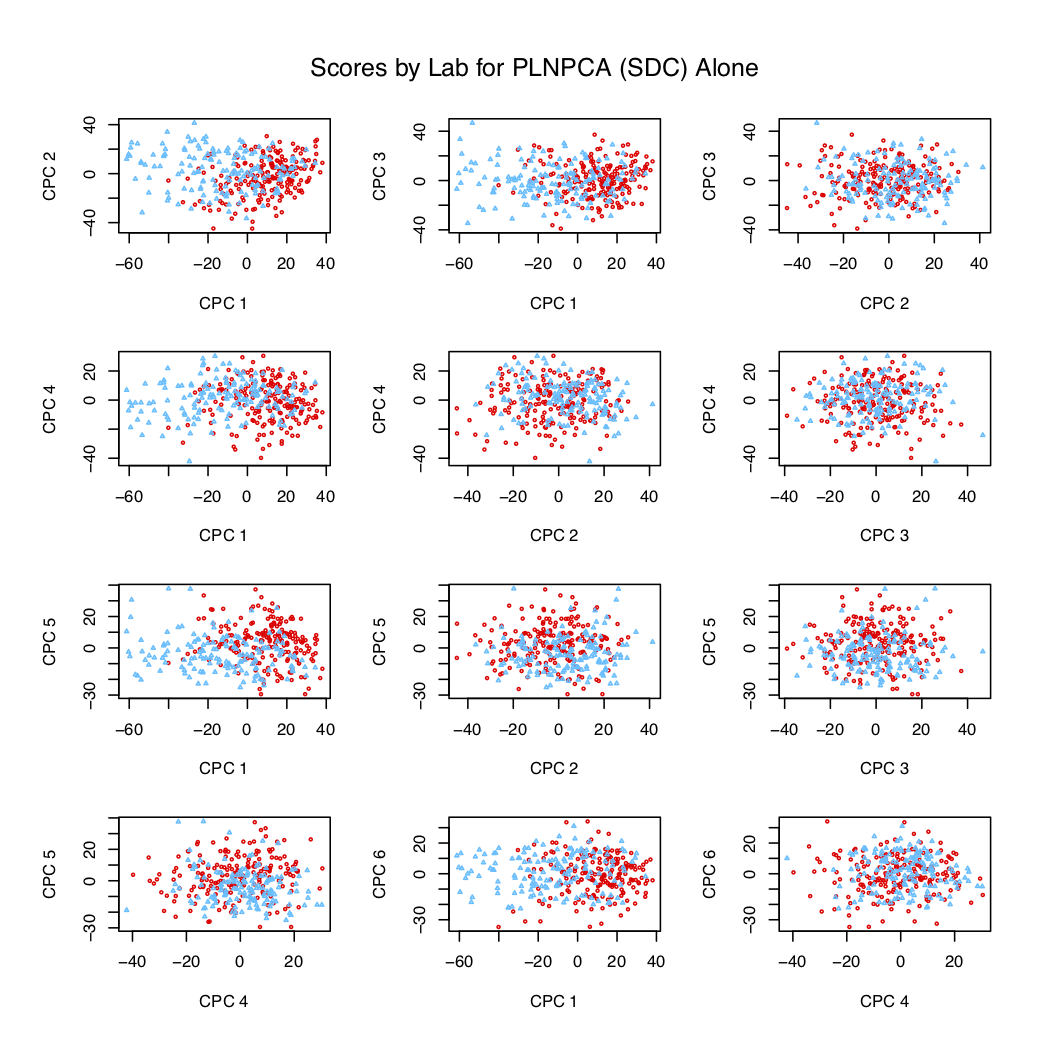} 
    \caption{Scores from PLNPCA alone with SDC by study of origin.}
    \label{app:fig:bylabpln}
\end{figure}

	          	 	    \begin{figure}
	    \includegraphics[scale=0.85]{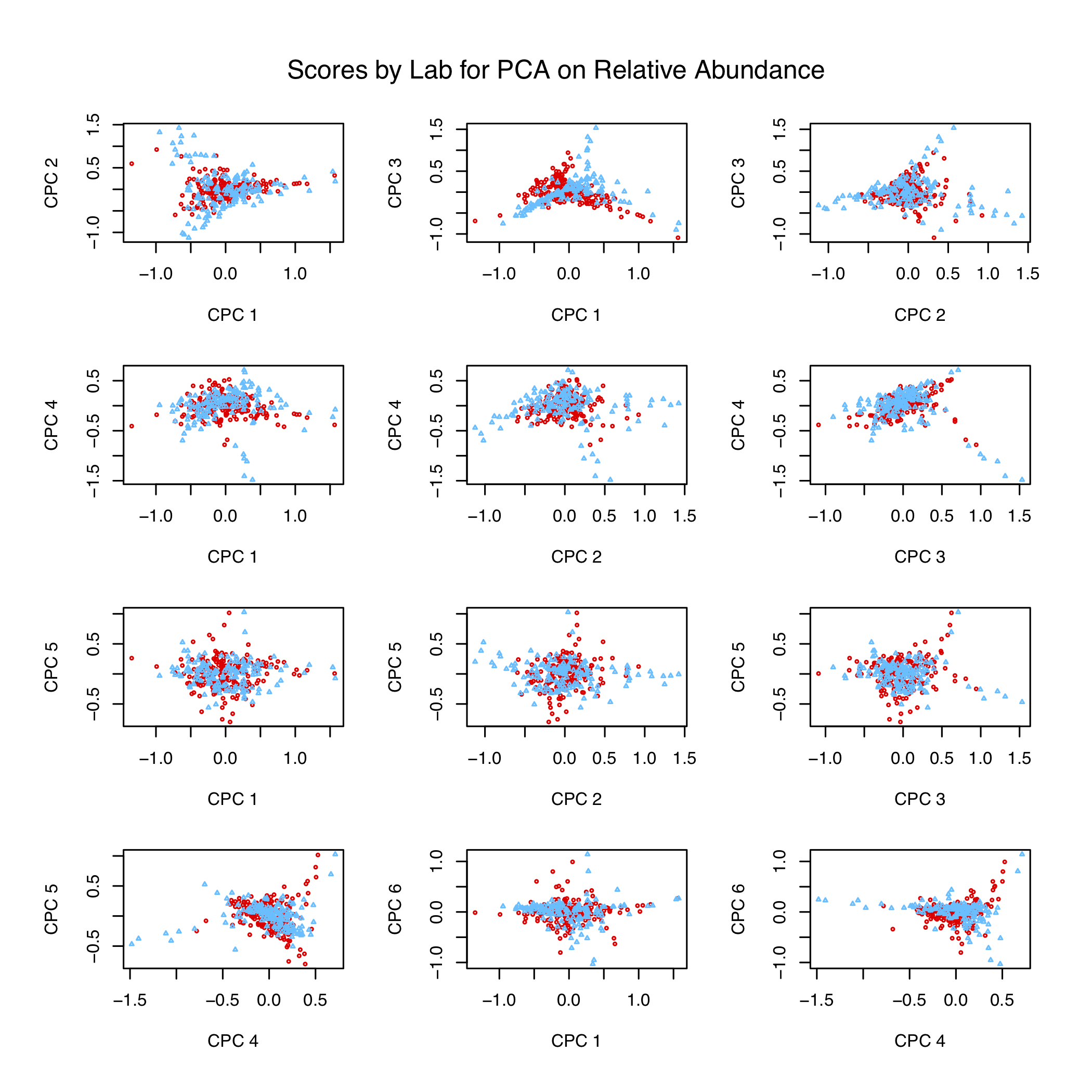} 
    \caption{Scores from PCA of relative abundance by study of origin.}
    \label{app:fig:bylabcent}
\end{figure}
	
	          	 	    \begin{figure}
	    \includegraphics[scale=0.58]{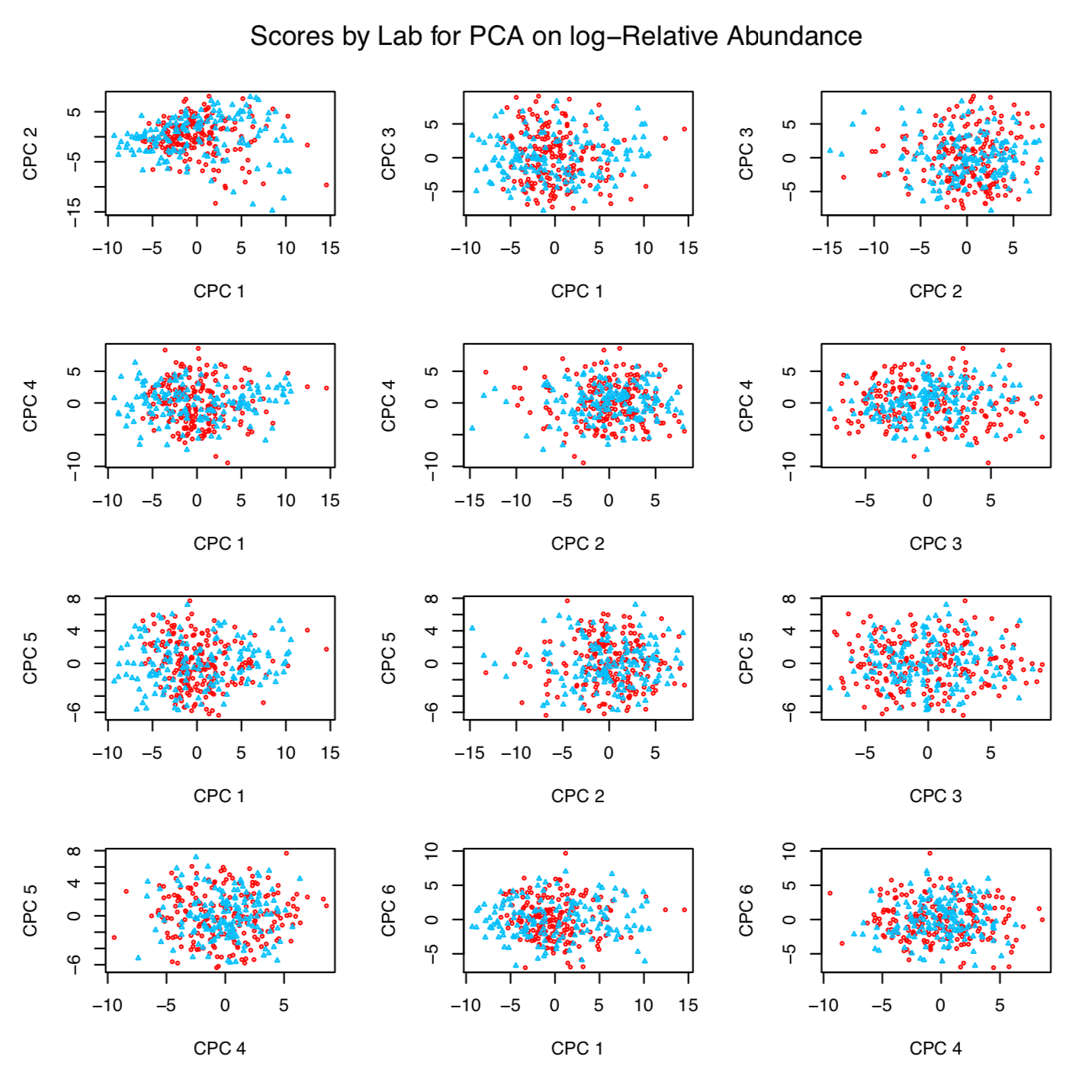} 
    \caption{Scores from PCA of log relative abundance by study of origin.}
    \label{app:fig:bylablogcent}
\end{figure}

	          	 	    \begin{figure}
	    \includegraphics[scale=0.57]{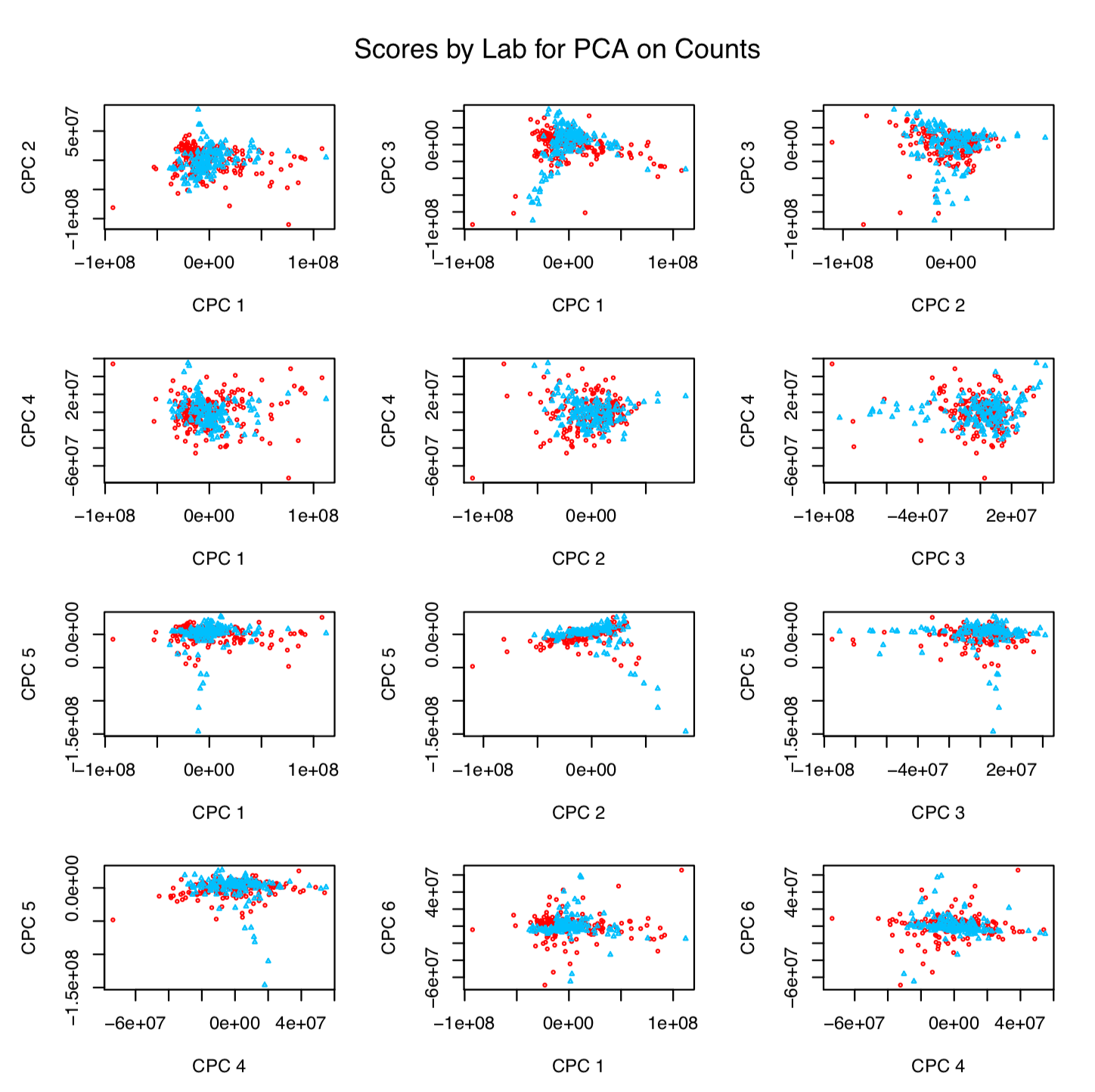} 
    \caption{Scores from PCA of counts by study of origin.}
    \label{app:fig:bylabcount}
\end{figure}
	
	          	 	    \begin{figure}
	    \includegraphics[scale=0.57]{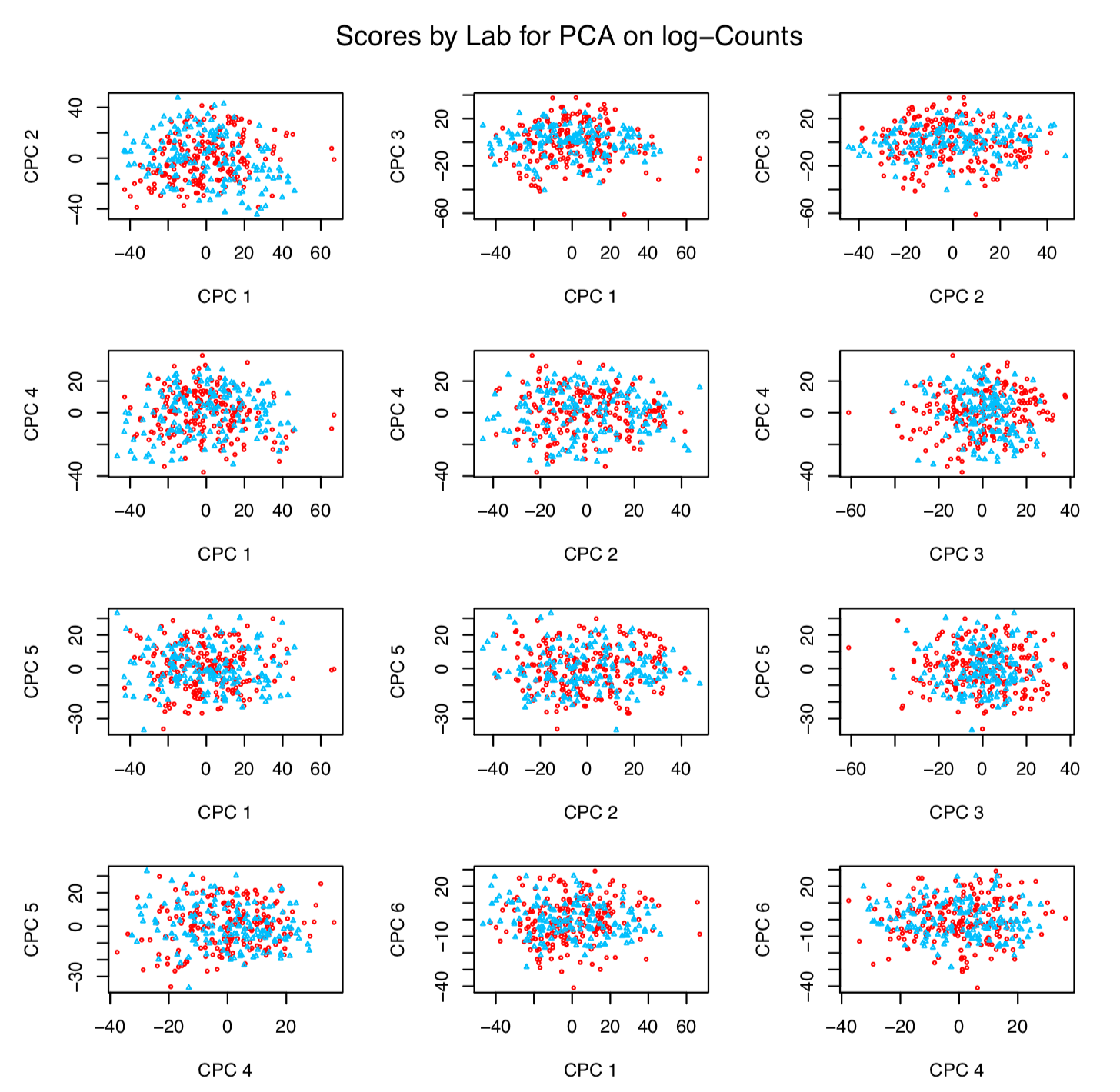} 
    \caption{Scores from PCA of log counts by study of origin.}
    \label{app:fig:bylabcountlog}
\end{figure}
	
\end{appendices}

\end{document}